\documentclass[arguments]{aastex63}

\submitjournal{The Astrophysical Journal}

\shorttitle{MSPES. V. Flux, component spectra \& emission States}
\shortauthors{Basu, Mitra \& Melikidze}

\begin{document}

\title{Meterwavelength Single-pulse Polarimetric Emission Survey. V. Flux density, component spectral variation and emission states}


\author[0000-0003-1824-4487]{Rahul Basu}
\affiliation{Inter-University Centre for Astronomy and Astrophysics, Pune, 411007, India.}
\affiliation{Janusz Gil Institute of Astronomy, University of Zielona G\'ora, ul. Szafrana 2, 65-516 Zielona G\'ora, Poland.}

\author[0000-0002-9142-9835]{Dipanjan Mitra}
\affiliation{National Centre for Radio Astrophysics, Tata Institute of Fundamental Research, Pune 411007, India.}
\affiliation{Janusz Gil Institute of Astronomy, University of Zielona G\'ora, ul. Szafrana 2, 65-516 Zielona G\'ora, Poland.}

\author[0000-0003-1879-1659]{George I. Melikidze}
\affiliation{Janusz Gil Institute of Astronomy, University of Zielona G\'ora, ul. Szafrana 2, 65-516 Zielona G\'ora, Poland.}
\affiliation{Evgeni Kharadze Georgian National Astrophysical Observatory, 0301 Abastumani, Georgia.}

\begin{abstract}
We present the flux density measurements of the pulsars observed in the 
Meterwavelength single-pulse polarimetric emission survey. The average flux 
densities were estimated in 113 pulsars at two frequencies of 325 and 610 MHz 
using interferometric imaging. The average profile and single pulse emission in
each pulsar were calibrated using the estimated flux density. We have used the 
flux calibrated average profile to study the variation of the spectral index
across the emission beam in 21 pulsars where the core, inner cone and the outer
conal components could be clearly identified. The central core component showed
a steeper increase in emission at the lower frequency compared with conal 
emission, with an average difference in spectral index 
$\delta\alpha_{core-cone}\sim-0.7$ between the core and the conal components in
this frequency range. In contrast the inner conal components had positive 
difference in their spectral index compared to the outer cones with average 
difference $\delta\alpha_{in-out}\sim+0.3$. The variation in the spectral index
across the pulse window should provide valuable inputs for constraining the 
radio emission processes. The single pulse emission showed the presence of 
emission mode changing in 12 pulsars with 3 cases where the phenomenon is being
reported for the first time. In addition we have also detected enhanced 
emission for short durations or flaring, in parts or across the entire emission
window in 14 pulsars. The sudden changes in the emission during mode changing 
as well as these bursting states are unrelated to the emission mechanism and 
suggest the presence of rapid and repetitive changes during the plasma 
generation process.
\end{abstract}


\section{Introduction}
\noindent
A systematic study of the radio emission properties from normal period pulsars 
($P>0.05$ seconds) has been initiated in the Meterwavelength Single-pulse 
Polarimetric Emission Survey \citep[MSPES,][]{MBM16}. In this survey polarized
single pulse emission from 123 pulsars were observed with high detection 
sensitivity at two frequencies, 325 and 610 MHz, using the Giant Meterwave 
Radio Telescope (GMRT). Several studies have used this large sample to 
investigate different emission behaviour, like the average profile polarization
behaviour \citep{MBM16}, subpulse drifting \citep{BMM16,BM18a,BMM19}, nulling 
and periodic modulations \citep{BMM17,BMM20}, and period and frequency 
evolution of the profile component widths \citep{SBM18}. In this work we focus 
on other aspects of the emission behaviour related to the flux density ($S$), 
the evolution of the spectral index ($\alpha$) across the emission beam and 
state transitions in the pulsar emission. We present a new measurement 
associated with the MSPES survey, specifically the interferometric estimation 
of the flux density of the pulsars. 

The flux density and the spectral properties provide important inputs 
into our understanding of the underlying physical processes. The majority of 
these measurements have been carried out using telescopes comprising of a 
single dish antenna where the emission is observed with high temporal 
resolution. The radio emission is usually seen as narrow pulses which occupy 
around 10\% of the period and a stable average profile is obtained after 
averaging the single pulses. The flux density of the pulsed emission is 
estimated by characterising the profile baseline level using the telescope 
system specifications. There are also estimates of pulsar flux from radio 
images where the pulsar is seen as an unresolved point source, primarily in 
large sky surveys \citep{KCA98,HT99,K00,FJM16}. In addition interferometric 
imaging is also used when the profile baselines are affected by the pulsed 
emission due to scattering and not clearly distinguished \citep{DBK15,BRK18}. 
The pulsar emission is affected by scintillation in the intervening medium 
which can cause strong fluctuations in the flux density over timescales of 
several minutes to months as well as within the frequency band \citep{S82,R90}.

Pulsars usually show steep power law spectra between 100 MHz and 10 GHz 
frequency range with an average spectral index of $\alpha\sim-1.6$ 
\citep[$S\propto\nu^\alpha$,][]{LYL95,MKK00,JvK18}. However, the spectral 
properties is not uniform across the pulse window. Detailed morphological
studies of the pulsar average profile have shown the emission beam to comprise 
of a central core region surrounded by one or two concentric rings of inner and
outer conal emission \citep{ET_R90,ET_R93,MD99,ET_MR02}. A primary 
distinguishing feature between the core and the conal components within the 
pulsar profile is the evolution of their respective intensities with frequency.
The core component has increasing intensity at lower frequencies suggesting a 
steeper spectra compared to the cones \citep{ET_R83}, but very little work has 
been done to characterise the exact nature of these variations. One notable 
exercise to understand the spectra of the individual components within the 
profile was carried out by \cite{K94}, and did not show any clear difference 
between different components. In this work the profiles were approximated with 
upto five separate Gaussian functions which were identified as components. This
made it difficult to have a direct connection between the core and conal 
emission and the fitted Gaussian components which were often displaced from the
profile components. The large sample in MSPES, with many profiles having both 
core and conal emission, provided an unique opportunity to explore their 
relative spectral evolution at the observing frequencies.

The single pulse emission also exhibit variability over multiple timescales 
due to intrinsic phenomena like mode changing, nulling, subpulse drifting, etc.
We have addressed the effects of nulling and subpulse drifting in the MSPES 
sample in earlier works and explore the mode changing behaviour here. In 
certain pulsars radio emission properties switches between two or more distinct
emission states and the phenomenon is known as mode changing. The most common 
form of mode changing involve change in the average profile, which assume 
different shapes in each mode. This phenomenon was first discovered in PSR 
B1237+25 by \citet{B70b}, which switched between a normal and an abnormal mode.
Other prominent examples of this behaviour was seen in PSR B0329+54 with four 
distinct modes \citep{L71,CWW11}, B0823+26 with the Bright and the Quiet modes 
lasting for hours at a time \citep{SYH15,HKB18,BM19,ROW20}, B1822--09 with the 
Bright mode showing periodic amplitude modulation and a Quiet mode, switching 
within minutes timescales \citep{FMW81,BMR10,LMR12,HKH17}, etc. The profile 
mode changing takes place rapidly, usually within a rotation period of the 
pulsar, however the average profile shape in the individual modes are highly 
stable and the pulsar can continue to reside in a particular mode over 
timescales of minutes to hours. In other cases the mode changing is seen when 
the subpulse drifting behaviour is transformed in the different modes. The 
examples of subpulse drifting associated mode changing include the pulsar 
B0031--07 which shows the presence of three distinct drift modes with different
periodicities \citep{HTT70,VJ97,SMK05,MBT17}, PSR B0943+10 shows the presence 
of a drifting Bright mode and a non-drifting Quiet mode \citep{BMR10,H13}, the 
three drifting modes and one non-drifting state seen in B2319+60 \citep{WF81,
RBMM21}, amongst others. Other forms of mode changing involve transition 
between a normal emission state and a bursting emission state similar to an 
RRAT emission, as seen in PSR J1752+2359 \citep{LWF04}, B0826--34 \citep{ELG05,
EAM12}, B0611+22 \citep{SLR14,RSL16}, as well as presence of flaring events 
preceding the pulse window as seen in PSR B0919+06 and B1859+07 \citep{RRW06,
WOR16,HHP16}.

We have estimated the flux densities in 113 of the 123 pulsars observed in 
MSPES by making detailed images of these sources from the interferometric data, 
which were recorded in conjunction with the high time resolution phased-array 
observations. The estimated flux densities were subsequently used to calibrate 
the intensity level of the single pulse emission as well as the average 
profile. We have investigated the spectral evolution of the core and the conal 
emission in 21 pulsars where both components were present and could be clearly 
separated. We have searched for the presence of mode changing in the single 
pulse emission and found 12 pulsars exhibiting this behaviour, with 3 pulsars 
where this was seen for the first time. We have also characterised an unique 
emission behaviour where either in parts or the entire pulse window the 
intensity is enhanced. This lasts for short durations, usually a single period 
and was seen in 14 pulsars of this sample. In section \ref{sec:obs} we present 
the details of the interferometric observations and the image analysis used to 
estimate the average pulsar flux. Section \ref{sec:flux} presents the estimated
flux density and the details of the single pulse and average profile 
intensities. We have carried out a detailed estimate of the spectral evolution 
across the different profile components in section \ref{sec:comp_spec}. The 
mode changing behaviour is addressed in section \ref{sec:mode}, while section 
\ref{sec:burst} characterises the short duration bursting emission. We conclude
the paper with a short discussion of the results in section \ref{sec:con}.

\section{Observation \& Analysis}\label{sec:obs}

\begin{deluxetable}{cccc|cccc}
\tablenum{1}
\tablecaption{Calibrators used in Imaging\label{tab:cal_list}}
\tablewidth{0pt}
\tablehead{ \multicolumn{4}{c}{\underline{325 MHz Observations}} & \multicolumn{4}{c}{\underline{610 MHz Observations}} \\ 
 \colhead{Flux Cal} & \colhead{Flux}  & \colhead{Phase Cal} & \colhead{Avg. Flux} & \colhead{Flux Cal} & \colhead{Flux} & \colhead{Phase Cal} & \colhead{Avg. Flux} \\
   & \colhead{(Jy)} &   & \colhead{(Jy)} &   &  \colhead{(Jy)} &   & \colhead{(Jy)}}

\startdata
 3C147 &  54.0 & 0025-260 & 24.7 & 3C147 & 39.0 & 0025-260 & 14.5 \\
 3C286 &  26.4 & 0116-208 & 11.5 & 3C286 & 21.7 & 0116-208 &  8.1 \\
 3C353 & 153.4 & 0318+164 &  8.4 & 3C48  & 30.1 & 0318+164 &  9.4 \\
 3C48  &  43.9 & 0323+055 &  7.3 &   &   & 0521-207 & 4.4 \\
   &   & 0521-207 &  8.9 &   &   & 0521+166 & 13.7 \\
   &   & 0521+166 & 15.6 &   &   & 0706-231 &  6.1 \\
   &   & 0804+102 &  5.6 &   &   & 0804+102 &  3.5 \\
   &   & 0837-198 & 12.0 &   &   & 0837-198 &  5.8 \\
   &   & 1021+219 &  6.1 &   &   & 1021+219 &  3.0 \\
   &   & 1033-343 &  4.8 &   &   & 1033-343 &  3.4 \\
   &   & 1156+314 &  5.3 &   &   & 1154-350 &  9.5 \\
   &   & 1311-222 & 16.7 &   &   & 1311-222 &  8.9 \\
   &   & 1510-057 &  7.4 &   &   & 1330+251 &  8.2 \\
   &   & 1625-311 &  3.3 &   &   & 1510-057 &  3.4 \\
   &   & 1822-096 &  3.6 &   &   & 1625-311 &  2.4 \\
   &   & 1830-360 & 20.7 &   &   & 1714-252 &  5.5 \\
   &   & 1924+334 &  6.5 &   &   & 1822-096 &  6.5 \\
   &   & 2047-026 &  6.1 &   &   & 1830-360 & 16.1 \\
   &   & 2137-207 &  8.3 &   &   & 1924+334 &  6.3 \\
   &   & 2225-049 & 16.8 &   &   & 2047-026 &  4.2 \\
   &   & 2251+188 &  7.9 &   &   & 2137-207 &  7.0 \\
   &   &   &   &   &   & 2219-279 & 4.6 \\
   &   &   &   &   &   & 2251+188 & 4.0 \\
\enddata
\end{deluxetable}

The GMRT consists of an Y-shaped array of 30 antennas of 45 meter diameter, 
with 14 antennas located within a central one square kilometer area and the 
remaining 16 antennas placed along the three arms which has a maximum 
distance of 27 kilometers. The MSPES observations were carried out at two 
different frequencies centered around 325 MHz and 610 MHz with a bandwidth of 
16.67 MHz at each frequency. The pulsars were observed simultaneously in the 
`phased-array' mode and the `interferometer' mode. In the phased-array mode 
the signal from each antenna was co-added in phase to generate high time 
resolution data (around 0.5 milliseconds) with four polarizations. A bright 
phase calibrator near the pulsar was observed before the start of the 
observations to correct for arbitrary phase shifts in each antenna. The time 
series observations from each pulsar were converted to a series of single 
pulses, averaged over the frequency band, after correcting for dispersion 
delays due to propagation in the interstellar medium and removing terrestrial 
radio frequency interference (RFI). The details of the analysis for obtaining 
the single pulse sequence from each pulsar is described in \citet{MBM16} and 
\citet{BMM16}. In the interferometric mode the cross-correlated visibility from
each antenna pair, called a baseline, was estimated which measured the 
different angular scales of the intensity pattern within the field of view of 
the telescope. The GMRT with 30 antennas can measure visibilities from 435 
baselines simultaneously in a co-ordinate system known as the $uv$-plane. The 
$uv$-plane gets filled up due to earth's rotation, which is known as aperture 
synthesis, and an image of the observed sky intensity distribution is obtained 
after adding the visibilities from a sufficiently filled $uv$-plane in their 
Fourier domain. Similar to the phased-array observations, bright nearby 
calibrators were observed at regular intervals to address the amplitude and 
phase variations in the antenna response. In the phased-array observations the 
measured signal had a baseline level which was a sum of the antenna systematics
(T$_{sys}$) as well as the sky background level (T$_{sky}$). In contrast the 
cross-correlation ensured that the mean background level of the images from the
interferometer was at zero, and the systematic fluctuations were reflected in 
the noise rms. As a result it was more reliable to estimate the flux 
density of the sources from the interferometric images.

The observations were carried out between February, 2014 and July, 2014 over
several sessions lasting usually 8-10 hours at a time. Standard flux 
calibrators 3C147, 3C48 and 3C286 were used during each observing session to 
estimate the flux scale. On one occasion the non-standard calibrator 3C353 was
also used due to the absence of the others within the observing window. These 
were used to calibrate the flux levels of the phase calibrators which were 
nearby the intended source and were observed for short durations of 1-2 minutes
before and after each pulsar observation. Table \ref{tab:cal_list} lists all 
the flux and phase calibrators used during these observations. The image 
analysis were carried out following standard imaging techniques using 
Astronomical Image Processing System (AIPS). The flux levels of the flux 
calibrators at each observing frequency were obtained from \cite{PB17} and are 
listed in table \ref{tab:cal_list}. The same phase calibrators were used for 
multiple nearby pulsars over different observing sessions and their average 
flux densities are also shown in table \ref{tab:cal_list}. In around 10-20 
percent of times at either frequency it was not possible to estimate the flux 
levels of the phase calibrators independently, due to unavailability of any 
flux calibrator within the observing duration, and the average flux level was 
used for calibrating the pulsar intensity. The details of the calibrators used 
during each observing session and the names of the corresponding pulsars they 
calibrated are reported in table \ref{tab:cal_flux} in appendix. The flux 
levels of the phase calibrators over different observing sessions showed 
variations that were within 10 percent of the mean level, which served as a 
consistency check for the flux calibration process. We were primarily 
interested in the flux measurement of pulsars which were unresolved point 
sources at the spatial resolutions of GMRT, 8"$\times$8" at 325 MHz and 
4"$\times$4" at 610 MHz, and had contributions at all baselines. A low baseline
cutoff of 2 k$\lambda$ in the $uv$-range was applied with only higher baselines
used for imaging process to reduce the effect of RFI which mostly affected 
shorter baselines. In some cases the pulsar intensity was affected by 
scintillation, that introduced phase errors around the source and increased 
noise rms.

\section{Average Flux Density and Calibrating the pulsed emission}\label{sec:flux}

\begin{figure}
\gridline{\fig{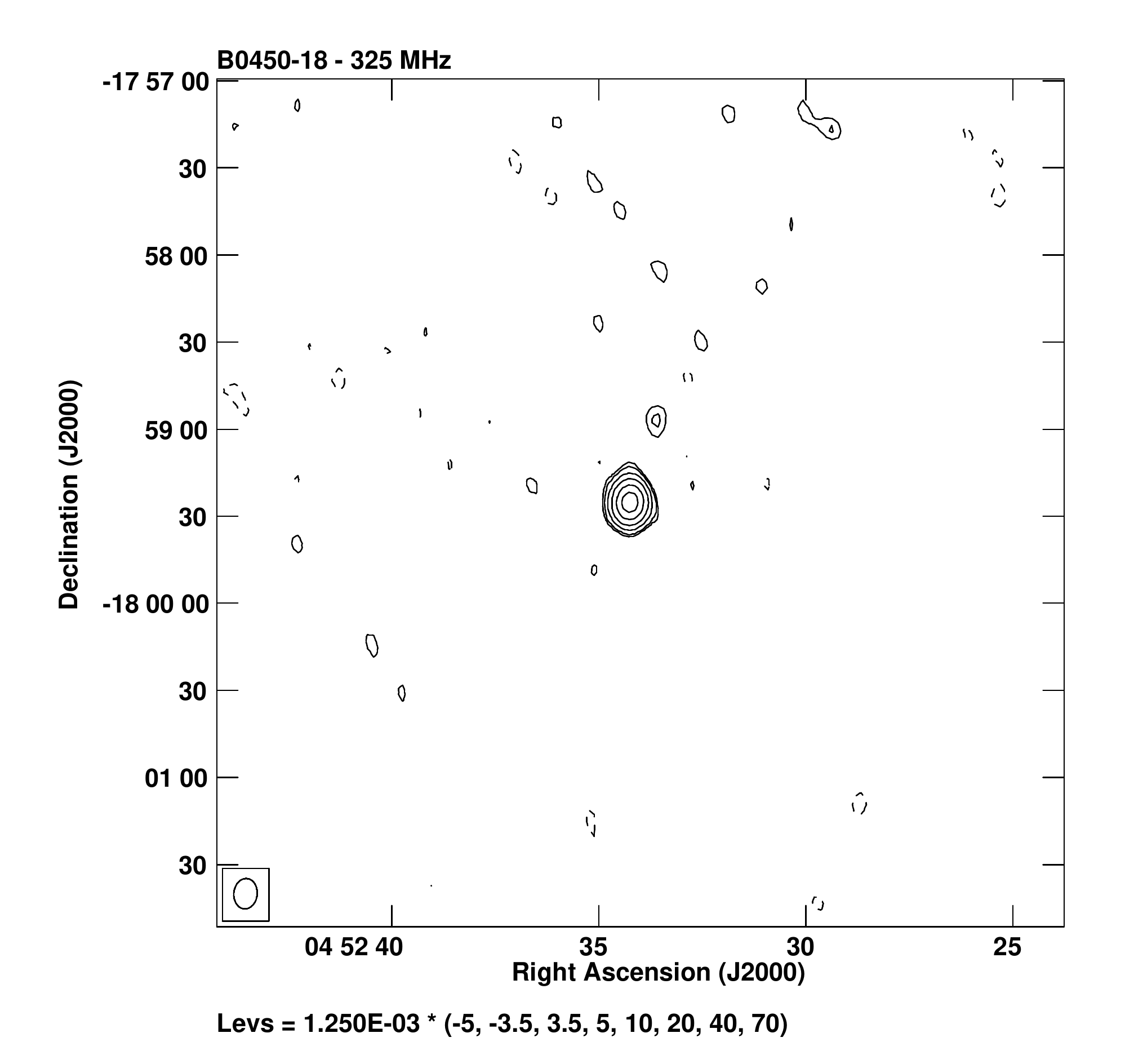}{0.45\textwidth}{(a)}
          \fig{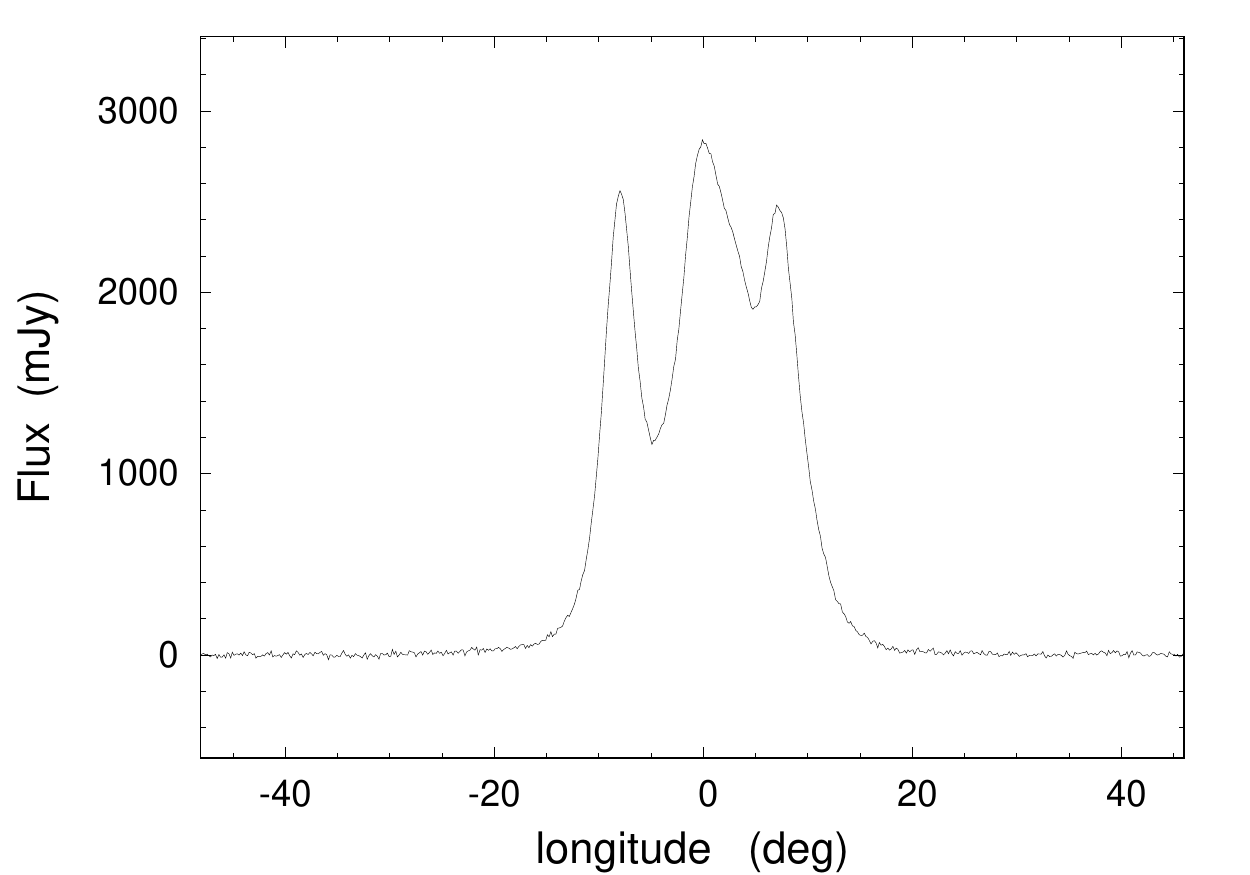}{0.45\textwidth}{(b)}
         }
\caption{(a) The contour image of the field containing the pulsar B0450-18
at 325 MHz, where the contour levels (Levs) are in units of Jy. The pulsar is 
seen as a point source in the field center with an estimated flux density of 
124.9$\pm$5.0 mJy. The synthesized beam corresponding to the spatial resolution
of the telescope is shown in the bottom left corner. (b) The average pulsar 
profile with three distinct components, where the intensity levels have been 
calibrated using the average flux density from the image. The flux density of 
the profile peak is 2841 mJy.
\label{fig:imgflux}}
\end{figure}

The image analysis was carried out in 113 pulsars, with 94 pulsars measured at
325 MHz, 97 pulsars at 610 MHz and 78 pulsars at both frequencies. The average 
flux densities measured from the maps are listed in table \ref{tab:flux}, along
with the noise rms levels around the pulsar. An example of the contour image 
corresponding to PSR B0450--18 is shown in figure \ref{fig:imgflux}a, where the
pulsar is seen as an unresolved point source in the center. There were seven
pulsars from the MSPES list where interferometric measurements were not carried
out, while an additional three pulsars J0905--4536, B1325--43, and B1634--45 
were not detected in the images. The duration of observations varied depending 
on the pulsar period, with around 2100 single pulses observed in each case. The
noise rms in the maps were between 0.2 -- 11 mJy at 325 MHz with the typical 
value around 1 mJy. At 610 MHz the noise levels varied between 0.1 -- 3 mJy and
usually were around 0.5 mJy. The pulsars were mostly observed on a single 
occasion at either frequency. However, there were issues with the instrumental 
setup for single pulse polarization at the lower frequency which required 
re-observations in around 25 \% times (23 cases) at 325 MHz. In addition the two
pulsars B0950+08 and B1929+10 were used as polarization calibrators with one or
both observed during each observing session. PSR B0950+08 was observed on 6 
separate days for a total of 7 different measurements at 325 MHz and had 4 
different measurements across same number of days at 610 MHz. The pulsar 
B1929+10 was observed on 10 observing sessions for a total of 12 different 
measurements at 325 MHz and on 8 sessions for a total of 12 measurements at 610
MHz. 

The pulsar observations were affected by scintillation with diffractive 
scintillation dominating for pulsars with low dispersion measure which was the
case for the majority of the sample \citep{R90}. We estimated the approximate 
modulation index for the pulsars in our sample, over the observing durations 
and bandwidths, from the galactic free electron density model of \cite{CR02}, 
which were more than 1~in almost all cases. As a result the measured flux 
densities of the pulsars, as shown in table \ref{tab:flux}, can show large 
deviations from the previously reported values in the literature. This is 
further highlighted in the case of PSR B0950+08 and B1929+10 which were 
observed on multiple occasions and showed large variations. For example PSR 
B1929+10 had flux densities between 70 to 390 mJy at 325 MHz over the different
observations. Due to the expected flux variability we have not carried out any 
absolute spectral index measurements from these observations. 

The images measure the average flux over the duration of the observations. When
the profile baseline mean level is at zero, the period averaged intensity is 
equivalent to the flux density measured from the images. As a result once the 
baseline level is appropriately subtracted from the profile, a scaling relation
can be obtained for the measured counts in the phased-array. The pulsar profile
was calibrated in all cases where the flux density was available. The single 
pulse intensities were also calibrated to their flux levels. An example of the 
flux calibrated profile of PSR B0450--18 at 325 MHz is shown in 
figure \ref{fig:imgflux}b, where the average flux was 124.9$\pm$5.0 mJy and the
profile had a peak flux level of 2.8$\pm$0.1 Jy. The peak flux level was much 
higher than the average flux since the pulsed emission is only seen for a 
narrow window ($\sim$10\%) of the period. The flux levels of the profile peaks 
are reported in table \ref{tab:flux}. 

\begin{figure}
\gridline{\fig{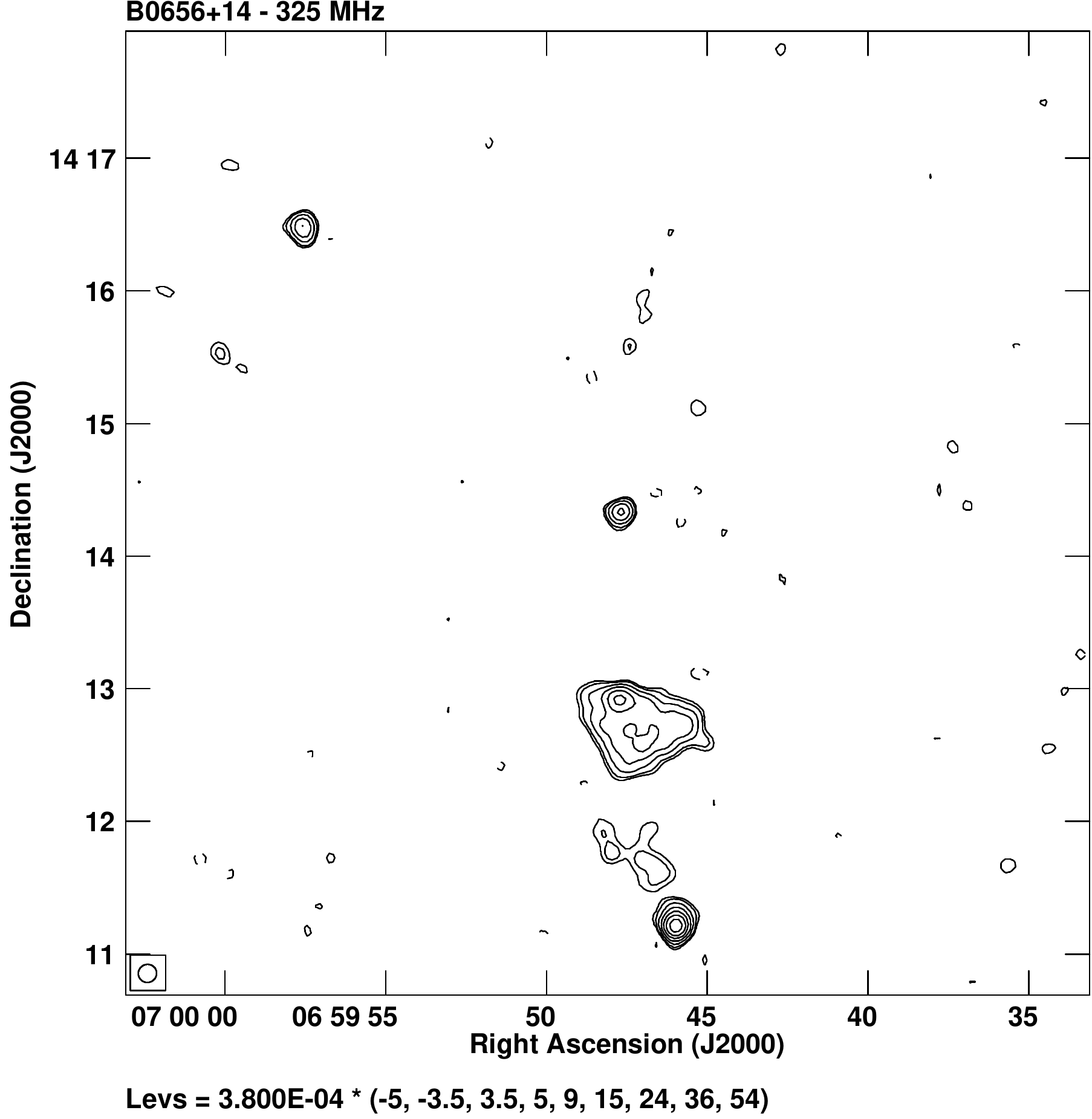}{0.34\textwidth}{(a)}
          \fig{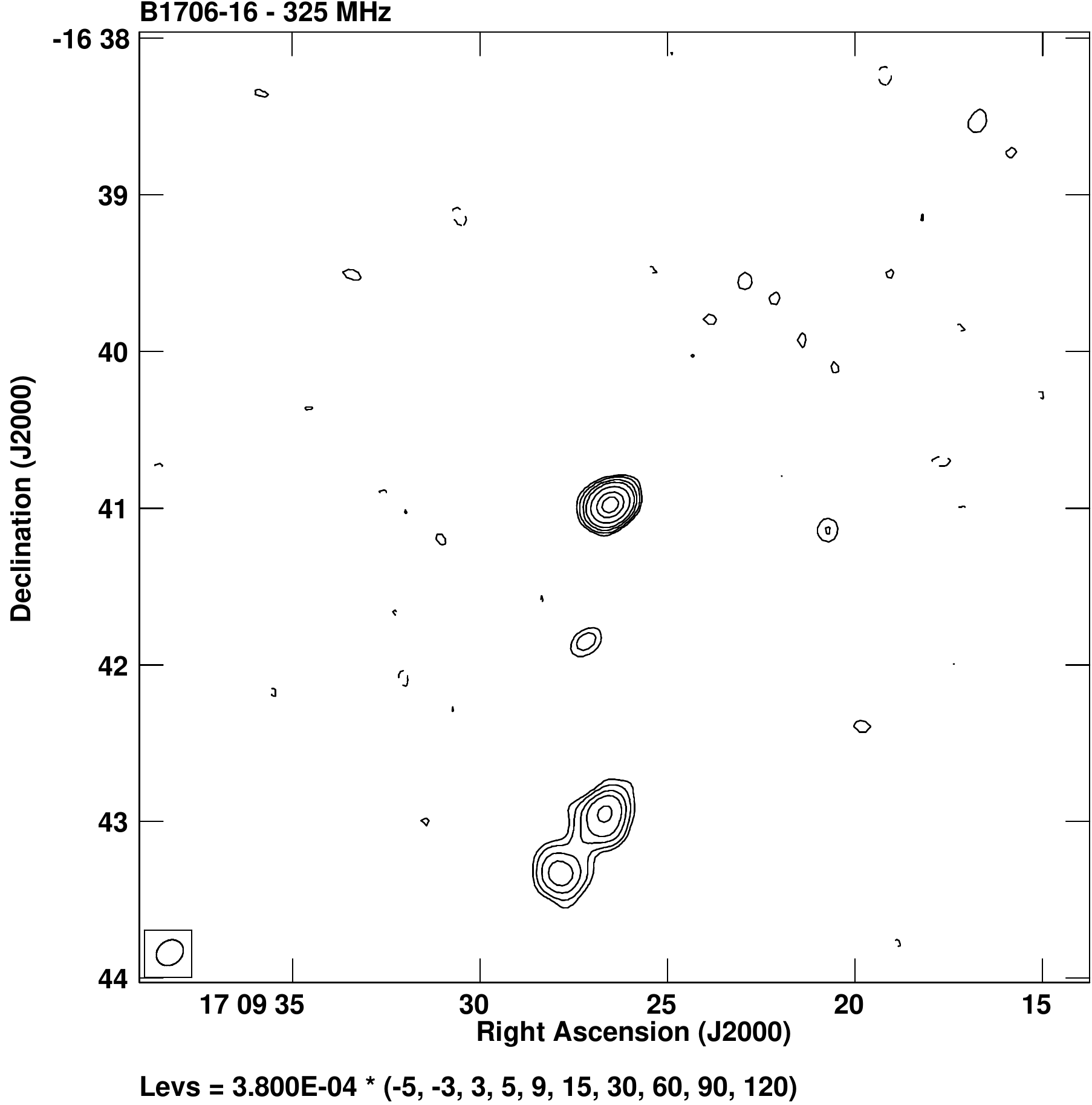}{0.34\textwidth}{(b)}
         }
\gridline{\fig{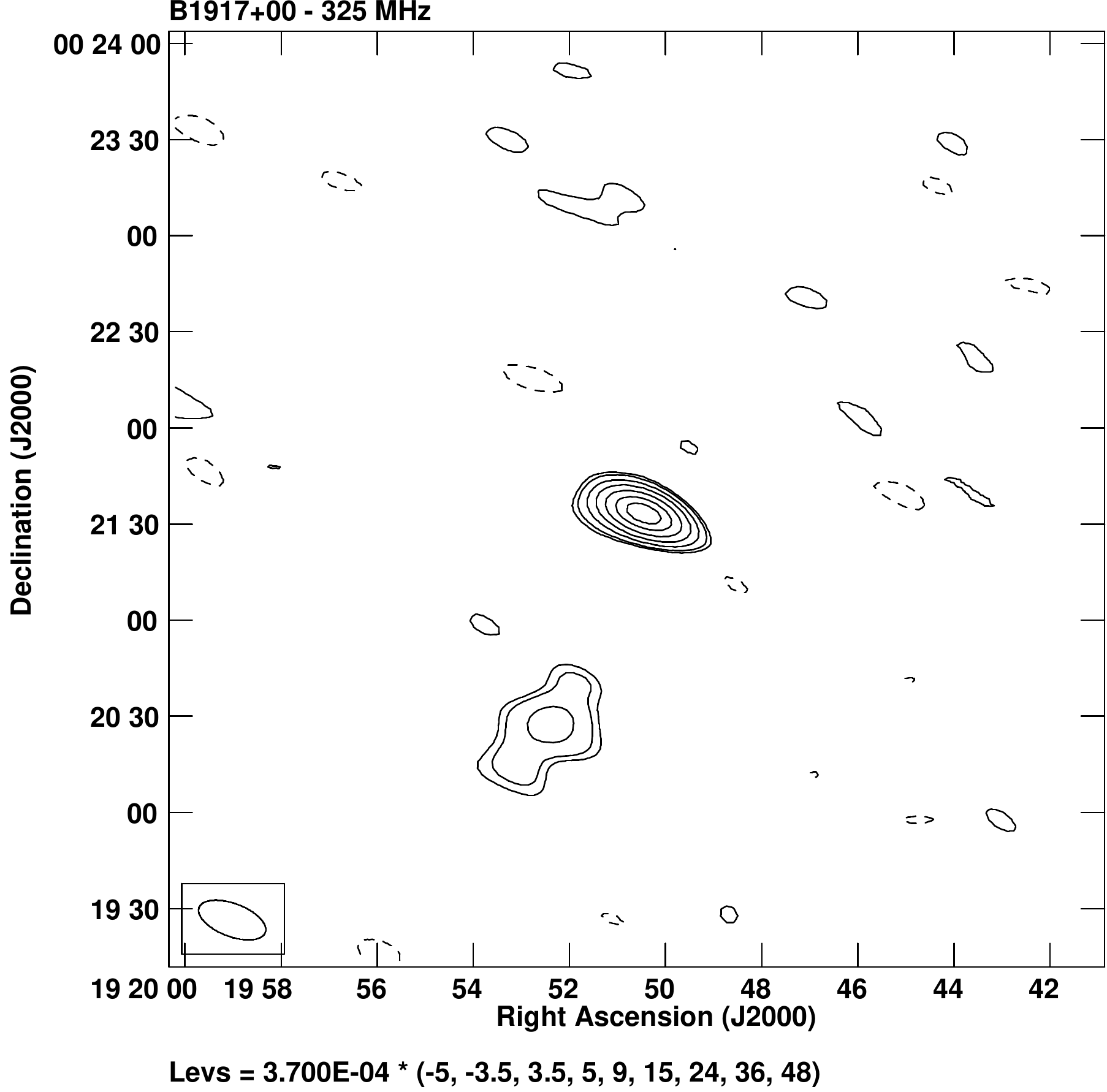}{0.34\textwidth}{(c)}
          \fig{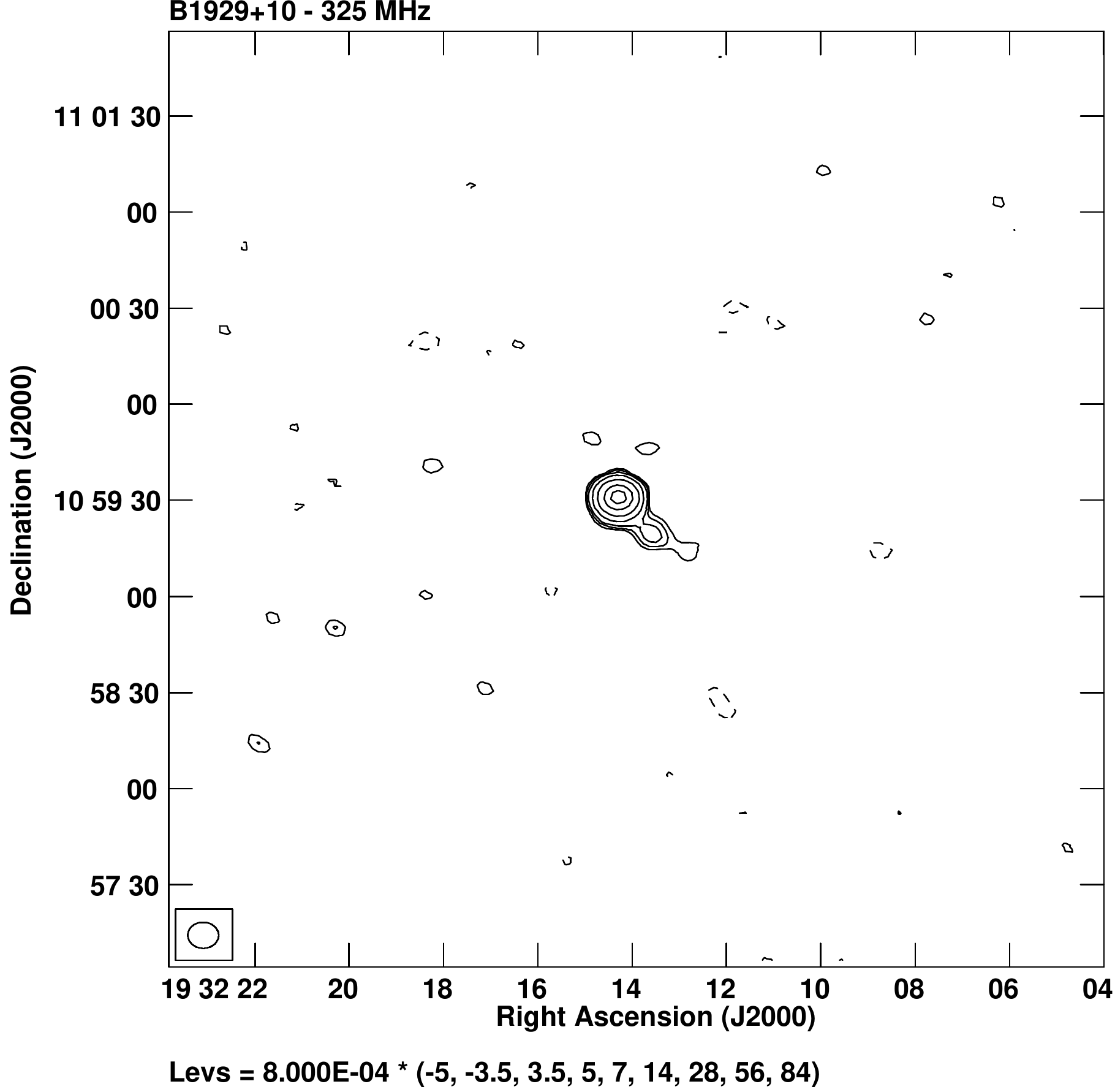}{0.34\textwidth}{(d)}
         }
\gridline{\fig{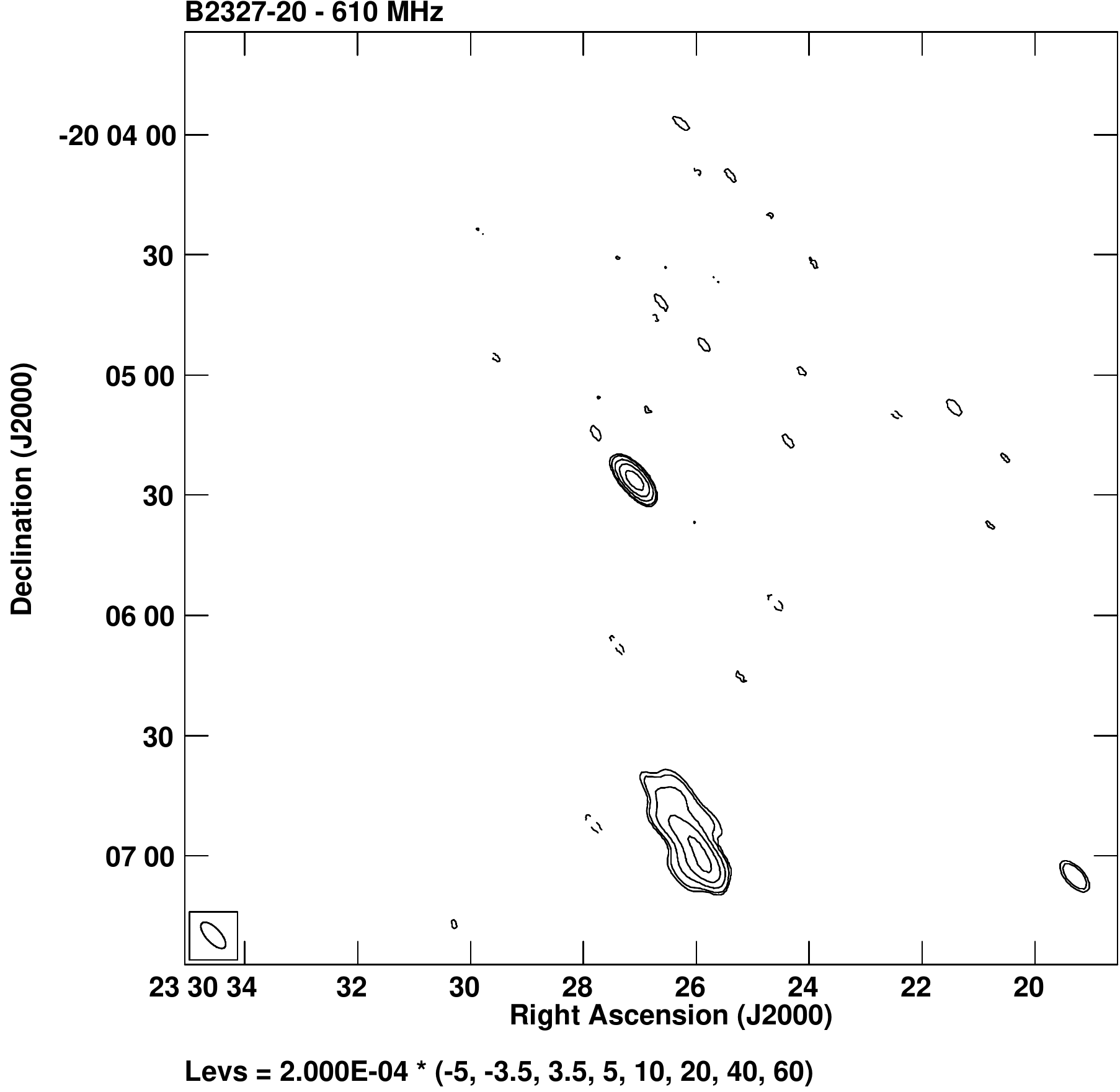}{0.34\textwidth}{(e)}
         }
\caption{The figure shows the fields of pulsars with interesting sources in 
their immediate environment. The total intensity contour levels (Levs) are
plotted in units of Jy and the pulsar is seen as a point source in field 
center. The telescope resolution is shown as the synthesised beam in the bottom
left corner, and is around 8"$\times$8" at 325 MHz and 4"$\times$4" at 610 MHz.
(a) PSR B0656+14 at 325 MHz. (b) PSR B1706-16 at 325 MHz. (c) B1917+00 at 325 
MHz. (d) B1929+10 at 325 MHz. (e) B2327-20 at 610 MHz.
\label{fig:psrenv}}
\end{figure}

The MSPES studies were primarily concerned with the single pulse behaviour of 
older and middle aged pulsars with characteristic ages of 10$^6$ years and 
older. These are mostly isolated pulsars that have moved away from the 
progenitor supernova remnant, most of which is likely to be dissipated. We have
searched for the presence of extended sources within a 20 arcminutes radius of 
the central pulsar. We found five pulsars with interesting emission features as
shown in figure \ref{fig:psrenv}. We briefly describe below the nature of the 
emission from these sources. It should be noted that due to the 2 k$\lambda$ 
cutoff in the $uv$-plane the images were not sensitive to detect large extended
structures with size in excess of 5 arcminutes at 325 MHz. \\

\emph{B0656+14} : An extended emission feature was seen below the pulsar which
had been identified as a pulsar wind nebula in previous studies \citep{CHMM89}.
However, no clear connection could be made between this extended source and the
pulsar \citep{BKBM99}. The emission was broadly divided into two parts, a 
triangular structure with a bright head, resembling a bow shock, located around
85 arcseconds below the pulsar and a narrow tail ending in a bright point like 
structure which maybe an unrelated background source. The separation between 
the bright regions at the head and tail of the source was around 105 
arcseconds. The flux density of the extended emission at 325 MHz was 167.5 mJy 
with the point source at the end of the tail having flux density of 34.3 mJy. 
The tail connecting the point source at the end to the main body was not seen 
at 610 MHz. The extended emission had flux density of 62.9 mJy at 610 MHz while
the point source at the end had flux density of 21.6 mJy. \\

\emph{B1706--16} : An extended double lobed structure was seen at a distance of 
around 120 arcseconds below the pulsar. The extended emission had physical size
of around 20 arcseconds and flux density at 325 MHz was 58.4 mJy. The 
interferometric observations were not carried out at 610 MHz for this pulsar.\\

\emph{B1917+00} : An elongated diffuse emission feature was detected at both 
frequencies near the bottom left of the pulsar and pointing towards it. The 
nearest detectable emission was at a distance of 51 arcseconds from the pulsar
at both frequencies. The emission was only partially seen at 325 MHz due to the
higher noise levels in the map and was between 30-40 arcseconds in size. The 
total flux density was around 8.9 mJy. At 610 MHz the emission had physical 
extent of 55-60 arcseconds and the estimated flux density was 9.2 mJy. \\

\emph{B1929+10} : The presence of a diffuse elongated tail opposite to the 
transverse motion of the pulsar had been reported in previous studies. These 
detections were primarily in X-rays with no clear counterparts seen at radio 
wavelengths \citep{WLB93,BKJT06}. The primary difficulty with detecting diffuse
radio emission in the vicinity of this source is the inherently high flux 
density of the pulsed emission which makes the noise levels around the source 
high due to phase errors. The pulsar was used as a polarization calibrator and 
observed multiple times at both frequencies. However, due to the large flux 
density of the pulsar it was not possible to detect any clear diffuse emission 
associated with it on most days. On one or two occasions when the pulsar flux 
was lower due to scintillation, the elongated tail could be seen (see figure 
\ref{fig:psrenv}d). The tail was seen upto a distance of around 16.5 arcseconds
from the pulsar and had flux density of around 16.3 mJy at 325 MHz. On visual 
inspection it seemed that the orientation of the tail roughly matched the X-ray
counterpart. No clear measurements were possible at 610 MHz. \\

\emph{B2327-20} : An elongated source was seen below the pulsar at a distance 
of around 90 arseconds. The extended structure was roughly elliptical in shape 
with major axis of around 32 arcseconds and a minor axis of 16 arcseconds. The
estimated flux density was around 39.4 mJy at 610 MHz. Interferometric 
measurements of this pulsar was not carried out at 325 MHz. \\

\section{Spectral evolution across the emission beam}\label{sec:comp_spec}

\begin{figure}
\gridline{\fig{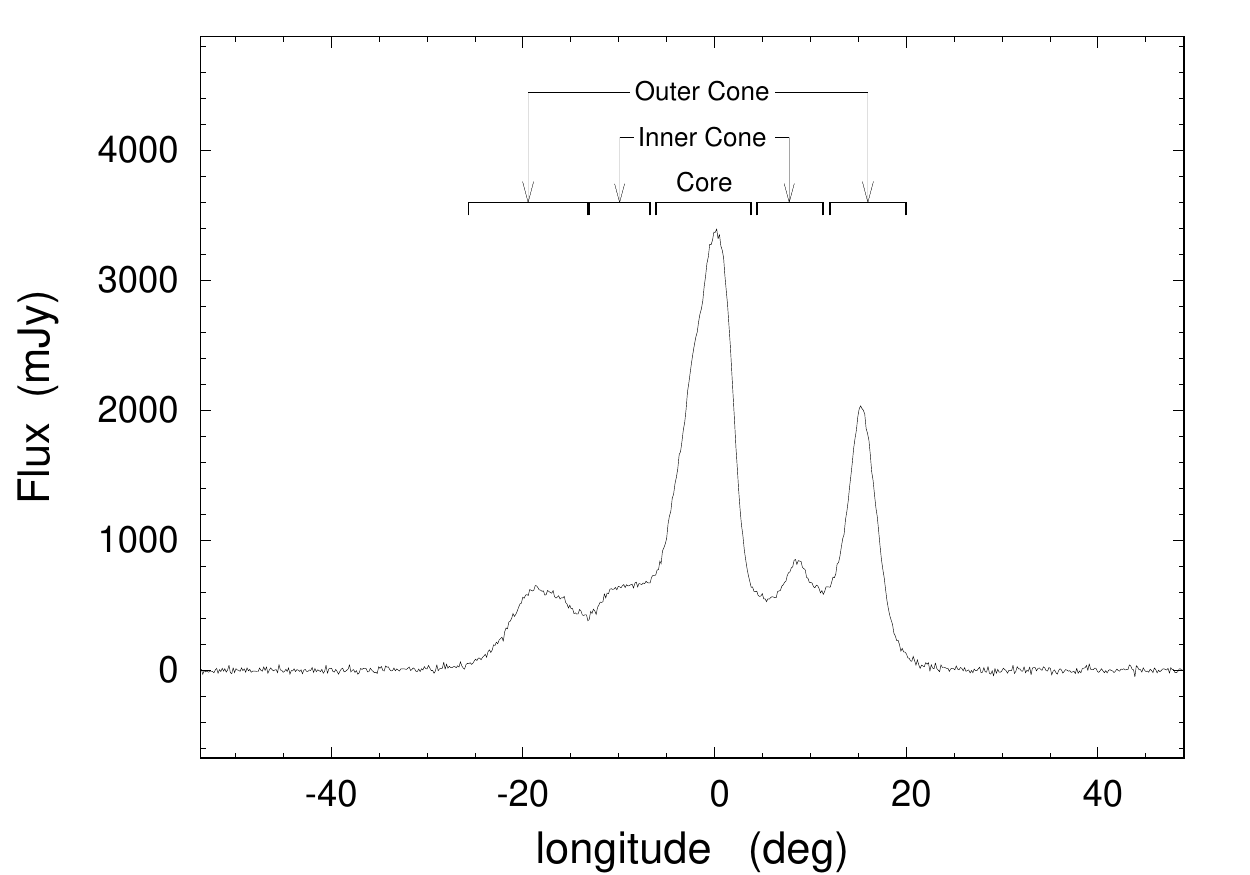}{0.5\textwidth}{(a)}
          \fig{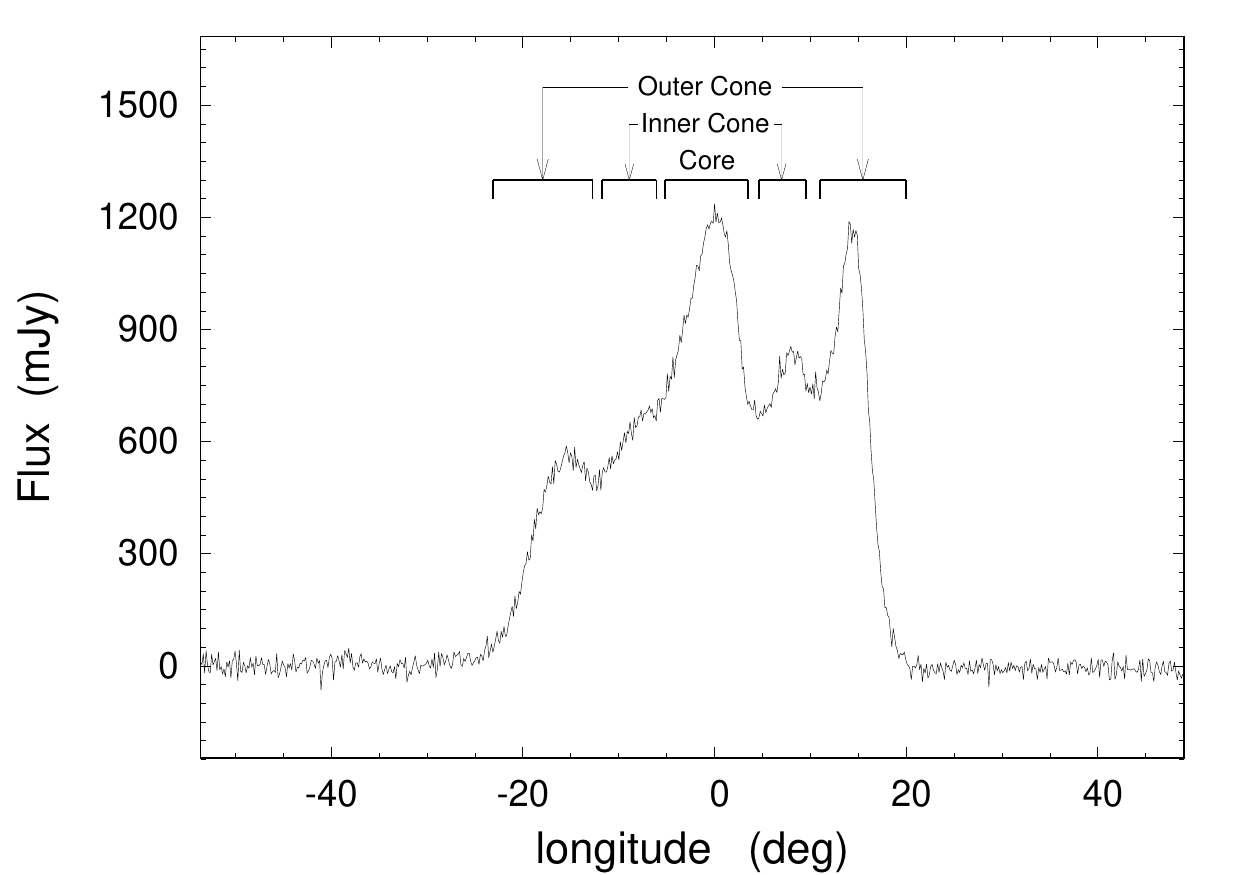}{0.5\textwidth}{(b)}
         }
\caption{The figure shows the average profile of PSR B1857--26 with five 
components at (a) 325 MHz and (b) 610 MHz. The central core emission, the inner
and outer cones are identified in each panel. The components show different 
spectral evolution, with the core emission having the steepest spectra, while 
the outer cone shows higher relative intensity at the lower frequency compared
to the inner cones.
\label{fig:compspec}}
\end{figure}

\begin{deluxetable}{ccccccc}
\tablenum{3}
\tablecaption{Spectral Difference across the profile\label{tab:comp_spect}}
\tablewidth{0pt}
\tablehead{  & \colhead{PSR} & \colhead{Type} & \colhead{$\Delta\alpha_{core-cone}$} & \colhead{$\Delta\alpha_{in-out}$} & \colhead{$\Delta\alpha_{core-in}$} & \colhead{$\Delta\alpha_{core-out}$}}

\startdata
 1 &  B0450--18 & T & -0.998$\pm$0.040 & --- & --- & --- \\
 2 &  B0626+24 & T$_{1/2}$ & -0.681$\pm$0.037 & --- & --- & --- \\
 3 &  B0844--35 & $_c$Q & --- & 0.61$\pm$0.14 & --- & --- \\
 4 &  B0919+06 & T & -0.501$\pm$0.044 & --- & --- & --- \\
 5 & J1034--3224 & $_c$Q & --- & 0.423$\pm$0.066 & --- & --- \\
 6 &  B1237+25 & M & -0.341$\pm$0.022 & 0.138$\pm$0.022 & -0.435$\pm$0.046 & -0.297$\pm$0.022 \\
 7 &  B1325--49 & M & -0.279$\pm$0.044 & 0.120$\pm$0.044 & -0.36$\pm$0.14 & -0.244$\pm$0.044 \\
 8 &  B1700--32 & T & -0.409$\pm$0.040 & --- & --- & --- \\
 9 &  B1732--07 & T & -0.911$\pm$0.040 & --- & --- & --- \\
10 &  B1737+13 & M & -0.813$\pm$0.051 & 0.568$\pm$0.069 & -1.169$\pm$0.057 & -0.601$\pm$0.051 \\
11 &  B1738--08 & $_c$Q & --- & 0.363$\pm$0.095 & --- & --- \\
12 &  B1745--12 & $_c$T/$_c$Q & --- & 0.112$\pm$0.033 & --- & --- \\
13 &  B1758--29 & T & -0.64$\pm$0.18 & --- & --- & --- \\
14 &  B1821+05 & T & -0.91$\pm$0.12 & --- & --- & --- \\
15 &  B1831--04 & M & -0.93$\pm$0.14 & 0.42$\pm$0.18 & -1.23$\pm$0.15 & -0.82$\pm$0.14 \\
16 &  B1857--26 & M & -1.043$\pm$0.062 & 0.319$\pm$0.073 & -1.223$\pm$0.062 & -0.904$\pm$0.062 \\
17 &  B1907+10 & T$_{1/2}$ & -0.825$\pm$0.040 & --- & --- & --- \\
18 &  B1914+09 & T$_{1/2}$ & -0.80$\pm$0.24 & --- & --- & --- \\
19 &  B1917+00 & T & -0.980$\pm$0.062 & --- & --- & --- \\
20 &  B2045--16 & T & -0.452$\pm$0.011 & --- & --- & --- \\
21 &  B2327--20 & T & -0.581$\pm$0.018 & --- & --- & --- \\
\enddata
\end{deluxetable}

The empirical theory of pulsar emission provides a framework for interpreting 
the radio emission beam, which comprises of a central core emission surrounded
by two concentric rings of inner and outer cones \citep{ET_R90,ET_R93,MD99}.
There are other competing models for the emission beam like the patchy 
beam \citep{LM88}, the hybrid patchy-conal beam \citep{KJ07}, the fan beam 
\citep{DRD10,WPZ14}, amonst others. However, the core-cone model is most widely
used in the literature with a wide variety of observational applications. The 
observed profile shape in this model depends on the line of sight (LOS) 
traverse across the emission beam and in case of more central cuts, where the 
LOS passes close to the magnetic axis, profiles with both core and conal 
components are seen. One primary difference reported in the behaviour of the 
different component types is their spectral difference, with the core expected 
to have a steeper spectra compared to the conal components. We intend to 
estimate the spectral variations across the emission beam from the MSPES sample
within the observed frequency range of 325 and 610 MHz. The primary profile 
types of interest for these estimates are as follows : \\
a. Triple (T) which is a three component profile with a central core and one 
pair of conal emission around it. In some cases one of the conal outriders 
maybe missing and the pulsar is classified as T$_{1/2}$. \\
b. Conal Triple ($_c$T) and conal Quadruple ($_c$Q) comprising of three/four 
component profiles corresponding to the inner and outer cones. \\
c. Multiple (M) with five components where the central core component is 
surrounded by two pairs of inner and outer cones. Figure \ref{fig:compspec} 
shows an example of a M type profile in PSR B1857--26.\\ 

It is not possible to obtain absolute spectral index of the components as the
pulsar emission is affected by scintillation. However, the scintillation are
expected to affect the emission across the entire profile in a similar manner.
Hence, the difference in the spectral index between the different component 
types are unaffected. The difference in spectral index between the core and the
conal components ($\Delta\alpha_{core-cone}$) can be obtained as :
\begin{eqnarray}
\Delta\alpha_{core-cone} & = & \log{\left(\frac{S_{core}^{\nu_1}}{S_{core}^{\nu_2}}\right)}/\log{\left(\frac{\nu_1}{\nu_2}\right)} - \log{\left(\frac{S_{cone}^{\nu_1}}{S_{cone}^{\nu_2}}\right)}/\log{\left(\frac{\nu_1}{\nu_2}\right)}, \nonumber \\
   & = & \log{\left(\frac{S_{core}^{\nu_1}}{S_{cone}^{\nu_1}}\right)}/\log{\left(\frac{\nu_1}{\nu_2}\right)} - \log{\left(\frac{S_{core}^{\nu_2}}{S_{cone}^{\nu_2}}\right)}/\log{\left(\frac{\nu_1}{\nu_2}\right)}.
\end{eqnarray}
Here $S_{core}^{\nu}$ and $S_{cone}^{\nu}$ gives estimates of the flux density 
of the core and conal component at a given frequency ($\nu$), and 
$S_{core}^{\nu}$/$S_{cone}^{\nu}$ removes the effect of scintillation. The 
spectral index difference between the inner and outer conal components 
($\Delta\alpha_{in-out}$), the core and inner cone 
($\Delta\alpha_{core-in}$) and the core and outer cone 
($\Delta\alpha_{core-out}$), can also be estimated in a similar manner.
\begin{eqnarray}
\Delta\alpha_{in-out} & = & \log{\left(\frac{S_{in}^{\nu_1}}{S_{out}^{\nu_1}}\right)}/\log{\left(\frac{\nu_1}{\nu_2}\right)} - \log{\left(\frac{S_{in}^{\nu_2}}{S_{out}^{\nu_2}}\right)}/\log{\left(\frac{\nu_1}{\nu_2}\right)}, \nonumber \\
\Delta\alpha_{core-in} & = & \log{\left(\frac{S_{core}^{\nu_1}}{S_{in}^{\nu_1}}\right)}/\log{\left(\frac{\nu_1}{\nu_2}\right)} - \log{\left(\frac{S_{core}^{\nu_2}}{S_{in}^{\nu_2}}\right)}/\log{\left(\frac{\nu_1}{\nu_2}\right)}, \nonumber \\
\Delta\alpha_{core-out} & = & \log{\left(\frac{S_{core}^{\nu_1}}{S_{out}^{\nu_1}}\right)}/\log{\left(\frac{\nu_1}{\nu_2}\right)} - \log{\left(\frac{S_{core}^{\nu_2}}{S_{out}^{\nu_2}}\right)}/\log{\left(\frac{\nu_1}{\nu_2}\right)}. 
\end{eqnarray}
Here $S_{in}^{\nu}$ is the flux density of the inner conal components, 
$S_{out}^{\nu}$ for the outer conal components.

There were 21 pulsars in the MSPES list where we were able to separate the 
different component types within the same profile. Table \ref{tab:comp_spect} 
shows the estimates of the spectral index difference for these sources. The 
list included 17 pulsars with M, T and T$_{1/2}$ profiles and showed the core 
emission to have a much steeper spectra in all cases, with an average 
$\Delta\alpha_{core-cone}$ = -0.71 between 325 and 610 MHz. In addition to the 
5 pulsars with M type profiles with all three types of components, there were 
an additional 4 pulsars with $_c$Q profiles where the inner and the outer conal
components could be discerned. The outer conal components showed a steeper 
spectra compared to the inner cone with the average $\Delta\alpha_{in-out}$ = 
0.34. The difference of the spectral behaviour is also illustrated in the 
profiles shown in Fig. \ref{fig:compspec}, where the core emission shows a 
much higher intensity at the lower frequency compared to both the inner and 
outer cones, while the outer cone is brighter at the lower frequency compared 
to the inner cone. In PSR B1237+25 there were two different emission modes with
very different profile shapes (see below) and the spectral estimates were 
carried out only for the `normal' mode.

\section{Emission Modes}\label{sec:mode}
We have searched for the presence of mode changing in the single pulse emission
of the MSPES sample by careful visual inspection of the single pulse 
sequence. The limitation of a survey setup, where the pulsars were observed 
for short durations of 2100 pulses, was not conducive for detecting longer 
duration mode changes with typical durations of hours timescales. We have 
identified 12 pulsars in our observations where the emission showed transition 
to different modes, 3 of these PSRs B0525+21, J1625--4048 and B1758--29, had no 
previous report of this phenomenon. In two other pulsars, B1819--22 and 
B2003--08, mode changing was first observed in the MSPES observations. More 
detailed studies have been carried out subsequently in these two sources to 
characterise the mode properties. We have measured the different mode behaviour
including relative abundance, number of modes, average length, flux density and
profile widths, that are reported in Table \ref{tab:mode}. In some cases 
the pulsar also showed the presence of nulling, lower intensity pulses or mixed
states which could not be clearly classified within the individual modes. A 
short description of mode changing in each pulsar is presented below.

\startlongtable
\begin{deluxetable}{ccccccccccc}
\tablenum{4}
\tablecaption{Emission Mode Characteristics\label{tab:mode}}
\tablewidth{0pt}
\tablehead{  
   & \colhead{PSR} & \colhead{Freq}  & \colhead{Npulse} & \colhead{Mode} & \colhead{Avg. Flux} & \colhead{Abundance} & \colhead{Nmode} & \colhead{Avg. Length} & \colhead{$W_{10}$} & \colhead{$W_{5\sigma}$} \\
   &   & \colhead{(MHz)} & \colhead{($P$)} &   & \colhead{(mJy)} & \colhead{(\%)} &   & \colhead{($P$)} & \colhead{($\degr$)} & \colhead{($\degr$)}
}

\startdata
  1 & B0031--07 & 325 & 2099 & A & 116.4$\pm$4.4 & 12.2 & 6 & 51.0 & 40.3$\pm$0.3 & 42.8$\pm$0.3 \\
   &   &   &   & B & 173.5$\pm$4.3 & 39.5 & 23 & 36.0 & 39.1$\pm$0.3 & 52.4$\pm$0.3 \\
   &   &   &   & C & 142.4$\pm$4.9 & 1.7 & 1 & 35.0 & --- & 43.6$\pm$0.3 \\
   &   &   &   &   &   &   &   &   &   &   \\
  2 & B0525+21 & 325 & 424 & A & 189.9$\pm$5.0 & 51.2 & 5 & 43.4 & 19.9$\pm$0.3 & 23.5$\pm$0.3 \\
   &   &   &   & B & 93.9$\pm$5.1 & 30.7 & 6 & 21.7 & 21.5$\pm$0.3 & 23.8$\pm$0.3 \\
   &   &   &   &   &   &   &   &   &   &   \\
  3 & B0844--35 & 325 & 2149 & A & 14.1$\pm$1.6 & 90.8 & 12 & 162.7 & 33.7$\pm$0.5 & 34.2$\pm$0.5 \\
   &   &   &   & B & 18.1$\pm$1.7 & 9.2 & 11 & 17.9 & --- & 30.9$\pm$0.5 \\
   &   & 610 & 2149 & A & 10.2$\pm$0.4 & 95.0 & 6 & 340.2 & --- & 28.1$\pm$0.5 \\
   &   &   &   & B & 12.2$\pm$0.8 & 5.0 & 5 & 21.6 & --- & 18.1$\pm$0.5 \\
   &   &   &   &   &   &   &   &   &   &   \\
  4 & B1237+25 & 325 & 4294 & Normal & --- & 91.3 & 6 & 653.3 & 15.6$\pm$0.2 & 18.8$\pm$0.2 \\
   &   &   &   & Abnormal & --- & 8.7 & 4 & 93.3 & 16.6$\pm$0.2 & 18.2$\pm$0.2 \\
   &   & 610 & 867 & Normal & 24.3$\pm$0.9 & 100 & --- & --- & 14.4$\pm$0.2 & 15.9$\pm$0.2 \\
   &   &   &   &   &   &   &   &   &   &   \\
  5 & J1625--4048 & 610 & 1524 & Bright & 7.1$\pm$0.4 & 27.8 & 26 & 16.3 & --- & 20.9$\pm$0.3 \\
   &   &   &   & Quiet & 4.9$\pm$0.4 & 24.2 & 19 & 19.4 & --- & 20.7$\pm$0.3 \\
   &   &   &   & Mixed & 5.8$\pm$0.4 & 19.8 & 10 & 30.2 & --- & 21.4$\pm$0.3 \\
   &   &   &   &   &   &   &   &   &   &   \\
  6 & J1727--2739 & 610 & 2127 & A & 8.7$\pm$0.6 & 10.2 & 7 & 31.0 & --- & 30.4$\pm$0.6 \\
   &   &   &   & B & 10.8$\pm$0.4 & 12.4 & 17 & 15.5 & --- & 32.2$\pm$0.6 \\
   &   &   &   &   &   &   &   &   &   &   \\
  7 & B1758--29 & 325 & 2107 & Bright & 13.7$\pm$0.5 & 49.1 & 30 & 34.5 & --- & 22.7$\pm$0.3 \\
   &   &   &   & Quiet & 7.1$\pm$0.6 & 37.7 & 33 & 24.1 & --- & 22.1$\pm$0.3 \\
   &   & 610 & 2106 & Bright & 15.5$\pm$0.6 & 47.3 & 27 & 36.9 & --- & 20.9$\pm$0.6 \\
   &   &   &   & Quiet & 9.6$\pm$0.7 & 39.3 & 27 & 30.6 & --- & 20.4$\pm$0.6 \\
   &   &   &   &   &   &   &   &   &   &   \\
  8 & B1819--22 & 325 & 2112 & A & 20.4$\pm$0.5 & 52.0 & 12 & 91.6 & 22.7$\pm$0.8 & 27.6$\pm$0.8 \\
   &   &   &   & B & 10.6$\pm$0.5 & 37.3 & 12 & 65.6 & --- & 28.0$\pm$0.8 \\
   &   &   &   & Trans-A & 14.6$\pm$0.7 & 5.5 & 6 & 19.3 & --- & 23.7$\pm$0.8 \\
   &   & 610 & 2189 & A & 20.2$\pm$0.7 & 49.4 & 14 & 77.3 & 15.1$\pm$0.8 & 17.8$\pm$0.8 \\
   &   &   &   & B & 13.6$\pm$0.7 & 44.0 & 13 & 74.2 & --- & 18.1$\pm$0.8 \\
   &   &   &   & Trans-A & 20.9$\pm$1.2 & 2.6 & 3 & 18.7 & --- & 15.9$\pm$0.8 \\
   &   &   &   &   &   &   &   &   &   &   \\
  9 & B1839--04 & 610 & 1957 & Bright & 26.8$\pm$0.7 & 24.9 & 5 & 97.6 & --- & 75.4$\pm$0.8 \\
   &   &   &   & Quiet & 16.8$\pm$0.7 & 72.5 & 5 & 283.6 & --- & 78.5$\pm$0.8 \\
   &   &   &   &   &   &   &   &   &   &   \\
 10 & B1944+17 & 325 & 5569 & A & 145.9$\pm$2.3 & 7.1 & 8 & 49.5 & 34.1$\pm$0.8 & 49.8$\pm$0.8 \\
   &   &   &   & B & 119.9$\pm$2.6 & 2.1 & 4 & 29.0 & 35.0$\pm$0.8 & 40.6$\pm$0.8 \\
   &   &   &   & C & 114.2$\pm$2.3 & 11.1 & 69 & 8.0 & 32.5$\pm$0.8 & 45.0$\pm$0.8 \\
   &   &   &   & D & 48.9$\pm$2.3 & 12.6 & 65 & 10.8 & 36.2$\pm$0.8 & 41.4$\pm$0.8 \\
   &   & 610 & 2179 & A & 168.2$\pm$2.4 & 12.6 & 7 & 39.1 & 36.6$\pm$0.8 & 42.6$\pm$0.8 \\
   &   &   &   & B & 151.0$\pm$4.3 & 1.0 & 1 & 21.0 & --- & 28.5$\pm$0.8 \\
   &   &   &   & C & 158.9$\pm$2.3 & 13.8 & 36 & 13.8 & 32.9$\pm$0.8 & 43.0$\pm$0.8 \\
   &   &   &   & D & 71.6$\pm$2.4 & 9.9 & 19 & 11.4 & --- & 37.0$\pm$0.8 \\
   &   &   &   &   &   &   &   &   &   &   \\
 11 & B2003--08 & 325 & 2134 & A & 66.1$\pm$2.3 & 6.8 & 3 & 48.7 & --- & 70.1$\pm$0.9 \\
   &   &   &   & B & 46.0$\pm$2.1 & 28.4 & 3 & 202.3 & --- & 69.9$\pm$0.9 \\
   &   &   &   & C & 33.6$\pm$2.6 & 2.1 & 5 & 9.0 & --- & 70.5$\pm$0.9 \\
   &   &   &   & D & 62.9$\pm$2.1 & 22.4 & 40 & 11.9 & --- & 69.0$\pm$0.9 \\
   &   & 610 & 1491 & A & 43.3$\pm$1.6 & 18.9 & 2 & 141.0 & --- & 58.9$\pm$0.9 \\
   &   &   &   & B & 27.9$\pm$1.6 & 20.2 & 2 & 150.5 & --- & 58.5$\pm$0.9 \\
   &   &   &   & C & 16.3$\pm$1.8 & 7.4 & 9 & 12.3 & --- & 62.8$\pm$0.9 \\
   &   &   &   & D & 43.6$\pm$1.7 & 10.0 & 17 & 8.8 & --- & 61.9$\pm$0.9 \\
   &   &   &   &   &   &   &   &   &   &   \\
 12 & B2303+30 & 325 & 1699 & Bright & 30.4$\pm$0.9 & 54.4 & 23 & 40.2 & 8.4$\pm$0.6 & 11.8$\pm$0.6 \\
   &   &   &   & Quiet & 15.7$\pm$1.0 & 34.7 & 27 & 21.8 & 10.4$\pm$0.6 & 12.1$\pm$0.6 \\
\enddata
\end{deluxetable}

\emph{B0031--07} : The radio emission in this pulsar has been investigated in 
detail and shows the presence of three distinct drift modes A, B and C as well
as nulling \citep{HTT70,VJ97,SMK05,MBT17}. The 610 MHz observations were 
affected by RFI and could not be used for these studies. At 325 MHz we detected
all three drift modes, with mode B being the most abundant, seen around 40 
percent of the total duration, mode A was seen for around 12 percent of times, 
while mode C was seen only for a single sequence of 35 periods. The average 
profiles in each mode is shown in the appendix, Fig.\ref{fig:appB0031}, which 
highlights the emission during mode A to be restricted near the trailing edge 
with a narrower width (see Table \ref{tab:mode}). The pulsar was around 50 
percent brighter during mode B in comparison with mode A. Mode C seemed to have
a double peaked structure, but there were too few pulses to form a clear idea. 
The drifting behaviour in the three modes were consistent with previous reports
in the literature.\\

\begin{figure}
\gridline{\fig{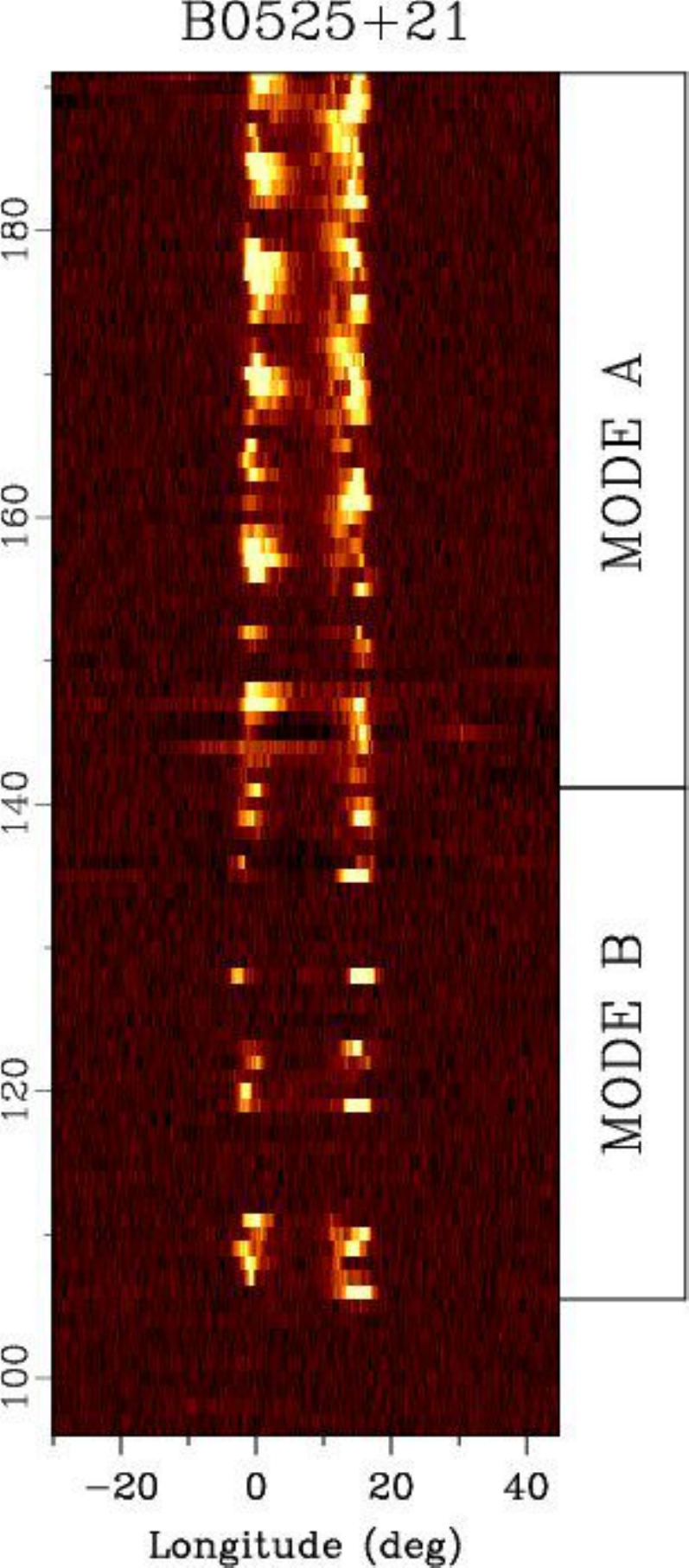}{0.25\textwidth}{(a)}
          \fig{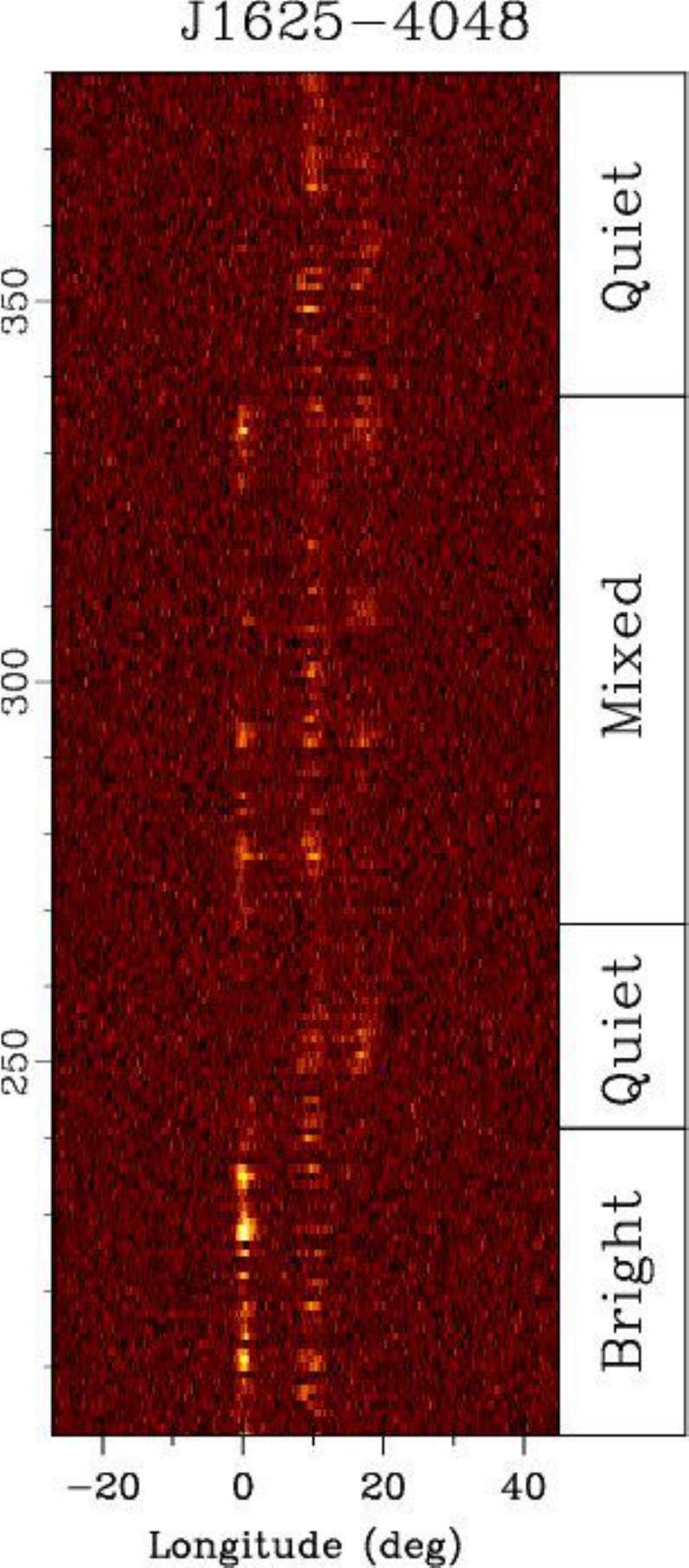}{0.25\textwidth}{(b)}
          \fig{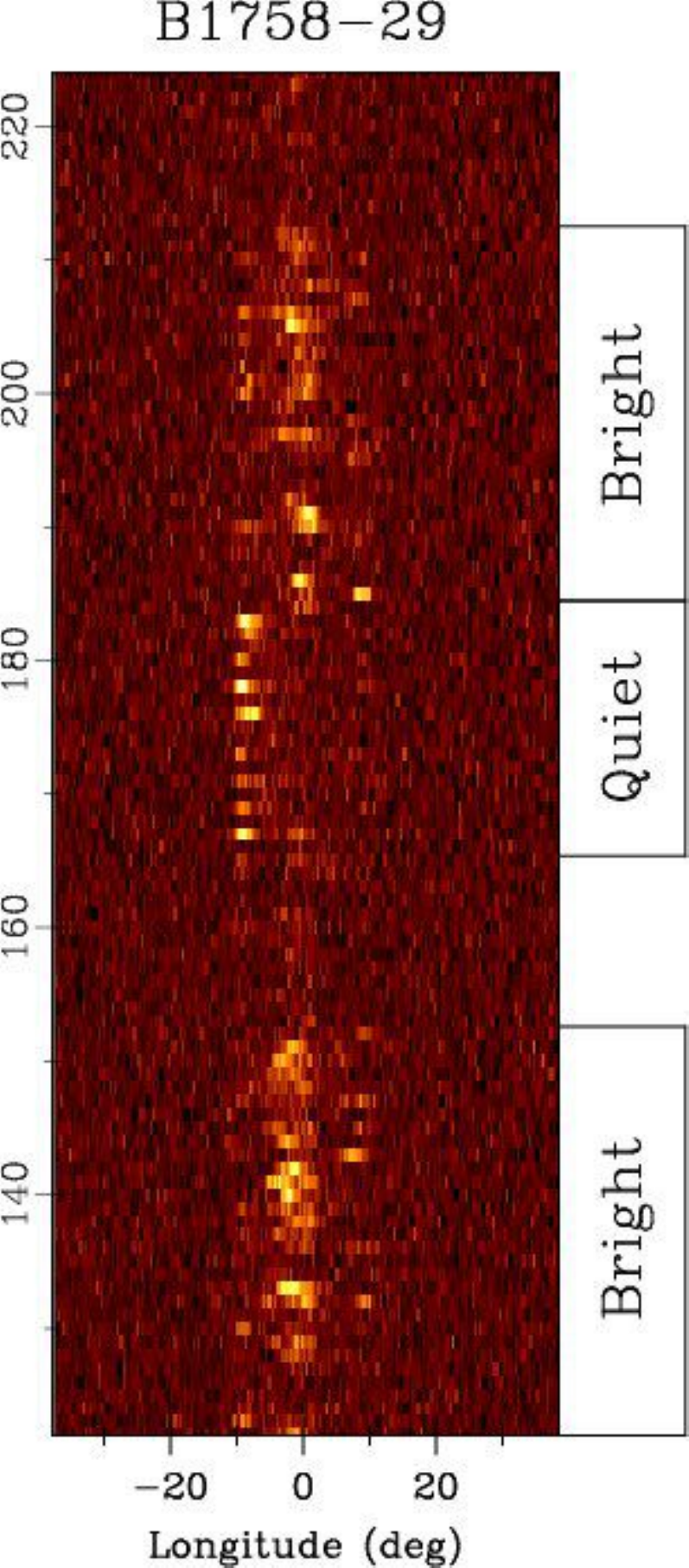}{0.25\textwidth}{(c)}
         }
\caption{The figure shows the mode changing behaviour seen in the single pulse
emission of the three pulsars where this phenomenon is reported for the 
first time. (a) The single pulses between pulse 95 and 190 from the start of 
the observation in PSR B0525+21 at 325 MHz. The transition from mode B to mode 
A is seen along with null pulses at the beginning as well as interspersed 
within the modes. (b) The transitions between the Bright, Weak and Mixed modes 
in PSR J1625--4048 at 610 MHz. The figure shows the single pulses between pulse
200 and 380 from the start of the observation. (c) The transitions between the 
Bright and Quiet modes in PSR B1758--29 at 325 MHz. The single pulses between 
121 and 223 from the start of the observation are shown in the figure which 
also shows the presence of nulls. 
\label{fig:modesingl}}
\end{figure}

\emph{B0525+21} : The pulsar has relatively longer period ($\sim3.7$ seconds) 
and was observed for around 800 pulses at 325 MHz. The later half of the 
observations were affected by RFI and only 424 single pulses could be used for 
these studies. The pulsar has a classical double conal profile with two clearly
separated components connected by a bridge emission and shows the presence of 
nulling. We have detected the presence of two distinct states in the single 
pulse emission, as shown in Fig.\ref{fig:modesingl}a, which has not been 
reported earlier. Mode A consisted of bunches of bright emission in both the 
leading and trailing components. In contrast the emission during mode B was 
mostly seen as single bright pulses, primarily towards the trailing edge 
followed by low level emission or nulls. The average profiles corresponding to 
the two modes (see Fig.\ref{fig:appB0525} in appendix) show the leading 
component to be brighter in mode A with the bridge emission between the 
components also being more enhanced compared to mode B. The flux density as 
well as the average duration of mode A was twice as large compared to mode B. 
The pulsar was seen for around 50 percent of times in mode A and around 30 
percent of the duration in mode B, but longer observations are required to 
develop better statistics of the modes. \\

\emph{B0844--35} : The pulsar has a four component conal profile of $_c$Q type,
with the third component being most prominent at both frequencies. The presence
of mode changing was reported in \citet{WMJ07}, with the central parts of the 
profile becoming bright during a short duration B mode. We have detected the 
presence of the centrally bright B mode for around 5-10 percent of times at 
both observing frequencies. Mode B was seen for short durations of around 20 
periods while the much longer duration mode A was seen for hundreds of periods 
at a time. The second component was comparable to the much brighter third 
component in Mode B (see Fig.\ref{fig:appB0844} in appendix). \\

\emph{B1237+25} : The pulsar has a five component, M type, profile with well 
resolved core, inner and outer cones. The single pulse emission has been 
studied in detail and shows the presence of subpulse drifting, nulling and two 
modes, characterised as normal and abnormal \citep{B70b,B70a,SR05,SRM13,MD14}. 
The pulsar was observed on two separate occasions at 325 MHz, while at 610 MHz 
the observations lasted 867 pulses. The mode changing was seen at 325 MHz where
the pulsar existed in the abnormal mode for around 9 percent of the total 
duration. Only the normal mode was seen during the shorter 610 MHz 
observations. The leading part of the profile, comprising of the first three 
components, became brighter during the abnormal mode while the trailing part 
was much weaker, compared to the normal mode (see Fig.\ref{fig:appB1237} in 
appendix). \\

\emph{J1625--4048} : The average profile shows the presence of three well 
separated components, with the central component identified as a core. 
The pulsar was only observed at 610 MHz and the single pulse emission showed 
the presence of three different modes, as shown in Fig.\ref{fig:modesingl}b, 
that have not been reported earlier. The core emission remained roughly 
constant during the three modes, which showed variations in the leading and 
trailing components. In case of the bright mode the emission in the leading 
component was enhanced to around four times the central core level while the 
trailing component had very little intensity (see Fig.\ref{fig:appJ1625} in 
appendix). In contrast during the quiet mode the leading component was 
mostly missing but the trailing component was more prominent and comparable to 
the central core. In the third mixed mode all three components had similar 
intensities. All three modes were seen for similar durations, between 20-30\% 
of the observing time (see Table \ref{tab:mode}), with the pulsar being in the 
null state for the remaining time. \\ 

\emph{J1727--2739} : The pulsar has a double conal profile with two well 
separated components. The single pulse emission has been studied in 
\citet{WWY16} and showed the pulsar to exist mostly in the null state with 
relative abundance of around 70 percent. The radio emission during the burst 
state showed two distinct drift modes A and B, with periodicities of around 
$10P$ and $5P$ respectively, and a non-drifting mode C. The detection 
sensitivity at 325 MHz was insufficient for these studies and the modal 
behaviour was estimated at 610 MHz. We detected the two drifting modes A and B
which were present for around 10 percent and 12 percent times, respectively. 
The third non-drifting mode C was not visible during these observations.\\

\emph{B1758--29} : We have detected the presence of mode changing in this 
pulsar with a three component T type profile and a central core. The pulsar 
showed the presence of a Bright and Quiet mode as shown in 
Fig.\ref{fig:modesingl}c. The Bright mode had higher intensity in the core, 
while the core vanished in the Quiet mode. The average profiles of the two
modes are shown in Fig.\ref{fig:appB1758}. The pulsar was seen around 50 
percent of time in the Bright mode and roughly 40 percent of time in the Quiet 
mode. The single pulse emission also showed the presence of nulls and low 
intensity level where the modal identification was not possible. \\

\emph{B1819--22} : The single pulse emission of this conal pulsar has been 
studied in \citet{BM18b}, and showed the presence of three drifting modes and a
non-drifting mode. The most abundant mode A had prominent drifting features, 
while it was absent during mode B. The short duration Trans-A was seen in some 
cases during transitions from mode A, and had a different drifting periodicity.
A rare mode C with distinct behaviour was absent during the MSPES observations.
The profiles of modes A, B and Trans-A at both observing frequencies are shown 
in Fig.\ref{fig:appB1819} in appendix, while the different modal 
characteristics are reported in Table \ref{tab:mode}. \\

\emph{B1839--04} : The pulsar has a double component conal profile with a 
prominent bridge emission between the two components. The presence of two 
emission modes, classified as the Bright and Quiet modes, was reported in this 
pulsar \citep{W16}. The more abundant Q mode showed regular drifting behaviour,
which was absent in the B mode. The 325 MHz observations were affected by 
scattering which made them unsuitable for identifying the emission modes. The 
modal analysis was carried out at 610 MHz and the characteristics of the two 
modes are reported in Table \ref{tab:mode}, while the average profiles are 
shown in Fig.\ref{fig:appB1839} in appendix. \\

\emph{B1944+17} : The single pulse emission of this conal pulsar showed the 
presence of three drifting modes A, B and C, and one non-drifting mode D 
\citep{DCH86,KR10}. The pulsar emission was dominated by nulling with more than
70 percent of time the single pulses were in the null state. The modes could be 
distinguished visually due to the different drift behaviour, with mode A having 
periodicity of around $14P$ and mode B around $6P$. Mode C seemed to have 
longer periodicity which could not be estimated due to the emission being 
frequently interrupted by nulls. The details of the mode behaviour at both 
observing frequencies are presented in Table \ref{tab:mode}. We have also 
estimated the average profiles in each mode which are shown in 
Fig.\ref{fig:appB1944} (see appendix).\\

\emph{B2003--08} : This five component pulsar with a M type profile showed the 
presence of four different modes as well as nulling in the single pulse 
emission \citep{BPM19}. The two modes A and B had subpulse drifting with 
different periodicities, with the core emission being more prominent in mode A.
The core showed higher intensity during mode D which also had periodic nulling 
behaviour. In contrast the core was mostly absent during mode C which was also
interspersed with nulls. The modal statistics during the MSPES observations at 
both frequencies are shown in Table \ref{tab:mode} with the respective profile
shapes in Fig.\ref{fig:appB2003} (see appendix). \\

\emph{B2303+30} : The pulsar showed the presence of a Bright and Quiet mode in 
the single pulse emission \citep{RWR05}. Subpulse drifting was seen in each 
mode, with mode B having periodicity of around $2P$ and mode Q mode around 
$3P$. The pulsar was observed only at 325 MHz and we detected both the emission
modes. The B mode had twice the intensity of the Q mode and was also present 
for longer durations. The characteristics of the two modes are reported in 
Table \ref{tab:mode} and the profiles are shown in Fig.\ref{fig:appB2303} in 
appendix.\\

In addition to the pulsars discussed above the presence of mode changing has 
also been reported in the pulsar B1918+19 \citep{HW87,RWB13}, which was 
observed in MSPES, but the single pulses were detected with low sensitivity and
the mode identification was not possible.

\section{Bursting Emission}\label{sec:burst}
We have observed several instances where the pulsar showed enhanced emission 
within the pulse window. There have been several reports of such behaviour in 
pulsars, the most prominent being the giant pulses which are short bursts of
emission with energy greatly exceeding the average level \citep{LCU95,KP18}. In
other cases `flaring' has been found over a wider window in pulsars with 
asymmetric profile shapes \citep{MR11}. Increased emission has also been 
reported at the onset of the mode changing in both the main pulse and 
postcursor components of PSR B0823+26 \citep{BM19}. Unlike mode changing these
enhanced emission states are seen for short durations usually a single period 
and do not show any periodicity which distinguish them from periodic amplitude 
modulation \citep{BMM20}. It is worth mentioning that \citet{GM05} suggested a 
model for the enhanced emission based on the coherent curvature emission 
mechanism. They argued that the short observed time scale results from angular 
beaming due to the relativistic motion of localized sources along dipolar field
lines, which involves a well-known ``kinematic boosting" effect that shortens 
the apparent duration and consequently increases the apparent brightness of the 
observed signal.

In the description below we have categorised the pulsars into two groups, the 
first group consists of 7 pulsars where increased emission was seen mostly 
throughout the pulse window, and in the second group there are also 7 cases 
with the emission seen as flaring within a specific longitude range of the 
pulse window.

\subsection{Short duration enhanced emission}
\begin{figure}
\gridline{\fig{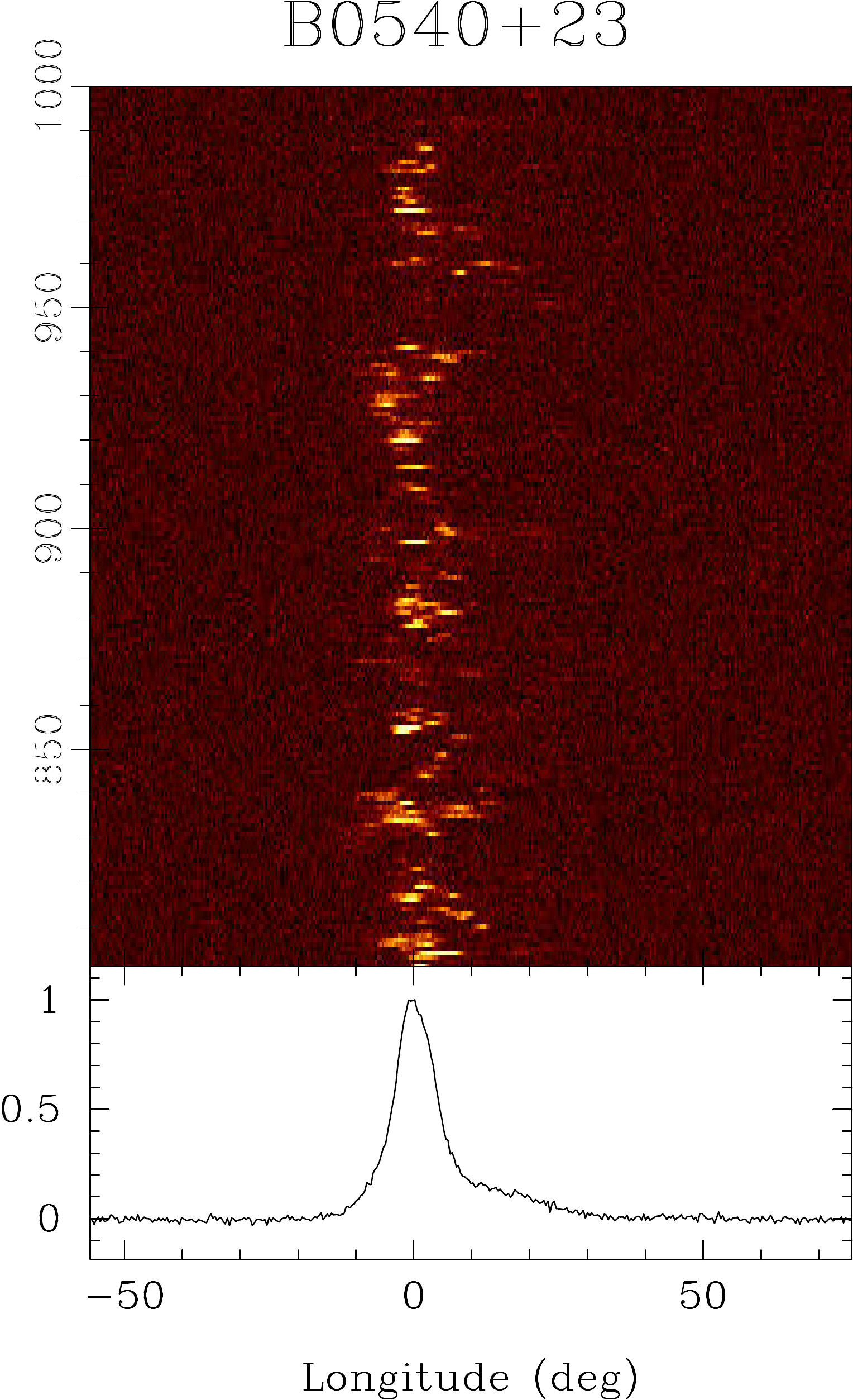}{0.25\textwidth}{(a)}
          \fig{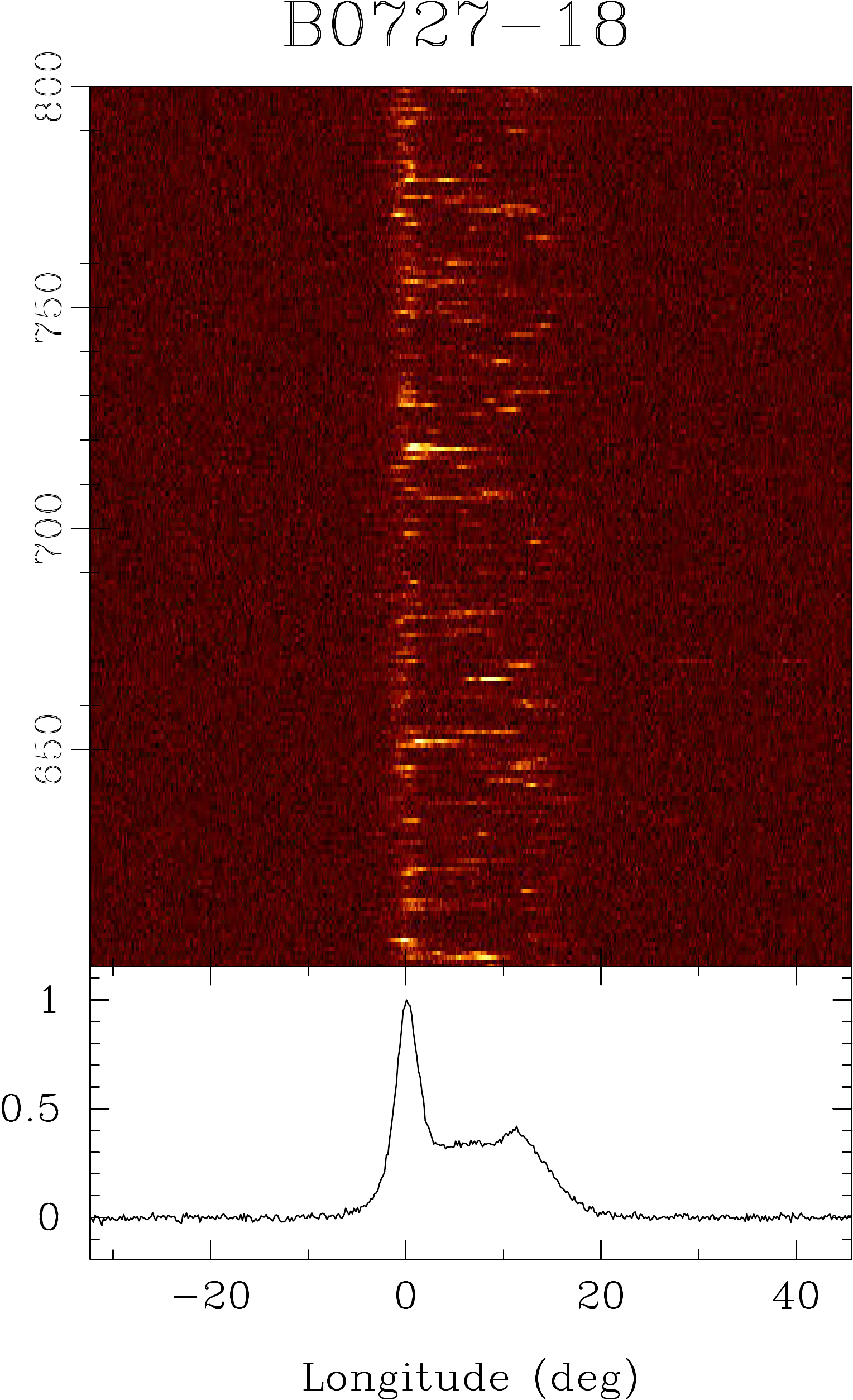}{0.25\textwidth}{(b)}
          \fig{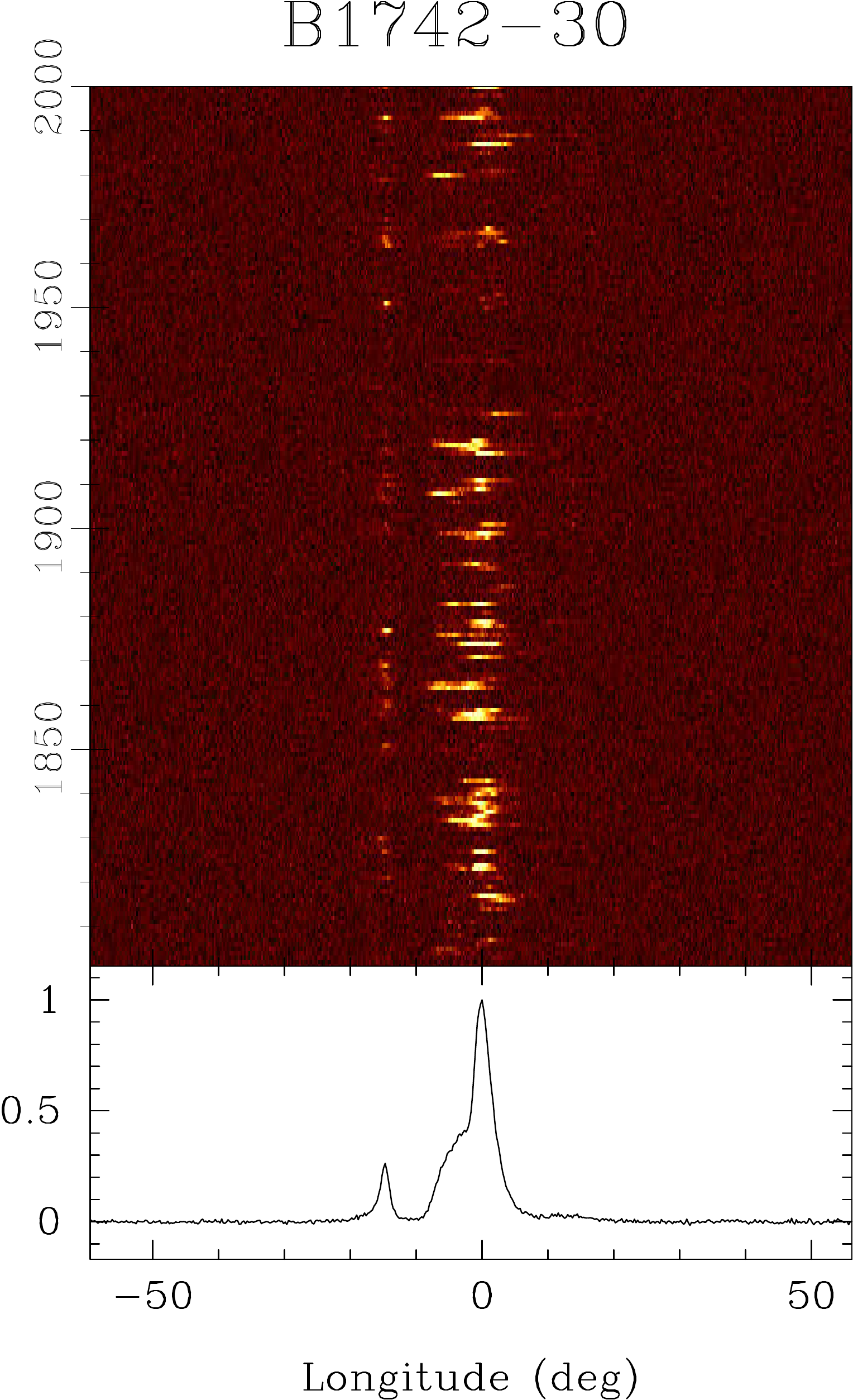}{0.25\textwidth}{(c)}
         }
\gridline{\fig{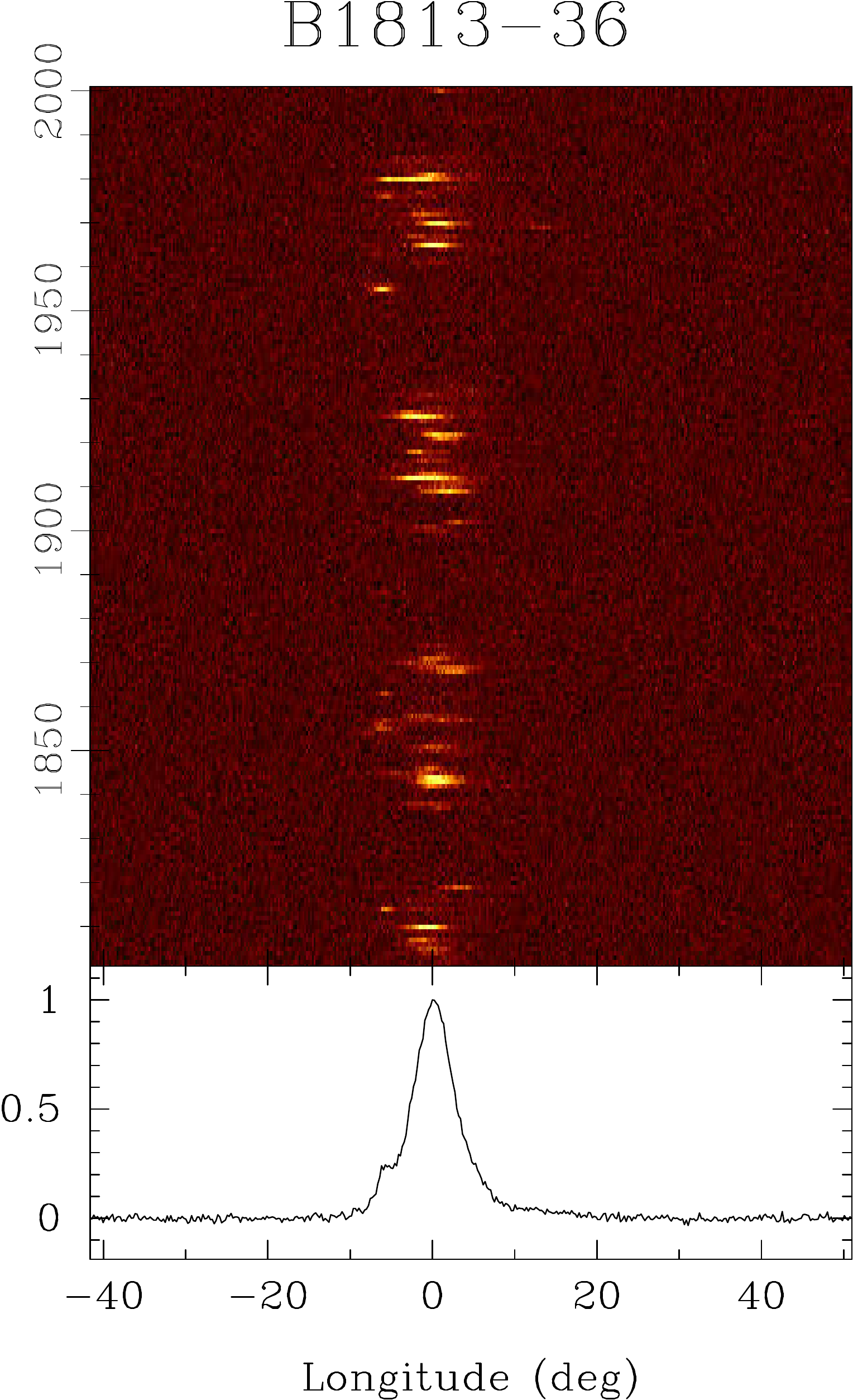}{0.25\textwidth}{(d)}
          \fig{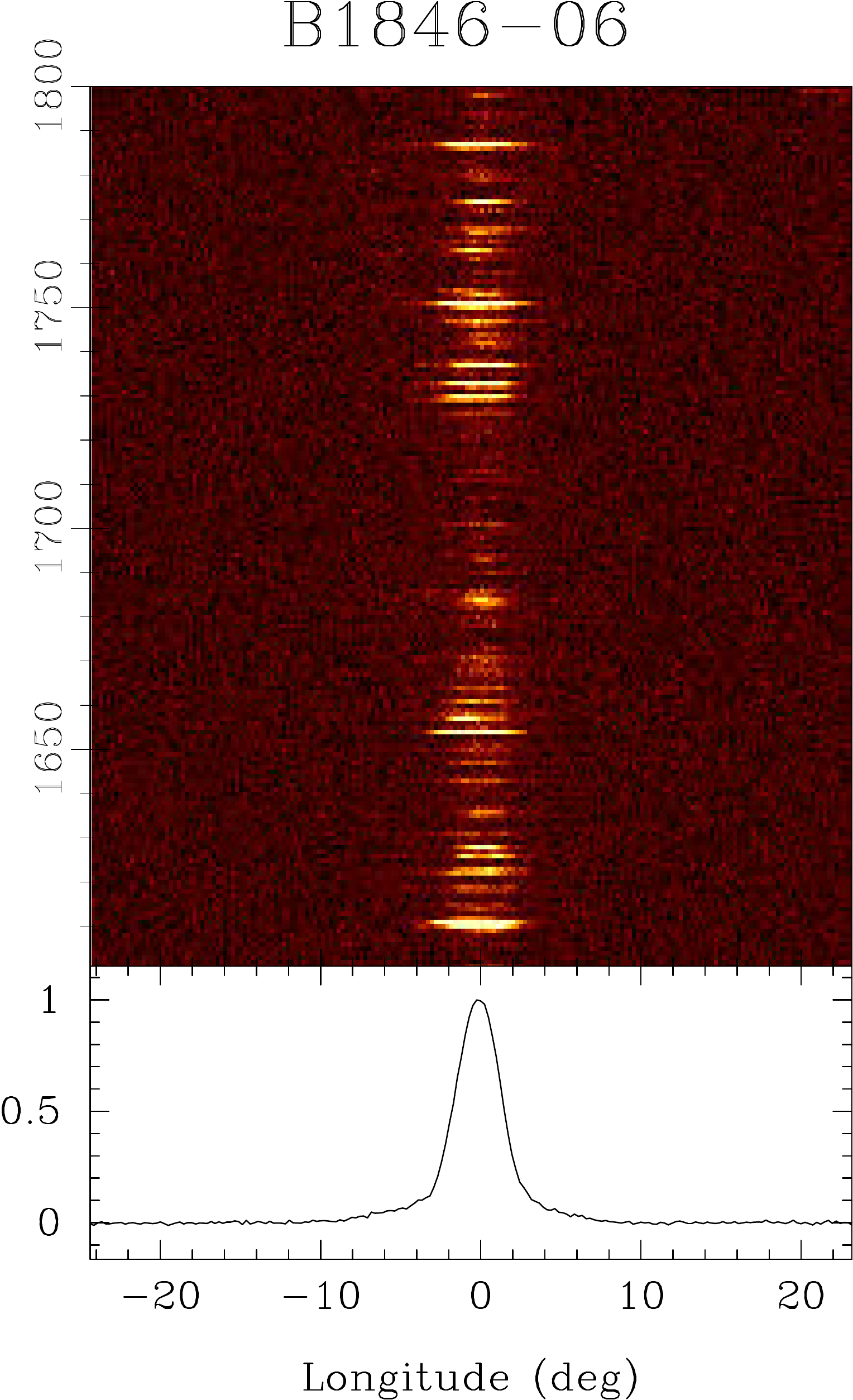}{0.25\textwidth}{(e)}
         }
\caption{The figure shows the bursting emission in the single pulses (top 
panel) and the average profile (bottom panel). (a) PSR B0540+23 at 325 MHz. 
(b) B0727--18 at 325 MHz. (c) PSR B1742--30 at 610 MHz. (d) B1813--36 at 325 
MHz. (e) B1846--06 at 610 MHz.
\label{fig:burstsingl}}
\end{figure}

\emph{B0540+23} : The pulsar profile has a single component with extended 
emission seen in the trailing edge. The radio emission was primarily in the 
form of bursts which were much narrower than the pulse window, but distributed 
throughout, as shown in Fig.\ref{fig:burstsingl}a. The peak flux level in the 
bursts reached more that 36 times the profile peak at 325 MHz and around a 
factor of 49 at 610 MHz. In 3-4 percent of pulses the peak flux were more than 
10 times the average profile level. \\

\emph{B0611+22} : The bursting emission in this pulsar has been observed in the
earlier studies of \citet{SLR14,RSL16}. The pulsar had a single component 
profile where the emission was seen throughout the pulse window. The burst 
state was seen primarily for short durations of few pulses as well as one 
relatively long stretch of around 50-60 pulses. The intensities of the bursting 
pulses were 5-10 times higher than the average value and were extended towards 
the trailing edge. The burst state was seen for around 15-20 percent of the 
observing duration at 325 MHz. The single pulses were detected with lower 
sensitivity at 610 MHz and the different states could not be distinguished. \\

\emph{B0656+14} : The presence of `spiky' emission has been reported in this 
pulsar, where bursts were seen within a narrow longitude range of the pulse 
window \citep{WWS06}. We detected pulses with peak intensities which exceeded 
the average profile peak level by a factor of 50~in around 2 percent of pulses,
with the maximum level being more than 240 times the average peak level at 325 
MHz. At 610 MHz the maximum intensity level was more than 125 times the average
profile peak and the intensities exceeded the average peak level by a factor of
40 around 2\% of the observing duration. \\

\emph{B0727-18} : The pulsar showed bursts of emission in the single pulses 
as shown in Fig.\ref{fig:burstsingl}b, with maximum intensity exceeding 
more than 35 times the average peak. The pulsar has a double component profile 
connected by a prominent central bridge. The bursts were seen in both 
components as well as the connecting bridge region. In around 2 percent of 
pulses the intensity level within the pulse window exceeded 10 times the 
profile peak level.\\

\emph{B1742-30} : The pulsar profile show the presence of a leading component
well separated from a double central structure which is followed by low level 
emission in the trailing side. The single pulses were seen primarily as short 
duration bursts as shown in Fig.\ref{fig:burstsingl}c. The maximum intensities 
in the single pulses were around a factor of 30 more than the average profile 
peak at both frequencies. In around 3 percent of the observing duration the 
peak pulse intensity was more than 10 times the peak flux density of the 
average profile. \\

\emph{B1813-36} : The pulsar show a single component profile with a narrow 
emission feature near the leading edge. The observations were carried out only 
at 325 MHz and Fig.\ref{fig:burstsingl}d shows a short section of the single 
pulses which were seen as short bursts across the pulse window. The maximum 
intensity in the single pulses was around a factor 24 higher than the profile 
peak, while strong bursts in excess of 5 times the average peak was seen in 7 
percent of the total observed pulses. \\

\emph{B1846-06} : The pulsar has a single component profile at 610 MHz, while 
the lower frequency is affected by scattering resulting in an elongated 
trailing edge. Bursts of short durations across the entire profile window were 
seen in conjunction with the lower intensity pulses (see 
Fig.\ref{fig:burstsingl}e). The maximum intensity in the bright pulses reached 
a level of more than 22 times the average profile peak, and in 5 percent of 
single pulses the peak intensity was more than 5 times the average level. \\

\subsection{Flaring within the pulse window}

\begin{figure}
\gridline{\fig{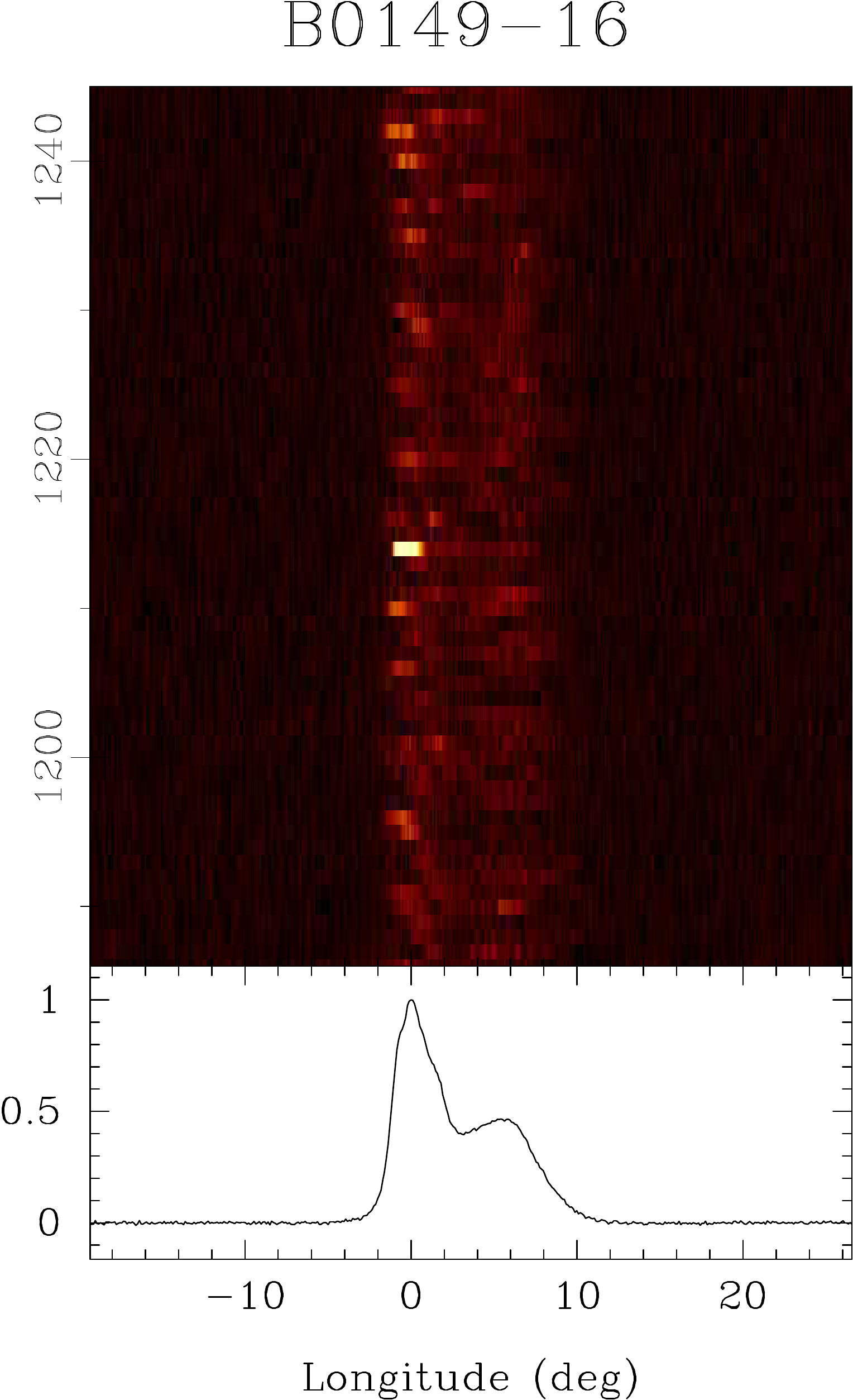}{0.23\textwidth}{(a)}
          \fig{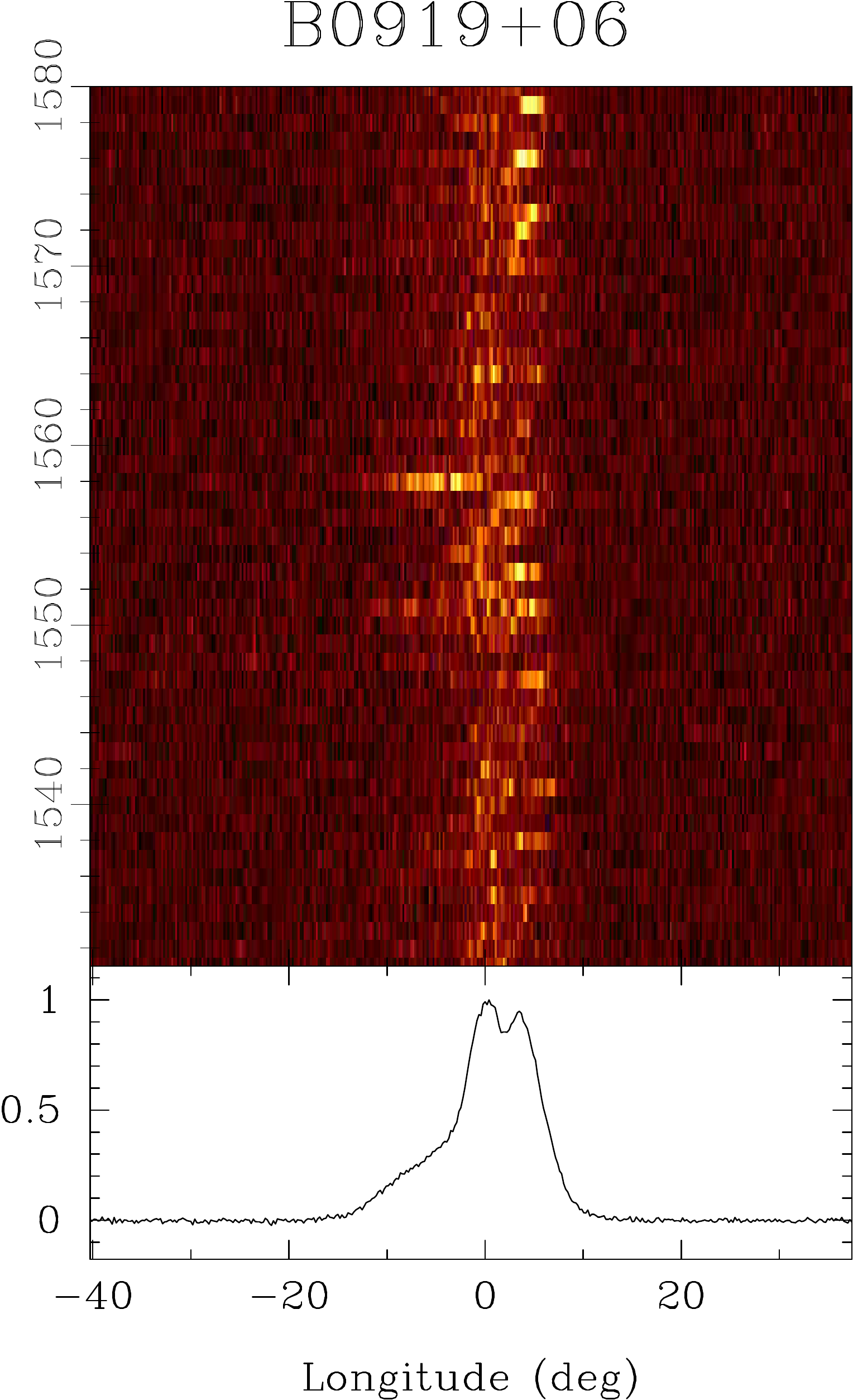}{0.23\textwidth}{(b)}
          \fig{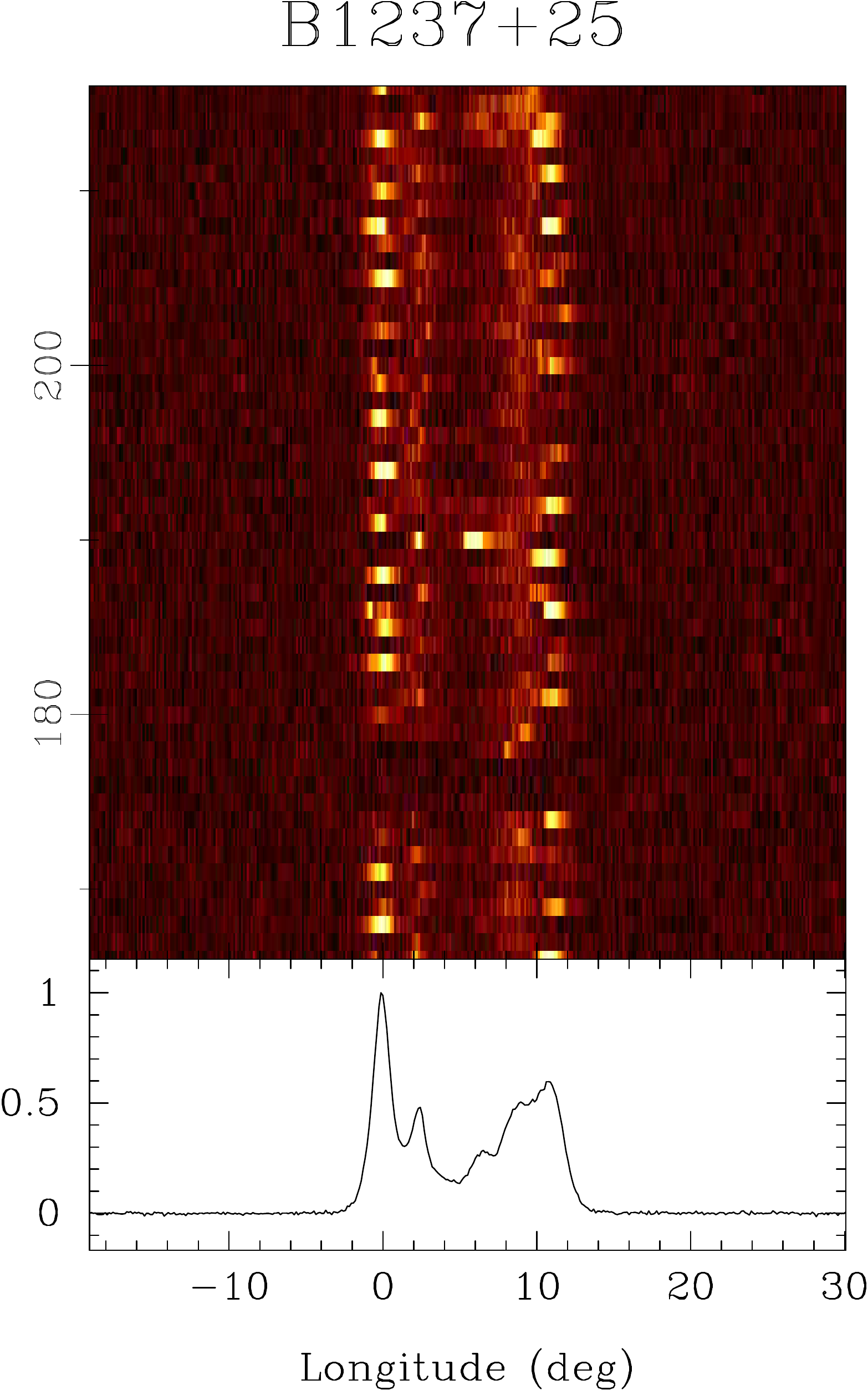}{0.23\textwidth}{(c)}
          \fig{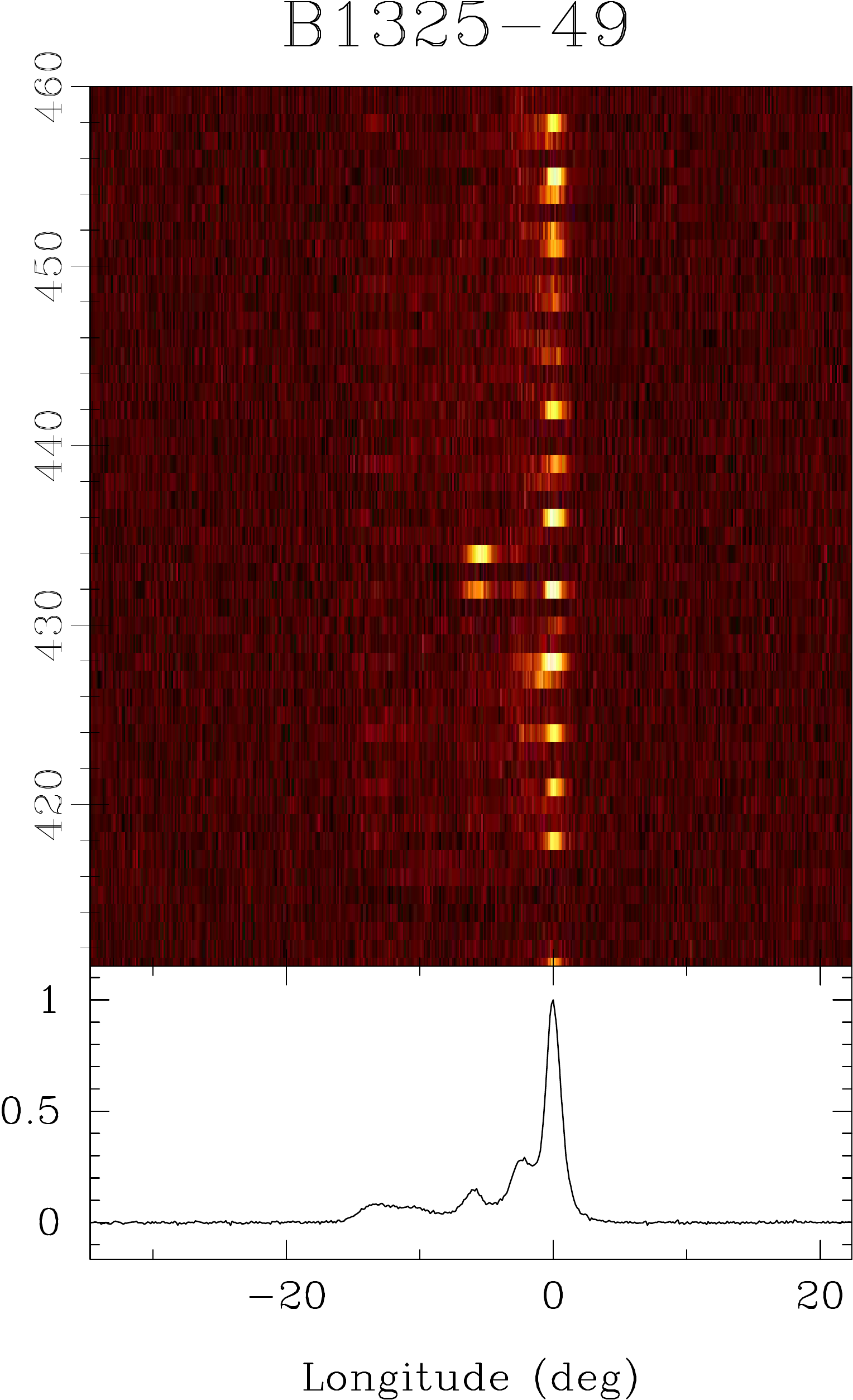}{0.23\textwidth}{(d)}
         }
\gridline{\fig{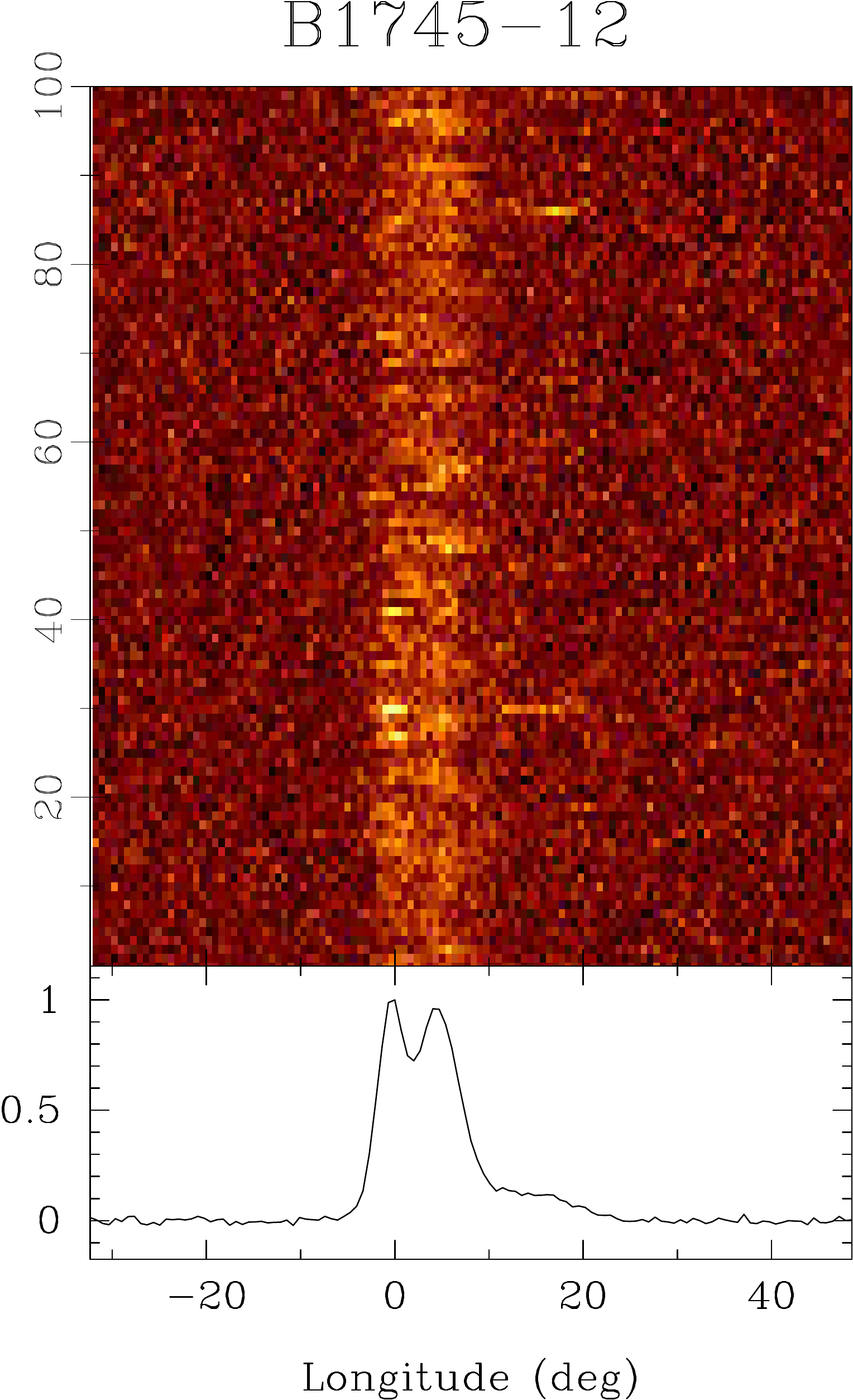}{0.23\textwidth}{(e)}
          \fig{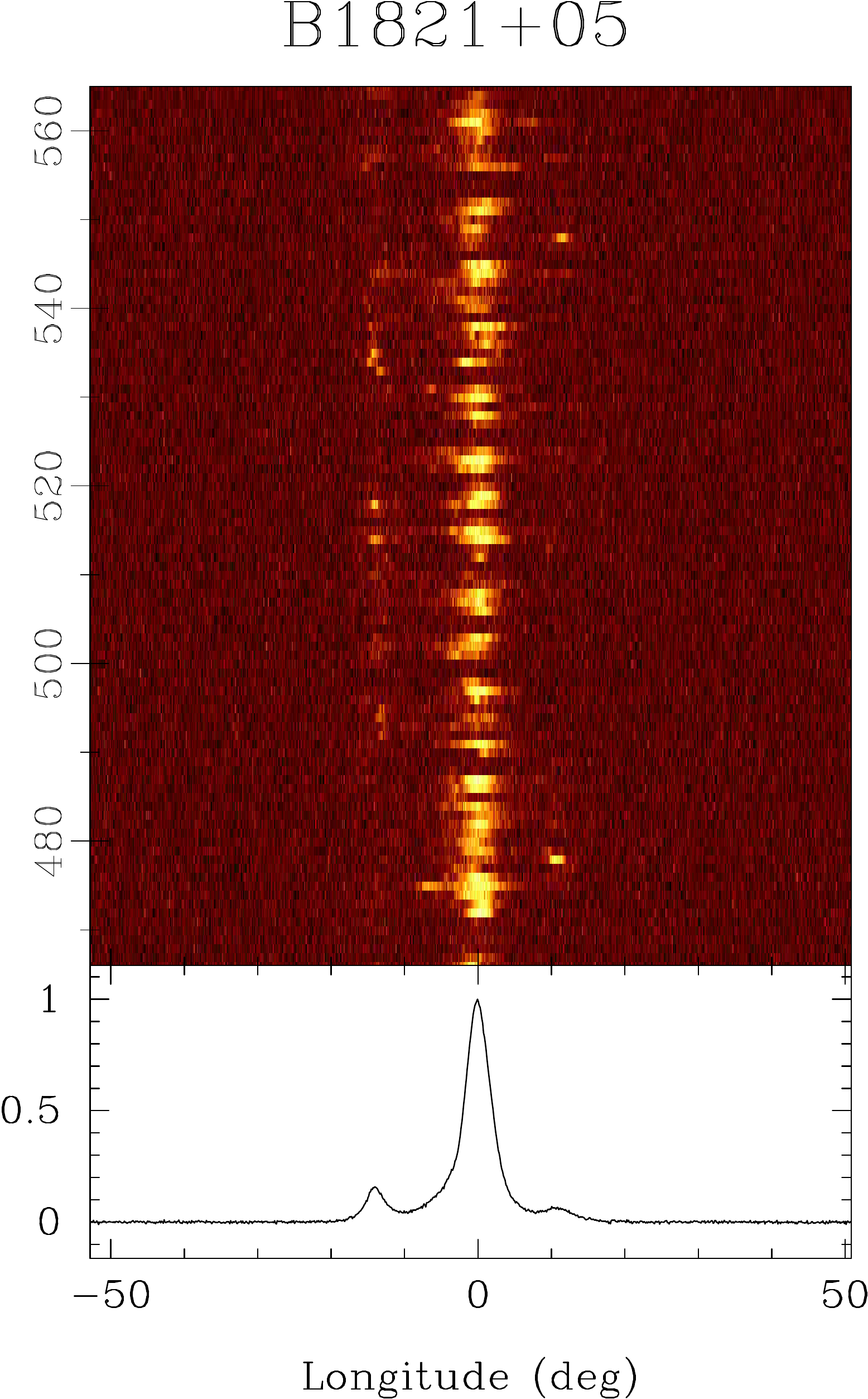}{0.23\textwidth}{(f)}
          \fig{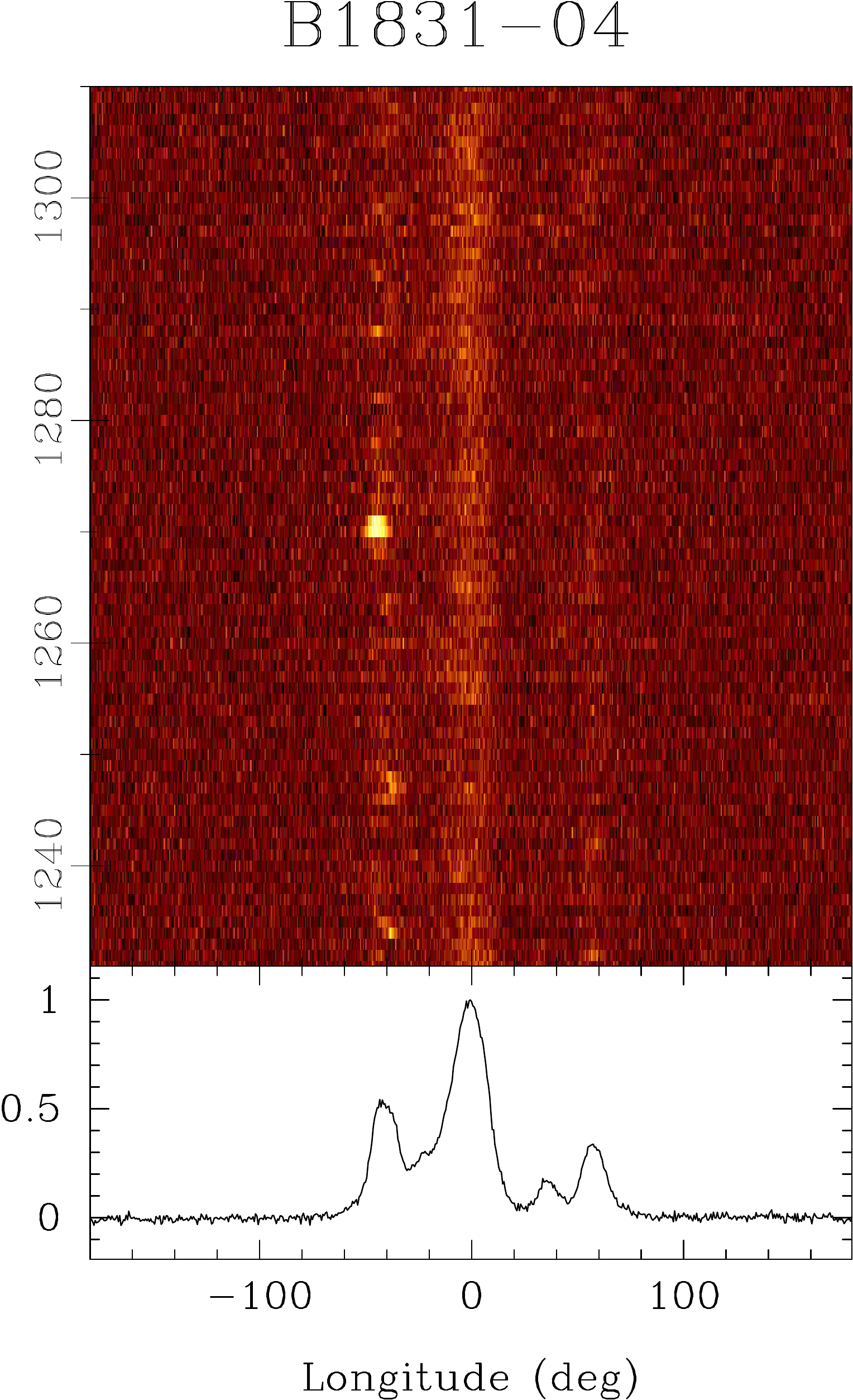}{0.23\textwidth}{(g)}
         }
\caption{The figure shows examples of enhanced emission within a narrow range 
of the pulse window (top panel) along with the average profile (bottom 
panel). (a) Flaring in the leading component of PSR B0149--16. (b) Flaring in 
the leading edge of PSR B0919+06. (c) Bursts of emission in central component 
of PSR B1237+25. (d) Bursts of emission in central component of PSR B1325--49. 
(e) Flaring in trailing edge of PSR B1745--12. (f) Bursts of emission in 
trailing component of PSR B1821+05. (g) Flaring in leading component of PSR 
B1831--04.
\label{fig:flaresingl}}
\end{figure}

\emph{B0149-16} : The pulsar has a double component profile and show the 
presence of subpulse drifting. Narrow flaring events were seen in leading 
component as shown in Fig.\ref{fig:flaresingl}a, where the intensity was 
greatly increased to a maximum level of more than 30 times the average energy.
Such events were however rare and seen in only around 0.8 percent of the total 
pulses. \\

\emph{B0919+06} : The pulsar profile show low level extended emission in the 
leading edge which connects with a higher intensity double peaked structure in 
the trailing side. Flaring was seen in the leading part of the pulse window as 
shown in Fig.\ref{fig:flaresingl}b. However, they were rarely seen in the 
single pulses, with less than 10 events recorded during 2100 pulses observed.
The intensities of the flares were a factor of 5-10 times more than the average
intensity. \\

\emph{B1237+25} : The pulsar has a five component profile with a central core 
emission. There are two emission modes in this pulsar, as reported above, the 
normal mode where the core is weaker than the conal emission and an abnormal 
mode with the central region becoming brighter. In the normal mode there were 
instances when the core emission showed flaring in intensity, around 10-20 
times the average level as shown in Fig.\ref{fig:flaresingl}c. These flaring 
events were rare and seen in less than 1 percent of the observed pulses. \\

\emph{B1325-49} : The pulsar has an asymmetric multi-component profile with the
trailing side being much more prominent than the leading part. In rare cases 
the central component showed flaring which had intensities of around 10-20 
times the mean level (see Fig.\ref{fig:flaresingl}d). These flares were seen in
less than 10 pulses during the observing sessions. \\

\emph{B1745-12} : The pulsar has two distinct components in the leading edge 
followed by a wide low level emission in the trailing side. There were flaring
events in the weak trailing side, as shown in Fig.\ref{fig:flaresingl}e, which 
were comparable to the leading emission and were a factor of 10-20 times 
greater than the average level. However, these were rare events seen in less 
than 1 percent of the total pulses. \\

\emph{B1821+05} : The pulsar has a triple component profile with a prominent 
central core component and weak leading and trailing cones. The emission in the
trailing side was mostly seen as short duration bursts (see 
Fig.\ref{fig:flaresingl}f) with maximum intensity more than 30 times the 
average profile level. In around 5 percent of times the emission in the 
trailing component had peak intensity in excess of 10 times the average 
value. \\

\emph{B1831-04} : The pulsar has a multi-component profile with a prominent 
core and outer cones as well as clearly distinguishable inner conal components.
The leading outer cone showed flaring events where the emission increased in 
intensity by a factor of 10-20 times the average level as shown in 
Fig.\ref{fig:flaresingl}g. However, these events were rare and seen in less 
than 20 pulses over the entire observing duration. \\

\section{Conclusion}\label{sec:con}
We have measured the flux density of 113 pulsars observed in the 
Meterwavelength Single-pulse Polarimetric Emission Survey, at 325 and 610 MHz, 
by making images from the interferometric observations of these sources. We 
have explored different emission features of these sources related to the 
variations in the average profile and single pulses. We have estimated the 
spectral variation across the emission beam in 21 pulsars with clearly 
identified core, and inner and outer conal components, where the steepest 
spectra was seen in the central core region, while the outer cones had steeper 
spectra compared to the inner cones. The radio emission is believed to 
originate in the relativistically outflowing plasma along the open magnetic 
field lines, where plasma instabilities result in charge bunching that gives 
rise to coherent curvature radiation along the field lines \citep{MGP00,GLM04,
LMM18,RMM20}. The observed radio emission is an incoherent addition of a large 
number of charge bunches, where each charge bunch emits coherently. Thus the 
observer cannot distinguish an individual charge bunch, but sees an average 
effect of the incoherently added power of a large number of individual charge 
bunches. The spectra of an individual charge bunch emitting curvature radiation
depends on the Lorentz factor, the amount of charge in the bunch, the size of 
the bunch and the radius of curvature of the magnetic field. If we assume that 
the size of the total charge and Lorentz factor of the charge bunch are 
similar, then the spectral difference between the core and conal emission can 
be explained by the change in radius of curvature. In this case the steep 
spectra of the core emission implies that the radius of curvature does not vary
much in the region where the core emission originates. Intuitively this seems 
to be reasonable, as the core emission arises from regions close to magnetic 
axis where the radius of curvature of the magnetic field lines do not vary 
much. Thus although there is incoherent averaging of the spectra, the spectra 
of all the charge bunches are similar and hence the resultant spectra closely 
resemble a single particle charge bunch spectra. The conal emission arises from
regions away from the magnetic axis, where across the conal emission regions 
variation of the radius of curvature is expected. This may lead to incoherent 
addition of spectra with slightly different spectral shape. This averaging can 
result in shallower spectra of the conal emission compared to the core 
emission. Alternatively, it is possible that the Lorentz factors or charge 
bunch variations can explain the change in spectral index across the window, 
however such effects can only be revealed by rigorous numerical simulations 
which is beyond the scope of this work.

Mode changing was found in two pulsars, B1819--22 and B2003--08, from the MSPES
observations and were separately studied in earlier works. We have investigated 
in detail the single pulse emission and found 12 pulsars ($\sim$10 percent of 
the sample) to have different emission modes, including the two mentioned 
above. In three of these pulsars, B0525+21, J1625--4048 and B1758--29, this 
phenomenon has not been reported earlier. The observations usually spanned 
roughly 2100 pulses and were sensitive to short duration mode changing. A 
different state change was also seen in 7 pulsars where the emission was seen 
as short duration bursts across the emission window. In another 7 pulsars 
flaring in intensity across a narrower longitude range of the pulsed region was
also characterised. These processes constitute changes in the steady state 
radio emission mechanism. In most cases discussed here the location of the 
radio emission within the magnetosphere appears to remain unaltered as 
suggested by the profile boundaries in the different states (see appendix 
\ref{app:mode}). As a result these state changes are likely to be driven by 
local changes of the magnetic field configuration near the polar cap, where the
outflowing plasma is generated, thereby affecting the plasma parameters 
\citep[a possible mechanism is presented in][]{GBM21}.

\section*{Acknowledgments}
We thank the referee for the comments that improved the paper. We thank the 
staff of the GMRT who have made these observations possible. The GMRT is run by
the National Centre for Radio Astrophysics of the Tata Institute of Fundamental
Research. DM acknowledges the support of the Department of Atomic Energy, 
Government of India, under project no. 12-R\&D-TFR-5.02-0700. DM acknowledges 
funding from the grant ``Indo-French Centre for the Promotion of Advanced 
Research - CEFIPRA" grant IFC/F5904-B/2018. This work was supported by the 
grant 2020/37/B/ST9/02215 of the National Science Centre, Poland.

\bibliography{MSPESV}{}
\bibliographystyle{aasjournal}

\startlongtable
\begin{deluxetable}{cccccccccc}
\tablenum{2}
\tablecaption{Flux Density measurement\label{tab:flux}}
\tablewidth{0pt}
\tablehead{  &     & \multicolumn{4}{c}{\underline{325 MHz}} & \multicolumn{4}{c}{\underline{610 MHz}}\\
  & \colhead{PSR} & \colhead{Date} & \colhead{Avg Flux} & \colhead{rms} & \colhead{Peak Flux} & \colhead{Date} & \colhead{Avg Flux} & \colhead{rms} & \colhead{Peak Flux}\\
  &   &   & (mJy) & (mJy) & (mJy) &   & (mJy) & (mJy) & (mJy)}

\startdata
   1 &  B0031-07  & 16/02/2014 & 92.7$\pm$4.3 & 2.0 & 1400$\pm$66 & 09/02/2014 & 22.4$\pm$4.2 & 1.2 & --- \\
   2 & J0134-2937 & 16/02/2014 & 8.1$\pm$1.3 & 0.7 & 225$\pm$36 & 09/02/2014 & 11.0$\pm$1.1 & 0.5 & 291$\pm$32 \\
   3 &  B0148-06  & 16/02/2014 & 8.4$\pm$2.8 & 1.4 & 184$\pm$61 & 23/03/2014 & 17.0$\pm$1.4 & 0.9 & 486$\pm$41 \\
  &   & 09/03/2014 & 11.0$\pm$3.1 & 1.7 & --- &   &   &   &   \\
   4 &  B0149-16  & 16/02/2014 & 37.4$\pm$1.5 & 0.6 & 2406$\pm$98 & 09/02/2014 & 5.5$\pm$0.4 & 0.2 & 311$\pm$24 \\
   5 &  B0203-40  & 28/06/2014 & 37.8$\pm$3.2 & 1.6 & 3090$\pm$264 & 23/03/2014 & ~1.1$\pm$0.5 & 0.3 & ~112$\pm$48 \\
   6 &  B0301+19  & 27/03/2014 & 59.2$\pm$3.8 & 2.2 & --- & 23/03/2014 & 28.3$\pm$0.9 & 0.2 & 1084$\pm$35 \\
  &   & 28/06/2014 & 35.2$\pm$1.6 & 0.6 & 1395$\pm$63 &   &   &   &   \\
   7 &  B0450-18  & 16/02/2014 & 124.9$\pm$5.0 & 1.5 & 2841$\pm$113 & 09/02/2014 & 74.6$\pm$2.1 & 0.4 & 1697$\pm$50 \\
   8 &  B0523+11  & 16/02/2014 & 22.4$\pm$2.1 & 1.1 & ~866$\pm$83~ & 09/02/2014 & 8.8$\pm$0.7 & 0.3 & 293$\pm$24 \\
   9 &  B0525+21  & 16/02/2014 & 88.5$\pm$5.0 & 1.8 & 3811$\pm$216 &   &   &   &   \\
  10 &  B0540+23  & 16/02/2014 & 29.0$\pm$2.0 & 0.9 & ~998$\pm$68~ & 09/02/2014  & 37.0$\pm$1.3 & 0.4 & 1318$\pm$46 \\
  11 &  B0611+22  & 16/02/2014 & 23.3$\pm$2.3 & 1.1 & 1036$\pm$106 & 09/02/2014  & 15.7$\pm$0.8 & 0.4 & 855$\pm$56 \\
  12 &  B0626+24  & 16/02/2014 & 70.1$\pm$3.1 & 1.1 & 3410$\pm$149 & 09/02/2014 & 40.0$\pm$1.2 & 0.2 & 1805$\pm$54 \\
  13 &  B0628-28  &   &   &   &   & 23/03/2014 & 97.0$\pm$2.8 & 0.7 & 1657$\pm$48 \\
  14 &  B0656+14  & 16/02/2014 & ~8.0$\pm$1.6 & 0.8 & ~148$\pm$33 & 09/02/2014 & ~8.0$\pm$0.5 & 0.2 & 180$\pm$15 \\
  &   & 09/03/2014 & ~4.0$\pm$0.6 & 0.4 & --- &   &   &   &   \\
  &   & 28/06/2014 & 10.6$\pm$0.8 & 0.5 & 226$\pm$21 &   &   &   &   \\
  15 &  B0727-18  & 28/06/2014 & 29.0$\pm$2.3 & 1.2 & 1298$\pm$104 & 23/03/2014 & 9.0$\pm$0.6 & 0.3 & 412$\pm$28 \\
  16 &  B0736-40  & 28/06/2014 & 318.5$\pm$9.2 & 1.7 & 1116$\pm$36 &   &   &   &   \\
  17 &  B0740-28  & 09/03/2014 & 482.2$\pm$16.8 & 6.0 & 13955$\pm$487 & 23/03/2014 & 104.3$\pm$3.4 & 1.4 & 4469$\pm$147 \\
  18 &  B0756-15  &   &   &   &   & 23/03/2014 & ~7.4$\pm$0.4 & 0.2 & 718$\pm$37 \\
  19 &  B0818-41  &   &   &   &   & 10/02/2014 & 22.7$\pm$0.7 & 0.2 & 121$\pm$8~ \\
  20 &  B0834+06  & 14/02/2014 & 135.1$\pm$4.4 & 1.6 & 10446$\pm$341 & 10/02/2014 & 40.5$\pm$1.2 & 0.4 & 2925$\pm$90 \\
  21 &  B0844-35  & 09/03/2014 & 21.7$\pm$3.3 & 1.7 & --- & 10/02/2014 & 10.3$\pm$0.4 & 0.1 & 396$\pm$17 \\
  &   & 28/06/2014 & 15.6$\pm$1.6 & 1.0 & 391$\pm$41 &   &   &   &   \\
  22 & J0905-5127 & 28/06/2014 & 5.9$\pm$1.3 & 0.7 & 202$\pm$46 &   &   &   &   \\
  23 &  B0919+06  & 28/06/2014 & 48.8$\pm$1.7 & 0.6 & 1690$\pm$61 & 20/03/2014 & 30.9$\pm$0.9 & 0.2 & 1490$\pm$45 \\
  24 &  B0942-13  & 28/06/2014 & 25.4$\pm$4.0 & 2.2 & 2416$\pm$379 &   &   &   &   \\
  25 &  B0950+08  & 14/02/2014 & 981$\pm$29 & 6.9 & 21037$\pm$613 & 05/03/2014 & 88.8$\pm$2.5 & 0.4 & 1919$\pm$55 \\
  &   & 16/02/2014 & 3813$\pm$105 & 11.9 & 63850$\pm$1754 & 07/03/2014 & 689$\pm$18 & 1.3 & 15693$\pm$429 \\
  &   & 09/03/2014 & 603$\pm$19 & 3.1 & 16319$\pm$494 & 20/03/2014 & 232.2$\pm$6.4 & 0.7 & 5194$\pm$144 \\
  &   & 19/03/2014 & 3019$\pm$84 & 11.7 & 60214$\pm$1691 & 23/03/2014 & 360$\pm$10 & 1.2 & 8407$\pm$232 \\
  &   & 30/03/2014 & 2592$\pm$72 & 11.2 & 52113$\pm$1450 &   &   &   &   \\
  &   & 28/06/2014 & 71.9$\pm$3.6 & 1.7 & 1562$\pm$78 &   &   &   &   \\
  &   & 28/06/2014 & 370$\pm$11 & 2.6 & 8216$\pm$250 &   &   &   &   \\
  26 & J1034-3224 & 14/02/2014 & 45.1$\pm$1.2 & 1.0 & 831$\pm$24 & 10/02/2014 & 43.2$\pm$1.2 & 0.2 & 511$\pm$15 \\
  27 &  B1114-41  &   &   &   &   & 10/02/2014 & 9.9$\pm$0.6 & 0.4 & 717$\pm$43 \\
  28 &  B1133+16  & 16/02/2014 & 334.2$\pm$10.8 & 2.9 & 24469$\pm$802 & 10/02/2014 & 66.1$\pm$2.1 & 0.6 & 5780$\pm$183 \\
  29 &  B1237+25  &   &   &   &   & 26/02/2014 & 24.3$\pm$0.9 & 0.5 & 1591$\pm$62 \\
  30 &  B1254-10  &   &   &   &   & 05/03/2014 & 3.8$\pm$0.3 & 0.2 & 310$\pm$23 \\
  31 &  B1325-49  & 16/02/2014 & 11.5$\pm$1.8 & 0.7 & 1387$\pm$216 & 05/03/2014 & 5.9$\pm$0.3 & 0.1 & 622$\pm$31 \\
  32 &  B1504-43  & 14/02/2014 & 8.9$\pm$2.5 & 1.3 & 509$\pm$142 & 05/03/2014 & 4.4$\pm$1.2 & 0.6 & 264$\pm$74 \\
  33 &  B1524-39  &   &   &   &   & 26/02/2014 & 4.3$\pm$0.3 & 0.2 & 390$\pm$27 \\
  &   &   &   &   &   & 26/02/2014 & 5.5$\pm$0.4 & 0.3 & --- \\
  34 & J1549-4848 & 14/02/2014 & 8.8$\pm$1.0 & 0.6 & 310$\pm$35 & 26/02/2014 & 3.8$\pm$0.7 & 0.7 & 114$\pm$23 \\
  35 &  B1552-31  & 16/02/2014 & 18.8$\pm$1.0 & 0.5 & 644$\pm$37 & 26/02/2014 & 19.9$\pm$0.8 & 0.5 & 620$\pm$27 \\
  36 & J1557-4258 & 02/03/2014 & 10.3$\pm$3.1 & 3.1 & 217$\pm$67 & 05/03/2014 & 13.4$\pm$2.4 & 1.3 & 836$\pm$152 \\
  37 &  B1556-44  & 14/02/2014 & 51.8$\pm$1.6 & 0.5 & 1767$\pm$56 &   &   &   &   \\
  38 &  B1558-50  & 16/02/2014 & 12.1$\pm$1.1 & 0.6 & 244$\pm$23 &   &   &   &   \\
  39 & J1603-2531 & 28/03/2014 & 6.9$\pm$1.9 & 1.2 & 624$\pm$186 & 27/03/2014 & 8.9$\pm$0.5 & 0.3 & 463$\pm$29 \\
  40 &  B1600-49  &   &   &   &   & 23/03/2014 & 23.3$\pm$5.2 & 2.9 & 1410$\pm$313 \\
  41 & J1625-4048 &   &   &   &   & 07/03/2014 & 5.2$\pm$0.3 & 0.3 & 348$\pm$23 \\
  42 &  B1642-03  &   &   &   &   & 07/03/2014 & 291.2$\pm$8.1 & 2.2 & 25953$\pm$721 \\
  43 & J1648-3256 & 30/03/2014 & $< 6.8$ & 1.4 & --- & 20/03/2014 & 3.1$\pm$0.7 & 0.5 & 201$\pm$47 \\
  44 & J1700-3312 & 19/03/2014 & $< 3.6$ & 0.8 & --- & 20/03/2014 & 5.4$\pm$0.5 & 0.5 & 308$\pm$31 \\
  45 &  B1700-32 & 02/03/2014 & 30.8$\pm$1.3 & 0.6 & 904$\pm$38 & 07/03/2014 & 29.7$\pm$0.9 & 0.2 & 984$\pm$28 \\
  46 & J1705-3423 & 02/03/2014 & 13.9$\pm$0.8 & 0.4 & 75$\pm$8 & 07/03/2014 & 15.0$\pm$0.5 & 0.3 & 279$\pm$13 \\
  &   & 19/03/2014 & 24.0$\pm$1.1 & 0.5 & 129$\pm$15 &   &   &   &   \\
  47 &  B1706-16  & 16/02/2014 & 66.7$\pm$0.9 & 0.4 & 3607$\pm$110 &   &   &   &   \\
  48 &  B1706-44  & 14/02/2014 & 9.6$\pm$1.1 & 0.6 & 99$\pm$18 & 20/03/2014 & 13.2$\pm$0.7 & 0.4 & 171$\pm$12 \\
  &   & 19/03/2014 & 14.0$\pm$1.6 & 0.9 & 137$\pm$24 &   &   &   &   \\
  49 &  B1717-29  &   &   &   &   & 07/03/2014 & 43.6$\pm$1.5 & 0.6 & 1075$\pm$43 \\
  50 &  B1718-32  &   &   &   &   & 10/02/2014 & 18.9$\pm$0.7 & 0.2 & 724$\pm$28 \\
  51 &  B1719-37  & 14/02/2014 & 11.3$\pm$1.3 & 0.7 & 327$\pm$40 & 10/02/2014 & 7.3$\pm$1.3 & 0.8 & 311$\pm$54 \\
  52 & J1727-2739 &   &   &   &   & 07/03/2014 & 2.2$\pm$0.2 & 0.1 & 56$\pm$7 \\
  53 &  B1727-47  & 14/02/2014 & 76.2$\pm$2.2 & 0.4 & 4551$\pm$131 & 20/03/2014 & 65.4$\pm$1.8 & 0.1 & 5100$\pm$140 \\
  54 &  B1730-22  & 31/05/2014 & 24.9$\pm$1.2 & 0.6 & 902$\pm$43 & 07/03/2014 & 35.4$\pm$2.6 & 2.0 & 904$\pm$69 \\
  55 &  B1730-37  & 02/03/2014 & 2.5$\pm$0.5 & 0.4 & --- & 20/03/2014 & 4.9$\pm$0.5 & 0.3 & 167$\pm$24 \\
  56 &  B1732-07  & 16/02/2014 & 55.8$\pm$1.6 & 0.7 & 3839$\pm$106 &   &   &   &   \\
  57 &  B1736-29  &   &   &   &   & 07/03/2014 & 6.4$\pm$0.3 & 0.2 & 170$\pm$12 \\
  58 &  B1737+13  & 12/07/2014 & 31.2$\pm$1.1 & 0.4 & 1419$\pm$50 & 26/02/2014 & 36.6$\pm$1.0 & 0.2 & 1265$\pm$37 \\
  59 &  B1737-39  & 19/03/2014 & 34.7$\pm$1.3 & 1.0 & 600$\pm$27 & 05/03/2014 & 21.2$\pm$0.7 & 0.2 & 1016$\pm$34 \\
  60 &  B1738-08  & 12/07/2014 & 10.7$\pm$0.4 & 0.2 & 354$\pm$16 & 23/03/2014 & 10.4$\pm$0.3 & 0.1 & 371$\pm$13 \\
  61 &  B1742-30  & 31/05/2014 & 9.4$\pm$0.6 & 0.3 & 276$\pm$18 & 10/02/2014 & 32.7$\pm$1.1 & 0.3 & 1804$\pm$59 \\
  62 &  B1745-12  & 16/02/2014 & 17.4$\pm$0.7 & 0.3 & 635$\pm$28 &   &   &   &   \\
  63 & J1750-3503 & 30/03/2014 & 10.4$\pm$0.9 & 0.5 & 120$\pm$17 & 27/03/2014 & 6.3$\pm$0.6 & 0.4 & 99$\pm$16 \\
  64 &  B1747-46  & 19/03/2014 & 68.9$\pm$2.2 & 0.7 & 3541$\pm$111 & 20/03/2014 & 34.3$\pm$1.0 & 0.2 & 2116$\pm$62 \\
  65 &  B1749-28  & 30/03/2014 & 238.1$\pm$6.5 & 0.6 & 17407$\pm$478 & 27/03/2014 & 311.5$\pm$8.7 & 2.2 & 27207$\pm$760 \\
  66 &  B1754-24  & 30/03/2014 & 4.6$\pm$1.1 & 0.7 & --- & 27/03/2014 & 6.8$\pm$0.8 & 0.6 & 149$\pm$22 \\
  &   & 27/07/2014 & 6.1$\pm$0.9 & 0.4 & --- &   &   &   &   \\
  67 &  B1758-03  & 28/03/2014 & 12.6$\pm$0.8 & 0.4 & 828$\pm$55 & 23/03/2014 & 5.4$\pm$0.3 & 0.1 & 470$\pm$26 \\
  68 &  B1758-29  & 30/03/2014 & 10.3$\pm$0.5 & 0.2 & 346$\pm$18 & 27/03/2014 & 10.5$\pm$0.3 & 0.2 & 388$\pm$19 \\
  69 &  B1804-08  & 28/03/2014 & 27.3$\pm$1.8 & 0.9 & 480$\pm$33 & 23/03/2014 & 40.4$\pm$1.6 & 0.7 & 1504$\pm$59 \\
  70 & J1808-0813 & 28/03/2014 & 8.8$\pm$1.4 & 0.7 & 183$\pm$31 & 23/03/2014 & 8.7$\pm$0.6 & 0.3 & 307$\pm$23 \\
  71 &  B1813-26  & 30/03/2014 & 12.0$\pm$0.8 & 0.4 & 194$\pm$16 & 27/03/2014 & 7.4$\pm$0.8 & 0.4 & 186$\pm$24 \\
  72 &  B1813-36  & 19/03/2014 & 16.2$\pm$0.9 & 0.7 & 755$\pm$42 &   &   &   &   \\
  73 & J1817-3837 & 30/03/2014 & 4.8$\pm$2.1 & 0.9 & 223$\pm$97 & 27/03/2014 & 3.4$\pm$0.6 & 0.4 & 231$\pm$40 \\
  74 &  B1818-04  & 28/03/2014 & 148.8$\pm$4.9 & 1.3 & 6245$\pm$205 & 27/03/2014 & 62.1$\pm$2.2 & 0.8 & 4069$\pm$142 \\
  75 &  B1819-22  & 31/05/2014 & 16.2$\pm$0.5 & 0.1 & 402$\pm$13 & 27/03/2014 & 17.2$\pm$0.6 & 0.2 & 585$\pm$23 \\
  &   &   &   &   &   & 27/03/2014 & 16.3$\pm$0.6 & 0.3 & 588$\pm$24 \\
  76 & J1823-0154 & 28/03/2014 & 4.0$\pm$1.1 & 0.6 & 201$\pm$58 & 27/03/2014 & 3.2$\pm$0.4 & 0.3 & 383$\pm$53 \\
  77 &  B1821+05  & 02/03/2014 & 25.0$\pm$1.2 & 0.5 & 1504$\pm$73 & 26/02/2014 & 8.4$\pm$0.6 & 0.3 & 309$\pm$25 \\
  &   & 12/07/2014 & 43.7$\pm$2.0 & 0.6 & 2512$\pm$118 &   &   &   &   \\
  78 &  B1820-31  & 19/03/2014 & 45.5$\pm$1.6 & 0.6 & 2143$\pm$77 & 23/03/2014 & 17.3$\pm$0.8 & 0.4 & 897$\pm$45 \\
  &   & 28/03/2014 & 46.7$\pm$2.1 & 0.9 & 2229$\pm$100 &   &   &   &   \\
  79 &  B1831-04  & 19/03/2014 & 60.9$\pm$2.5 & 1.1 & 518$\pm$23 & 20/03/2014 & 53.7$\pm$1.8 & 0.7 & 334$\pm$13 \\
  80 & J1835-1020 & 12/07/2014 & 3.1$\pm$1.3 & 0.7 & --- & 07/03/2014 & 2.8$\pm$0.6 & 0.3 & 131$\pm$27 \\
  81 & J1835-1106 & 12/07/2014 & 10.1$\pm$1.4 & 0.7 & 112$\pm$17 & 07/03/2014 & 5.2$\pm$0.7 & 0.5 & 175$\pm$27 \\
  82 &  B1839+09  & 02/03/2014 & 13.6$\pm$3.9 & 1.7 & 761$\pm$213 & 26/02/2014 & 9.8$\pm$0.7 & 0.4 & 498$\pm$37 \\
  83 &  B1839-04  & 19/03/2014 & 10.2$\pm$1.3 & 0.6 & --- & 20/03/2014 & 17.9$\pm$0.7 & 0.3 & 235$\pm$13 \\
  &   & 27/07/2014 & 11.4$\pm$0.7 & 0.3 & 72$\pm$8 &   &   &   &   \\
  84 & J1843-0000 & 19/03/2014 & 7.0$\pm$0.4 & 0.3 & --- & 20/03/2014 & 8.1$\pm$0.3 & 0.1 & 191$\pm$9 \\
  &   & 27/07/2014 & 7.7$\pm$0.5 & 0.2 & --- &   &   &   &   \\
  85 &  B1842+14  & 02/03/2014 & 36.0$\pm$1.1 & 0.3 & 1562$\pm$51 & 07/03/2014 & 7.5$\pm$0.3 & 0.2 & 313$\pm$16 \\
  &   & 30/03/2014 & 26.3$\pm$2.2 & 1.4 & 1188$\pm$107 &   &   &   &   \\
  86 &  B1844-04  & 19/03/2014 & 42.2$\pm$1.6 & 0.6 & 525$\pm$22 & 20/03/2014 & 15.6$\pm$0.5 & 0.2 & 586$\pm$22 \\
  87 &  B1845-01  & 21/06/2014 & 32.1$\pm$1.5 & 0.7 & 161$\pm$14 & 20/03/2014 & 26.8$\pm$1.3 & 0.7 & 657$\pm$34 \\
  88 & J1848-1414 & 02/03/2014 & $< 4.1$ & 0.8 & --- & 07/03/2014 & 2.0$\pm$0.7 & 0.5 & 60$\pm$23 \\
  &   & 27/07/2014 & $< 5.2$ & 1.0 & --- &   &   &   &   \\
  89 &  B1846-06  & 31/05/2014 & 20.2$\pm$0.6 & 0.2 & 736$\pm$25 &   &   &   &   \\
  &   & 21/06/2014 & 16.6$\pm$0.5 & 0.2 & 615$\pm$21 &   &   &   &   \\
  90 & J1852-2610 & 21/06/2014 & 18.8$\pm$2.1 & 1.1 & 619$\pm$70 &   &   &   &   \\
  91 &  B1857-26  & 12/07/2014 & 119.9$\pm$3.6 & 0.8 & 3397$\pm$103 & 09/02/2014 & 77.5$\pm$2.2 & 0.3 & 1236$\pm$38 \\
  92 & J1901-0906 & 27/07/2014 & 7.8$\pm$0.4 & 0.2 & 818$\pm$47 & 07/03/2014 & 5.7$\pm$0.3 & 0.2 & 526$\pm$28 \\
  93 &  B1907+10  & 02/03/2014 & 36.3$\pm$1.4 & 0.6 & 1699$\pm$66 & 26/02/2014 & 14.0$\pm$1.6 & 0.9 & 937$\pm$108 \\
  94 &  B1907+03  & 12/07/2014 & 28.8$\pm$1.1 & 0.4 & 343$\pm$15 & 05/03/2014 & 15.7$\pm$0.5 & 0.1 & 227$\pm$10 \\
  95 &  B1911-04  & 14/02/2014 & 81.5$\pm$2.3 & 0.4 & 8863$\pm$252 & 10/02/2014 & 78.2$\pm$2.1 & 0.2 & 8005$\pm$220 \\
  96 &  B1914+09  & 02/03/2014 & 8.4$\pm$0.8 & 0.5 & 381$\pm$39 & 26/02/2014 & 6.2$\pm$1.2 & 0.7 & 268$\pm$54 \\
  &   & 21/06/2014 & 10.6$\pm$1.3 & 0.5 & 488$\pm$59 &   &   &   &   \\
  97 &  B1915+13  & 02/03/2014 & 35.9$\pm$1.1 & 0.3 & 1536$\pm$51 & 26/02/2014 & 10.2$\pm$0.9 & 0.5 & 566$\pm$49 \\
  &   & 09/03/2014 & 41.5$\pm$1.6 & 0.5 & 1794$\pm$68 &   &   &   &   \\
  98 &  B1917+00  & 02/03/2014 & 9.3$\pm$0.9 & 0.5 & 887$\pm$82 & 05/03/2014 & 4.6$\pm$0.4 & 0.2 & 389$\pm$35 \\
  &   & 21/06/2014 & 18.7$\pm$1.1 & 0.5 & 1637$\pm$99 &   &   &   &   \\
  99 & J1919+0134 & 09/03/2014 & 2.4$\pm$0.5 & 0.4 & --- &   &   &   &   \\
 100 &  B1918+19  & 02/03/2014 & 19.8$\pm$1.0 & 0.5 & 326$\pm$18 & 26/02/2014 & 10.2$\pm$0.5 & 0.2 & 161$\pm$11 \\
 101 &  B1919+21  &   &   &   &   & 27/03/2014 & 31.1$\pm$0.9 & 0.3 & 1547$\pm$48 \\
 102 &  B1929+10  & 14/02/2014 & 127.8$\pm$3.9 & 1.0 & 4620$\pm$142 & 09/02/2014 & 137.2$\pm$4.4 & 1.6 & 4556$\pm$151 \\
  &   & 16/02/2014 & 195.0$\pm$6.5 & 2.2 & 6555$\pm$218 & 10/02/2014 & 130.5$\pm$3.7 & 0.8 & 4828$\pm$139 \\
  &   & 16/02/2014 & 390.9$\pm$11.4 & 2.6 & 13769$\pm$400 & 26/02/2014 & 34.6$\pm$1.2 & 0.5 & 1225$\pm$44 \\
  &   & 02/03/2014 & 312.3$\pm$9.1 & 2.7 & 10956$\pm$320 & 26/02/2014 & 22.2$\pm$1.3 & 0.7 & --- \\
  &   & 09/03/2014 & 173.7$\pm$5.6 & 1.7 & 6763$\pm$217 & 05/03/2014 & 139.9$\pm$4.1 & 0.8 & 5401$\pm$155 \\
  &   & 19/03/2014 & 139.9$\pm$4.6 & 1.7 & 5137$\pm$170 & 05/03/2014 & 244.1$\pm$7.4 & 2.3 & --- \\
  &   & 30/03/2014 & 344.6$\pm$14.6 & 7.7 & 12742$\pm$540 & 07/03/2014 & 123.2$\pm$3.4 & 0.5 & 4718$\pm$132 \\
  &   & 31/05/2014 & 111.2$\pm$3.8 & 1.6 & 4040$\pm$137 & 07/03/2014 & 53.5$\pm$1.9 & 0.9 & --- \\
  &   & 21/06/2014 & 353.3$\pm$10.4 & 2.5 & 12467$\pm$368 & 20/03/2014 & 231.4$\pm$6.4 & 0.9 & --- \\ 
  &   & 12/07/2014 & 200.2$\pm$6.2 & 1.7 & 7036$\pm$217 & 20/03/2014 & 261.8$\pm$1.2 & 0.8 & 10021$\pm$276 \\
  &   & 27/07/2014 & 243.5$\pm$8.0 & 3.2 & 8851$\pm$293 & 23/03/2014 & 129.9$\pm$3.8 & 1.0 & 4349$\pm$129 \\
  &   & 27/07/2014 & 73.1$\pm$2.6 & 1.0 & 2954$\pm$104 & 23/03/2014 & 81.2$\pm$2.5 & 0.8 & 5496$\pm$178 \\
  &   &   &   &   &   & 27/03/2014 & 58.5$\pm$1.8 & 0.8 & 2025$\pm$63 \\
 103 &  B1937-26  & 16/02/2014 & 7.7$\pm$1.9 & 0.9 & 523$\pm$128 &   &   &   &   \\
 104 &  B1944+17  & 16/02/2014 & 15.1$\pm$1.8 & 0.9 & 391$\pm$47 & 27/03/2014 &  51.6$\pm$1.6 & 0.5 & 996$\pm$34 \\
  &   & 12/07/2014 & 34.8$\pm$2.2 & 1.2 & 728$\pm$46 &   &   &   &   \\
  &   & 27/07/2014 & 30.3$\pm$1.6 & 0.7 & 594$\pm$31 &   &   &   &   \\
 105 &  B2003-08  & 14/02/2014 & 34.3$\pm$2.0 & 1.0 & 426$\pm$27 & 10/02/2014 & 18.5$\pm$1.5 & 0.8 & 247$\pm$24 \\
 106 &  B2043-04  & 14/02/2014 & 27.8$\pm$1.8 & 1.1 & 2278$\pm$147 & 10/02/2014 & 13.8$\pm$0.5 & 0.2 & 989$\pm$33 \\
  &   & 16/02/2014 & 40.5$\pm$4.7 & 0.5 & 2878$\pm$335 &   &   &   &   \\
 107 &  B2045-16  & 19/03/2014 & 221.5$\pm$6.3 & 1.3 & 10180$\pm$291 & 20/03/2014 & 75.4$\pm$2.3 & 0.9 & 4166$\pm$129 \\
 108 & J2144-3933 & 16/02/2014 & 21.6$\pm$1.2 & 0.6 & 8472$\pm$476 & 09/02/2014 & 3.6$\pm$0.4 & 0.2 & 1700$\pm$177 \\
 109 &  B2303+30  & 27/07/2014 & 23.0$\pm$0.9 & 0.3 & 1648$\pm$65 &   &   &   &   \\
 110 &  B2310+42  & 27/07/2014 & 90.8$\pm$3.4 & 1.5 & 4070$\pm$152 &   &   &   &   \\
 111 &  B2315+21  & 09/03/2014 & 37.2$\pm$1.4 & 0.4 & 2469$\pm$93 & 27/03/2014 & 4.2$\pm$0.2 & 0.1 & 309$\pm$17 \\
  &   & 21/06/2014 & 21.2$\pm$1.9 & 0.8 & 1635$\pm$148 &   &   &   &   \\
 112 &  B2327-20  &   &   &   &   & 23/03/2014 & 12.6$\pm$0.4 & 0.1 & 1515$\pm$48 \\
 113 & J2346-0609 & 09/03/2014 & 4.8$\pm$1.1 & 0.6 & --- & 23/03/2014 & 4.8$\pm$0.3 & 0.2 & 264$\pm$18 \\
  &   & 21/06/2014 & 10.3$\pm$1.6 & 0.8 & 399$\pm$61 &   &   &   &   \\
\enddata
\end{deluxetable}

\appendix

\section{Estimating the Flux Scale from Calibrators}

\startlongtable
\begin{deluxetable}{cccccl}
\tablenum{5}
\tablecaption{Calibrator Sources\label{tab:cal_flux}}
\tablewidth{0pt}
\tablehead{ \colhead{Date} & \colhead{Flux Cal} & \colhead{Flux} & \colhead{Phase Cal} & \colhead{Flux} & \colhead{Calibrated PSR} \\
   &   & \colhead{(Jy)} &   & \colhead{(Jy)} &  }

\startdata
   \multicolumn{6}{c}{\underline{325 MHz Observations}} \\
   &   &   &   &   &   \\
 14/02/2014 & 3C286 & 26.4 & 2047-026 & 5.9 & B2003-08, B2043-04 \\
   &   &   & 1822-096 & 3.7 & B1911-04, B1929+10 \\
   &   &   & 1830-360 & 20.7 & B1706-44, B1719-37, B1727-47 \\
   &   &   & 1625-311 & 3.5 & B1504-43, J1549-4848, B1556-44 \\
   &   &   & 1021+219 & 6.6 & B0950+08 \\
   &   &   & 1033-343 & 4.8 & J1034-3224 \\
   &   &   & 0804+102 & 5.6 & B0834+06 \\
   &   &   &   &   &   \\
 16/02/2014 & 3C48 & 43.9 & 0521+166 & 15.6 & B0523+11, B0525+21, B0540+23, \\
   &   &   &   &   & B0611+22, B0626+24, B0656+14 \\
   &   &   & 0521-207 & 8.9 & B0450-18 \\
   &   &   & 0116-208 & 10.8 & B0031-07, J0134-2937, B0148-06, B0149-16 \\
   & 3C286 & 26.4 & 1830-360 & 21.5 & B1937-26 \\
   &   &   & 2047-026 & 6.2 & B1929+10, B2043-04 \\
   &   &   & 1924+334 & 6.2 & B1929+10, B1944+17 \\
   &   &   & 1625-311 & 3.2 & B1552-31, B1558-50 \\
   &   &   & 1822-096 & 3.6 & B1706-16, B1732-07, B1745-12 \\
   &   &   & 1156+314 & 5.3 & B1133+16 \\
   &   &   & 1311-222 & 16.7 & B1325-49 \\
   &   &   & 1021+219 & 5.7 & B0950+08 \\
   &   &   & 2137-207 & 7.9 & J2144-3933 \\
   &   &   &   &   &   \\
 02/03/2014 & 3C286 & 26.4 & 1822-096 & 3.3 & B1821+05, B1917+00 \\
   &   &   & 1924+334 & 6.5 & B1839+09, B1842+14, B1907+10, B1914+09, \\
   &   &   &   &   & B1915+13, B1918+19, B1929+10 \\
   &   &   & 1625-311 & 3.1 & J1557-4258, B1700-32, J1705-3423, B1730-37 \\
   &   &   &   &   &   \\
 09/03/2014 & 3C48 & 43.9 & 0025-260 & 24.7 & J2346-0609 \\
   &   &   & 0116-208 & 11.5 & B0148-06 \\
   &   &   & 2251+188 & 7.9 & B2315+21 \\
   &   &   & 1924+334 & 6.5 & B1915+13, J1919+0134, B1929+10 \\
   &   &   & 0521+166 & 15.5 & B0656+14 \\
   &   &   & 1021+219 & 6.1 & B0950+08 \\
   &   &   & 0837-198 & 12.7 & B0740-28, B0844-35 \\
   &   &   &   &   &   \\
 19/03/2014 & 3C353 & 153.4 & 1830-360 & 20.7 & J1705-3423, B1706-44, B1737-39, \\
   &   &   &   &   & B1813-36, B1820-31 \\
   &   &   & 2137-207 & 8.7 & B1929+10, B2045-16 \\
   &   &   & 1822-096 & 3.6 & B1831-04, B1839-04, B1844-04, J1843-0000 \\
   &   &   & 1625-311 & 3.3 & J1700-3312, B1747-46 \\
   &   &   & 1021+219 & 6.1 & B0950+08  \\
   &   &   &   &   &   \\
 27/03/2014 & 3C147 & 54.0 & 0318+164 & 8.4 & B0301+19 \\
   &   &   &   &   &   \\
 28/03/2014 & 3C286 & 26.4 & 1830-360 & 19.9 & B1820-31 \\
   &   &   & 1822-096 & 3.8 & B1758-03, B1804-08, J1808-0813, \\
   &   &   &   &   & B1818-04, J1823-0154 \\
   &   &   & 1510-057 & 7.8 & J1603-2531  \\
   &   &   &   &   &   \\
 30/03/2014 & 3C286 & 26.4 & 1830-360 & 20.5 & J1648-3256, B1749-28, J1750-3503, B1754-24 \\
   &   &   &   &   & B1758-29, B1813-26, J1817-3837 \\
   &   &   & 1822-096 & 3.6 & B1842+14, B1929+10 \\
   &   &   & 1021+219 & 6.1 & B0950+08 \\
   &   &   &   &   &   \\
 31/05/2014 & 3C286 & 26.4 &   &   &   \\
   &   &   & 1822-096 & 3.8 & B1819-22, B1846-06, B1929+10 \\
   &   &   & 1830-360 & 21.0 & B1730-22, B1742-30  \\
   & 3C48 & 43.9 &   &   &   \\
   &   &   &   &   &   \\
 21/06/2014 & 3C48 & 43.9 & 1822-096 & 3.6 & B1845-01, B1846-06, J1852-2610, B1929+10 \\
   &   &   & 1924+334 & 6.5 & B1914+09, B1917+00 \\
   &   &   & 2225-049 & 16.8 & B2315+21, J2346-0609 \\
   &   &   &   &   &   \\
 28/06/2014 & 3C48 & 43.9 & 0116-208 & 12.1 & B0203-40 \\
   &   &   & 0323+055 & 7.3 & B0301+19 \\
   &   &   & 1021+219 & 6.1 & B0919+06, B0950+08 \\
   &   &   & 0521+166 & 15.7 & B0656+14 \\
   &   &   & 0837-198 & 11.4 & B0727-18, B0736-40, B0844-35, B0942-13 \\
   &   &   & 1033-343 & 4.8 & J0905-5127 \\
   &   &   &   &   &   \\
 12/07/2014 & 3C286 & 26.4 & 1822-096 & 3.5 & B1737+13, B1738-08, J1835-1106, B1857-26 \\
   &   &   & 1924+334 & 6.6 & B1821+05, J1835-1020, B1907+03, B1929+10 \\
   & 3C48 & 43.9 & 1924+334 & 6.6 & B1944+17 \\
   &   &   &   &   &   \\
 27/07/2014 & 3C286 & 26.4 &   &   &   \\
   &   &   & 1822-096 & 3.6 & B1754-24, B1839-04, J1843-0000, \\
   &   &   &   &   & J1848-1414, J1901-0906, B1929+10 \\
   &   &   & 1924+334 & 6.6 & B1944+17 \\
   &   &   & 2251+188 & 7.8 & B2303+30, B2310+42 \\
   & 3C48 & 43.9 &   &   &   \\
   \multicolumn{6}{c}{\underline{610 MHz Observations}} \\
   &   &   &   &   &   \\
 09/02/2014 & 3C147 & 39.0 & 0521+166 & 13.7 & B0523+11, B0540+23, B0611+22, \\
   &   &   &   &   & B0626+24, B0656+14 \\
   &   &   & 0521-207 & 4.4 & B0450-18 \\
   &   &   & 0116-208 & 7.9 & B0031-07, J0134-2937, B0149-16 \\
   &   &   & 1924+334 & 6.3 & B1929+10 \\
   &   &   & 2219-279 & 4.4 & J2144-3933 \\
   &   &   & 1830-360 & 15.8 & B1857-26 \\
   &   &   &   &   &   \\
 10/02/2014 & 3C286 & 21.7 & 2047-026 & 4.2 & B1911-04, B2003-08, B2043-04 \\
   &   &   & 1924+334 & 6.3 & B1929+10 \\
   &   &   & 1822-096 & 6.5 & B1718-32, B1719-37, B1742-30 \\
   &   &   & 1154-350 & 9.5 & B1114-41 \\
   &   &   & 1033-343 & 3.4 & J1034-3224 \\
   &   &   & 0804+102 & 3.5 & B0834+06 \\
   &   &   & 0837-198 & 5.8 & B0844-35, B0818-41 \\
   &   &   & 1021+219 & 3.0 & B1133+16 \\
   &   &   &   &   &   \\
 26/02/2014 & 3C286 & 21.7 & 1822-096 & 6.7 & B1737+13, B1821+05, B1839+09, B1929+10 \\
   &   &   & 1924+334 & 6.3 & B1907+10, B1914+09, B1915+13, \\
   &   &   &   &   & B1918+19, B1929+10 \\
   &   &   & 1625-311 & 2.5 & B1524-39, J1549-4848, B1552-31 \\
   &   &   & 1330+251 & 8.2 & B1237+25 \\
   &   &   &   &   &   \\
 05/03/2014 & 3C286 & 21.7 & 1021+219 & 3.1 & B0950+08 \\
   &   &   & 1311-222 & 8.9 & B1254-10, B1325-49 \\
   &   &   & 1625-311 & 2.3 & B1504-43, J1557-4258 \\
   &   &   & 1822-096 & 6.5 & B1907+03, B1917+00, B1929+10 \\
   &   &   & 1830-360 & 16.1 & B1737-39 \\
   &   &   &   &   &   \\
 07/03/2014 & 3C286 & 21.7 & 1021+219 & 3.0 & B0950+08 \\
   &   &   & 1510-057 & 3.4 & B1642-03 \\
   &   &   & 1625-311 & 2.4 & J1625-4048, B1717-29, B1730-22 \\
   &   &   & 1822-096 & 6.5 &  J1835-1020, J1835-1106, B1842+14, \\
   &   &   &   &   & J1848-1414, J1901-0906, B1929+10 \\
   &   &   & 1830-360 & 16.1 & B1700-32, J1705-3423, J1727-2739, B1736-29 \\
   &   &   &   &   &   \\
 20/03/2014 & 3C48 & 30.1 & 2137-207 & 7.0 & B1929+10, B2045-16 \\
   &   &   & 1822-096 & 6.2 & B1831-04, B1839-04, J1843-0000, \\
   &   &   &   &   & B1844-04, B1845-01, B1929+10 \\
   &   &   & 1830-360 & 16.5 & B1706-44, B1727-47, B1730-37, B1747-46 \\
   &   &   & 1625-311 & 2.2 & J1648-3256, J1700-3312 \\
   &   &   & 1021+219 & 3.0 & B0919+06, B0950+08 \\
   &   &   &   &   &   \\
 23/03/2014 & 3C286 & 21.7 & 1822-096 & 6.3 & B1738-08, B1758-03, B1804-08, \\
   &   &   &   &   & J1808-0813, B1820-31, B1929+10 \\
   &   &   & 1625-311 & 2.4 & B1600-49 \\
   & 3C48 & 30.1 & 2219-279 & 4.7 & B1929+10 \\
   &   &   & 0025-260 & 14.5 & B2327-20, J2346-0609 \\
   &   &   & 0116-208 & 8.2 & B0148-06, B0203-40 \\
   &   &   & 0318+164 & 9.4 & B0301+19 \\
   &   &   & 0706-231 & 6.1 & B0628-28, B0727-18, B0740-28, B0756-15 \\
   &   &   & 1021+219 & 3.0 & B0950+08 \\
   &   &   &   &   &   \\
 27/03/2014 & 3C48 & 30.1 & 1924+334 & 6.2 & B1919+21, B1944+17 \\
   &   &   & 1822-096 & 6.8 & J1817-3837, B1818-04, B1819-22, \\
   &   &   &   &   & J1823-0154, B1929+10 \\
   &   &   & 1625-311 & 2.5 & J1603-2531, B1749-28, J1750-3503, B1754-24 \\
   &   &   & 2251+188 & 4.0 & B2315+21 \\
   &   &   & 1830-360 & 16.1 & B1758-29, B1813-26 \\
\enddata
\end{deluxetable}

\section{Mode Changing in pulsars} \label{app:mode}
In this section we present the average profiles of different modes in the 12 
pulsars that show the mode changing behaviour.

\begin{figure}
\gridline{\fig{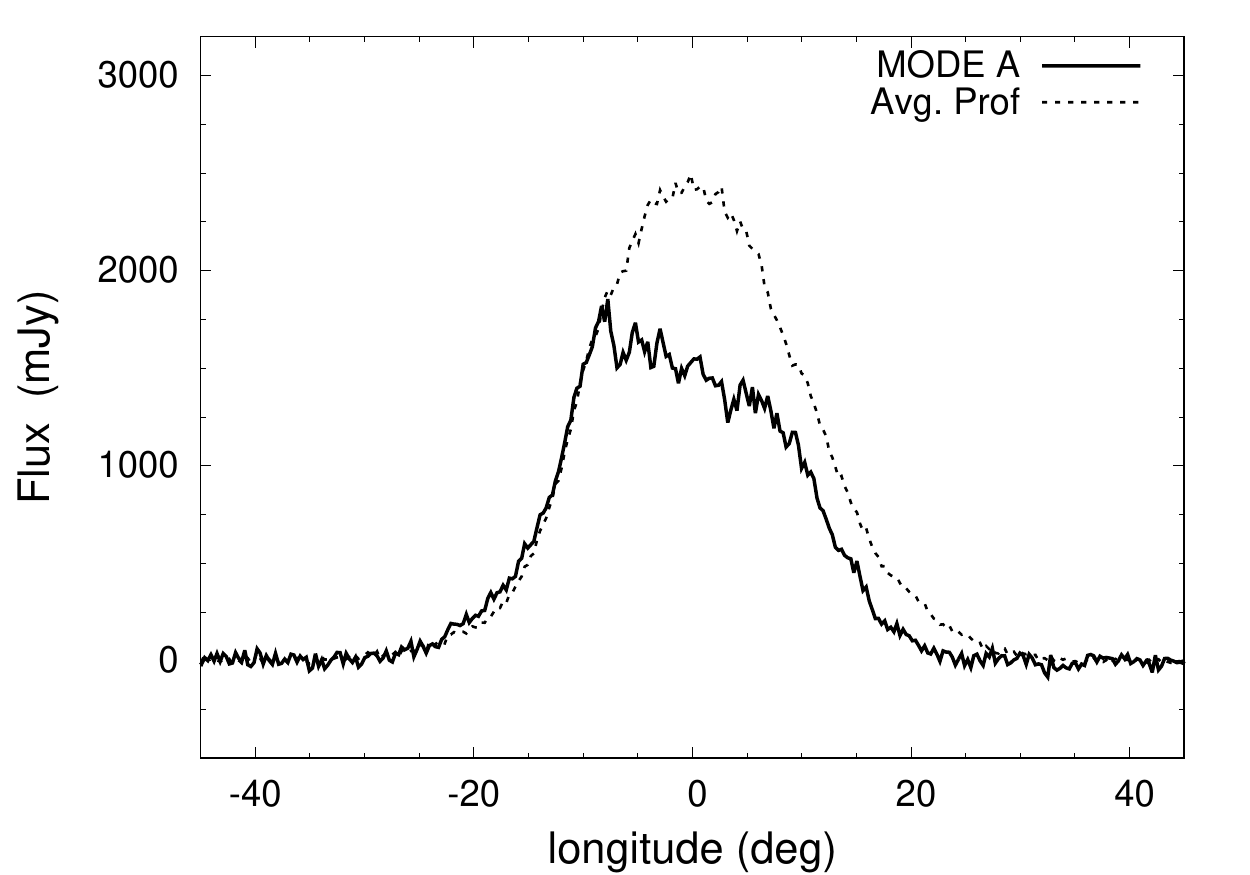}{0.3\textwidth}{(a)}
          \fig{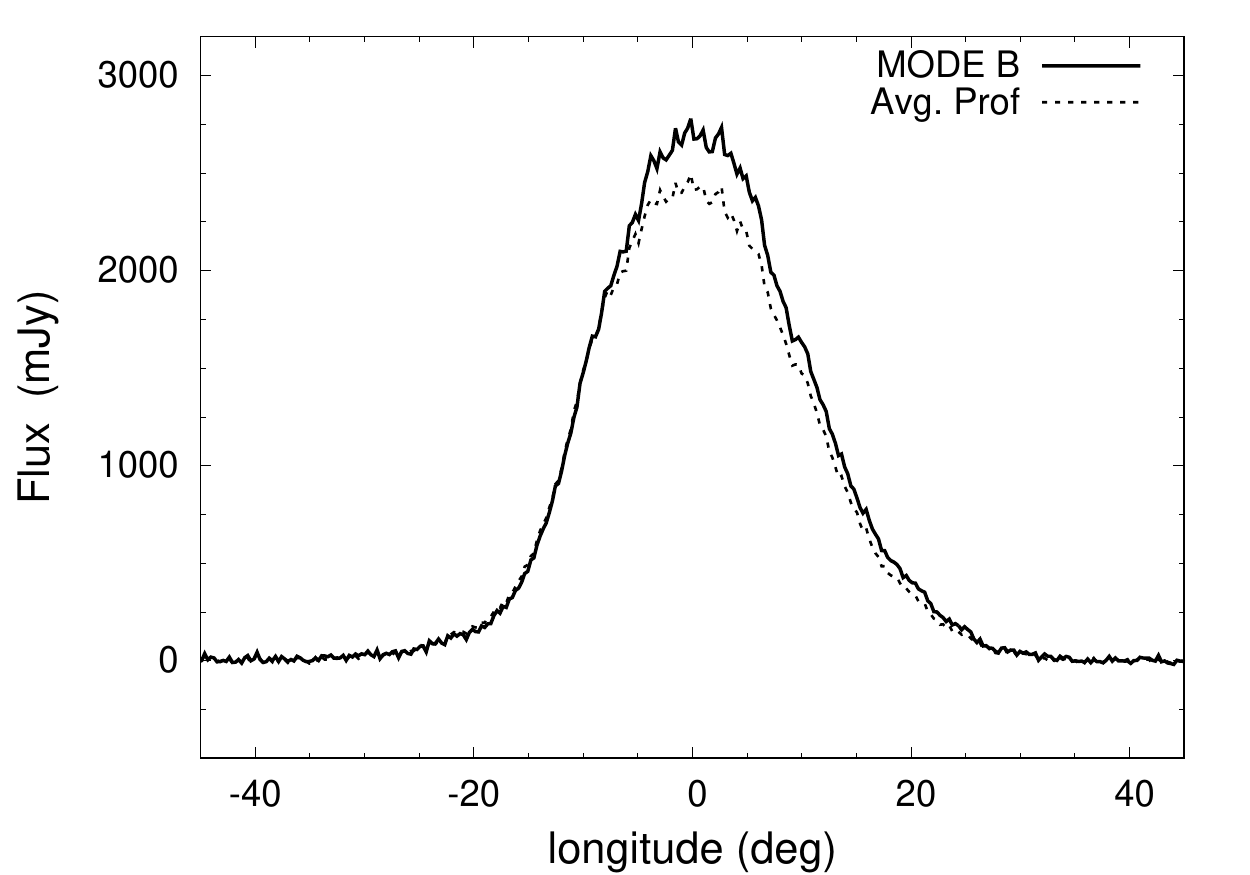}{0.3\textwidth}{(b)}
          \fig{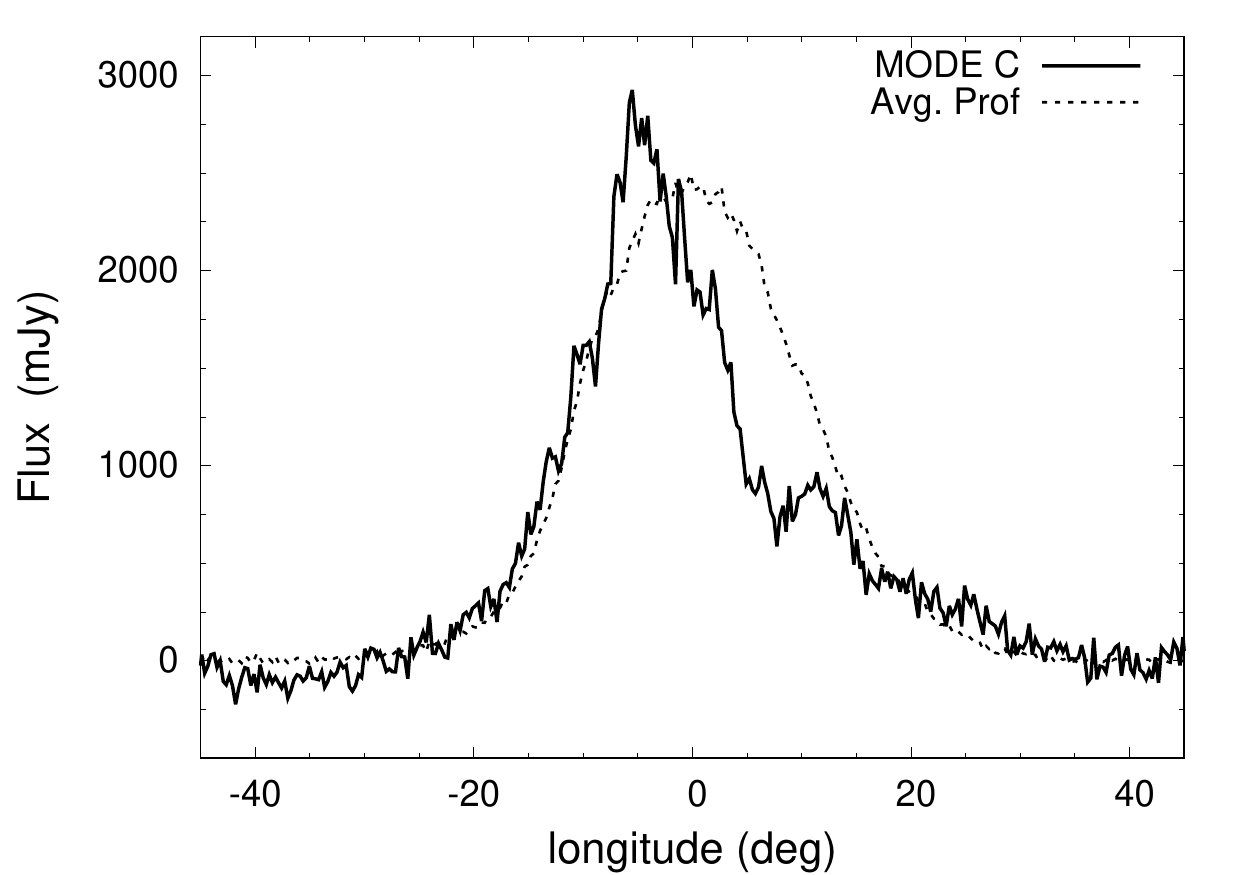}{0.3\textwidth}{(c)}
         }
\caption{The average profiles in PSR B0031--07 at 325 MHz, (a) Mode A, (b) Mode
B and (c) Mode C.
\label{fig:appB0031}}
\end{figure}

\begin{figure}
\gridline{\fig{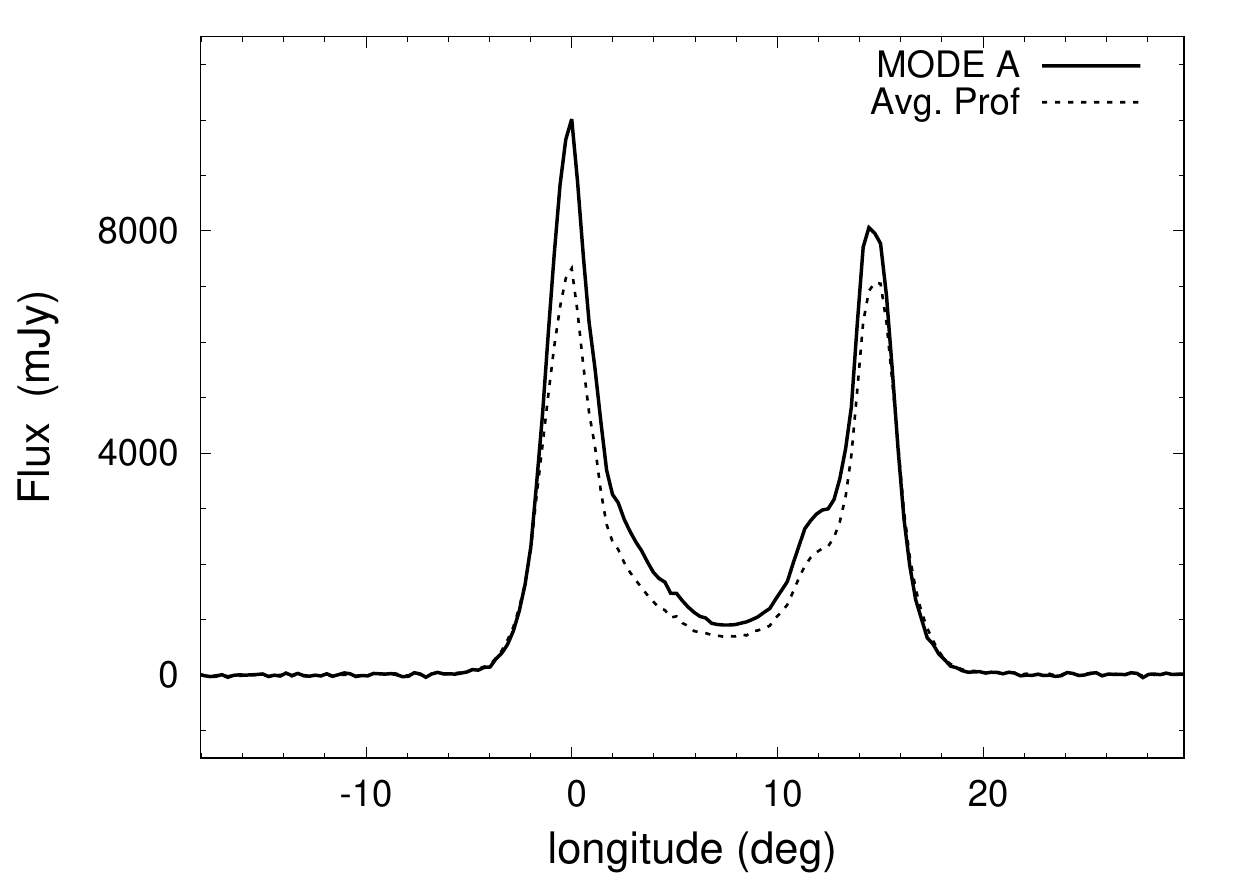}{0.3\textwidth}{(a)}
          \fig{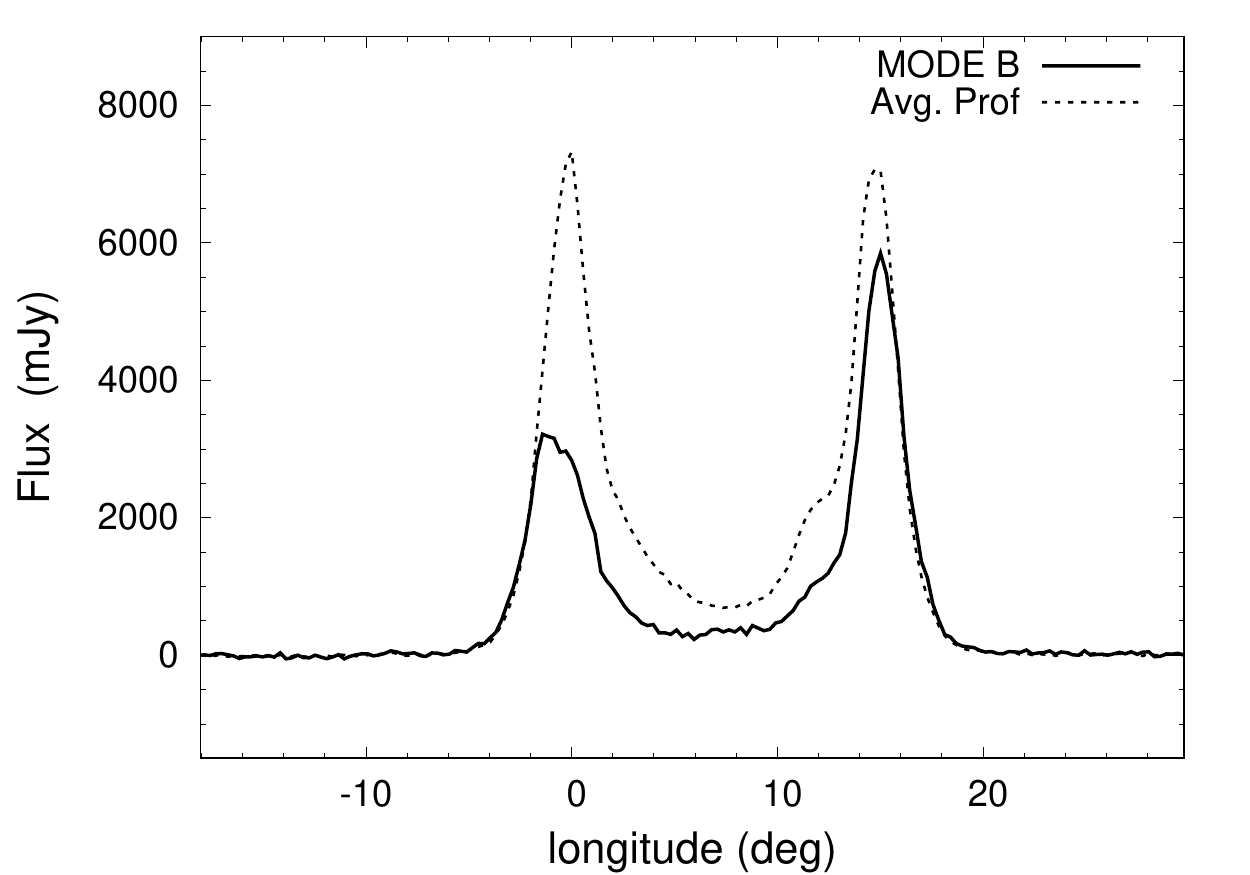}{0.3\textwidth}{(b)}
         }
\caption{The average profiles in PSR B0525+21 at 325 MHz, (a) Mode A and (b) 
Mode B.
\label{fig:appB0525}}
\end{figure}

\begin{figure}
\gridline{\fig{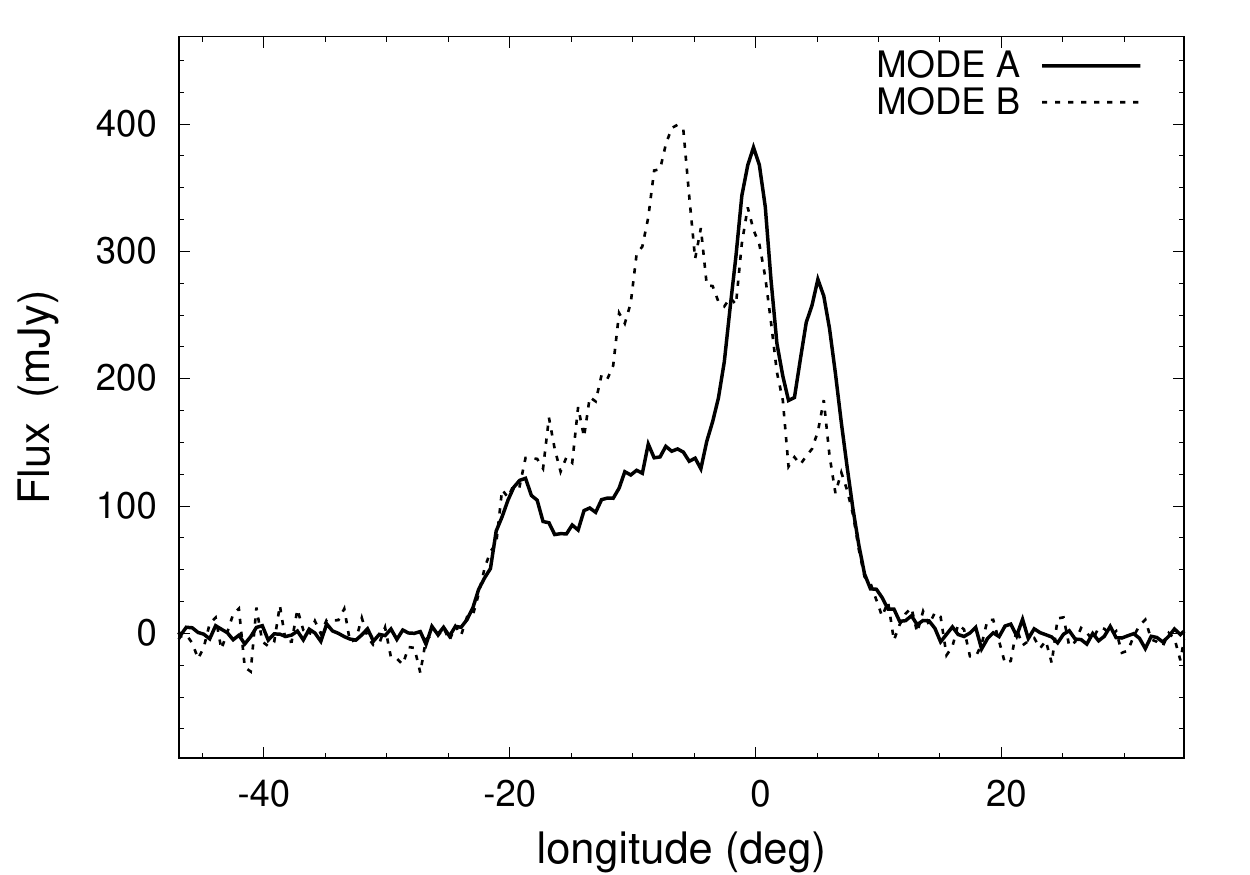}{0.3\textwidth}{(a)}
          \fig{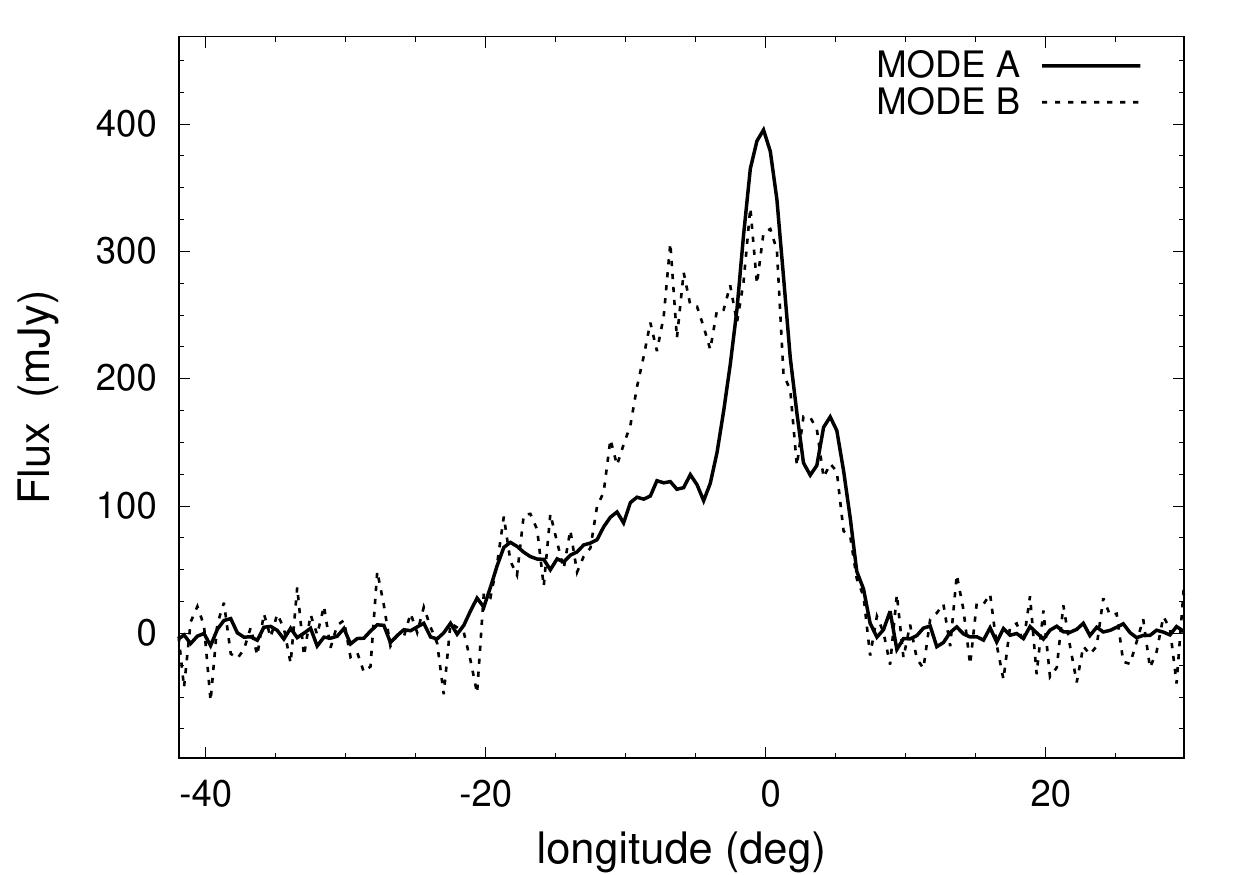}{0.3\textwidth}{(b)}
         }
\caption{The average profiles of the emission modes in PSR B0844-35, (a) 325 
MHz and (b) 610 MHz.
\label{fig:appB0844}}
\end{figure}

\begin{figure}
\gridline{\fig{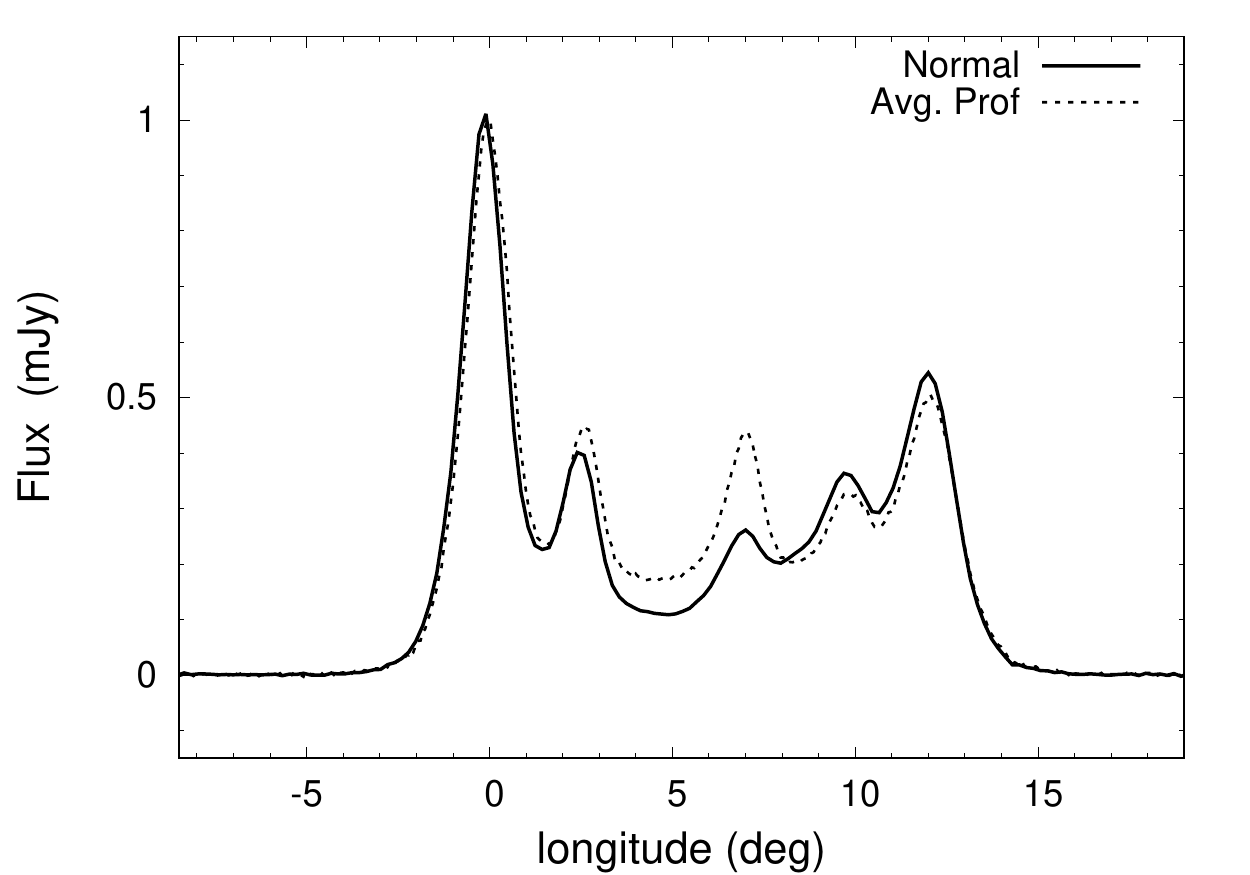}{0.3\textwidth}{(a)}
          \fig{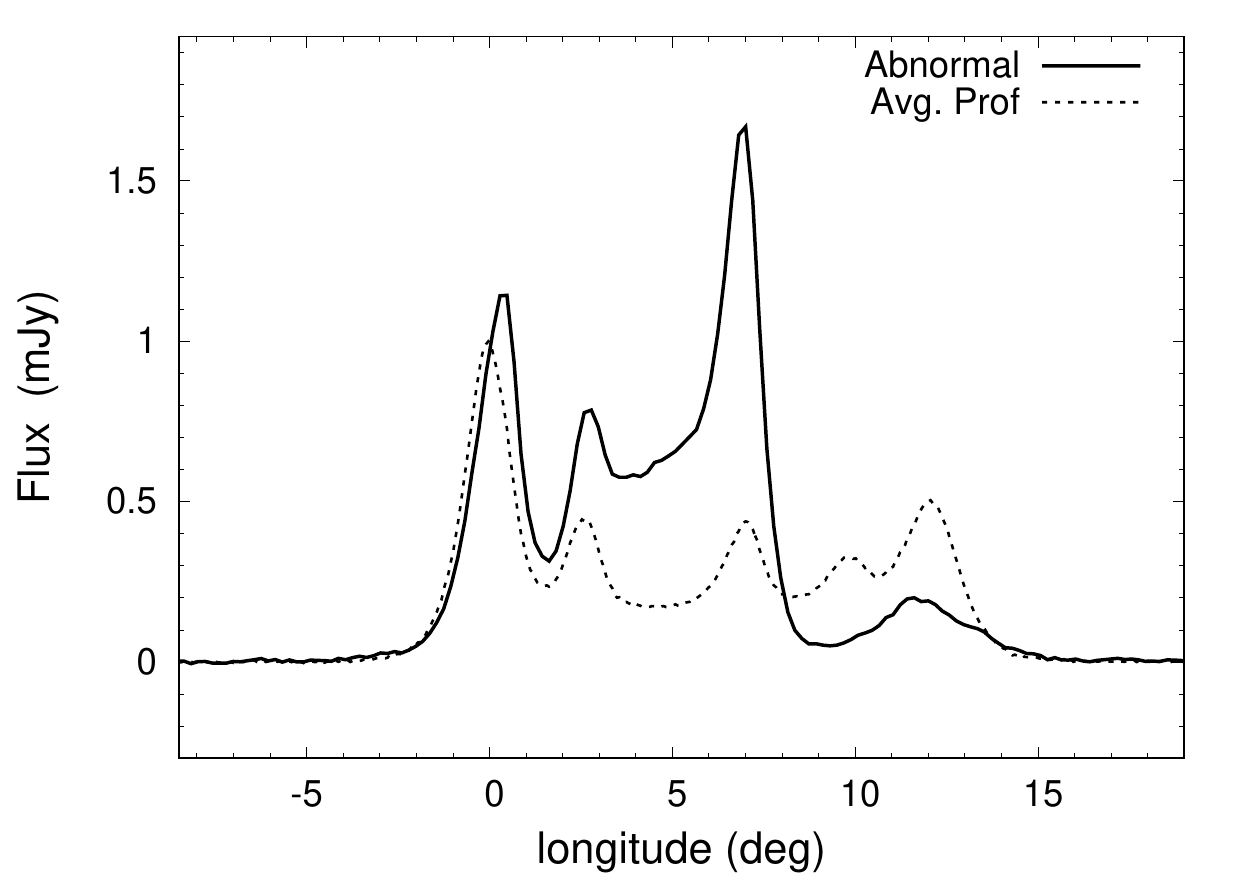}{0.3\textwidth}{(b)}
         }
\caption{The average profiles of the emission modes in PSR B1237+25 at 325 MHz,
(a) Normal mode and (b) Abnormal mode.
\label{fig:appB1237}}
\end{figure}

\begin{figure}
\gridline{\fig{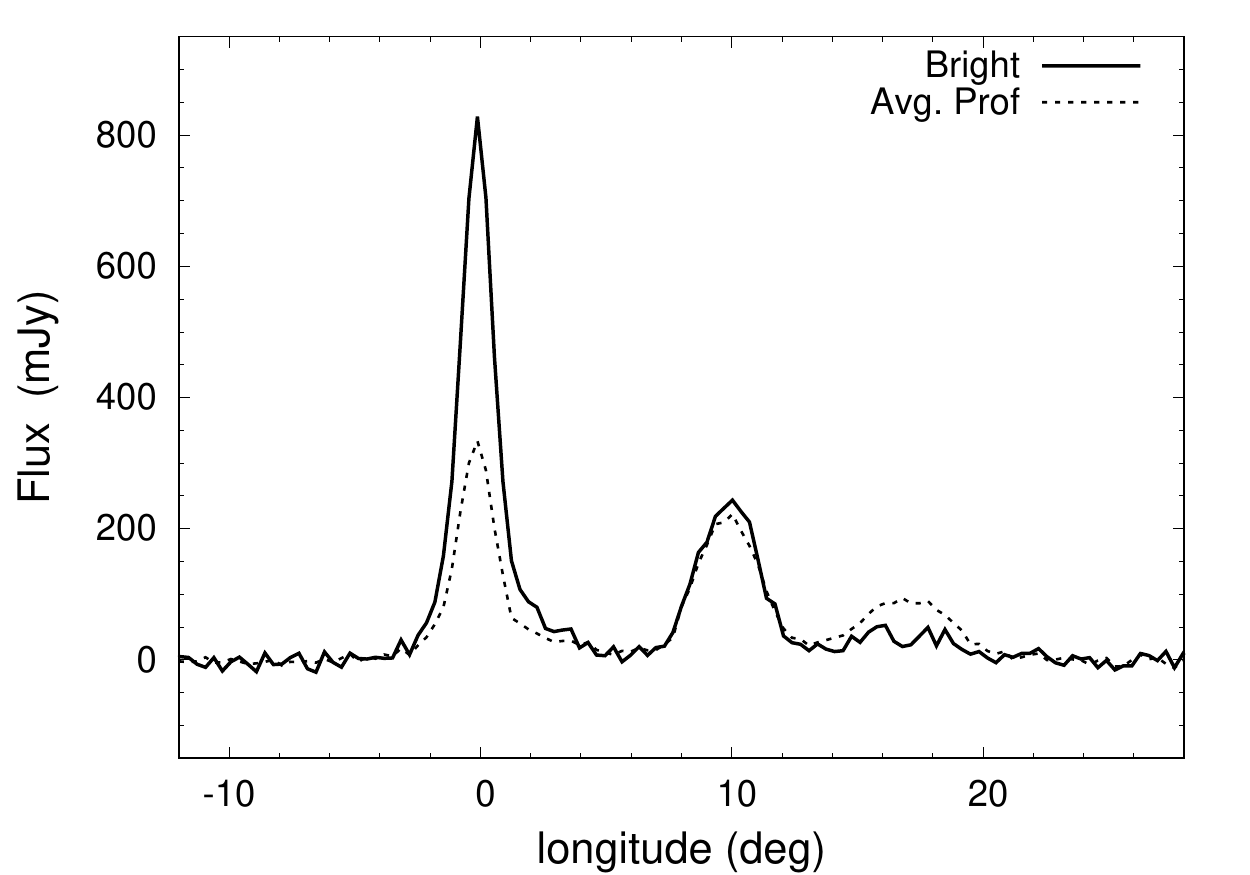}{0.3\textwidth}{(a)}
          \fig{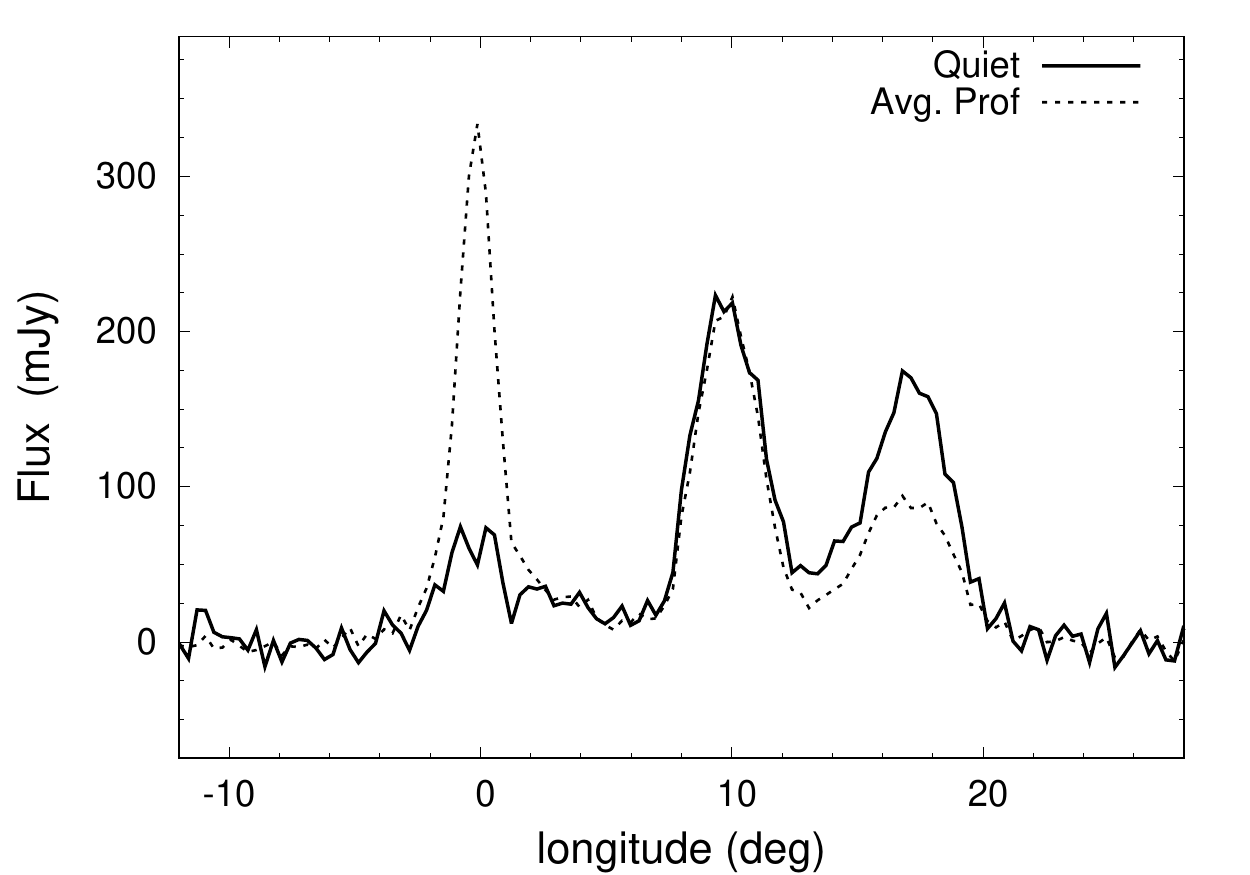}{0.3\textwidth}{(b)}
          \fig{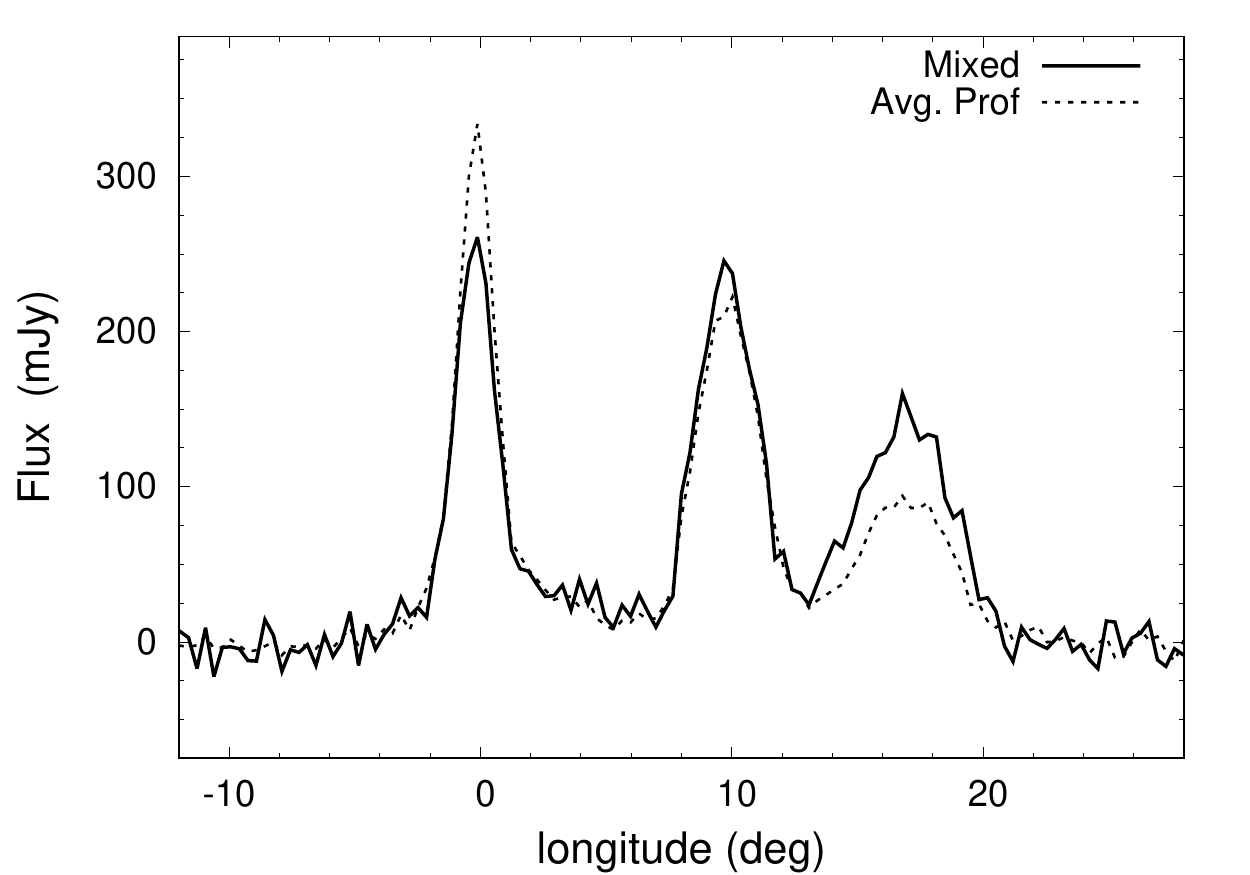}{0.3\textwidth}{(c)}
         }
\caption{The average profiles in PSR J1625--4048 at 610 MHz, (a) Bright mode, 
(b) Quiet mode, and (c) Mixed Mode.
\label{fig:appJ1625}}
\end{figure}

\begin{figure}
\gridline{\fig{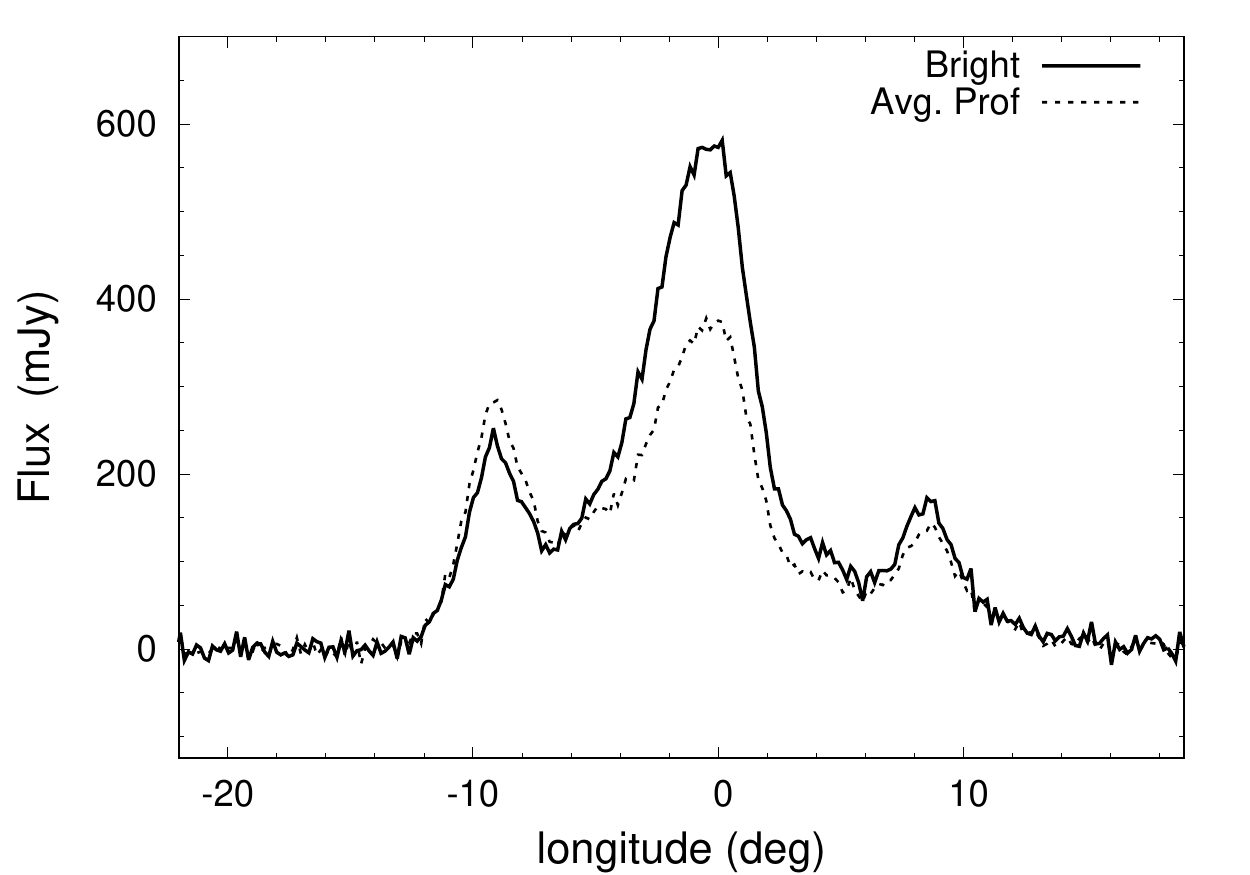}{0.3\textwidth}{(a)}
          \fig{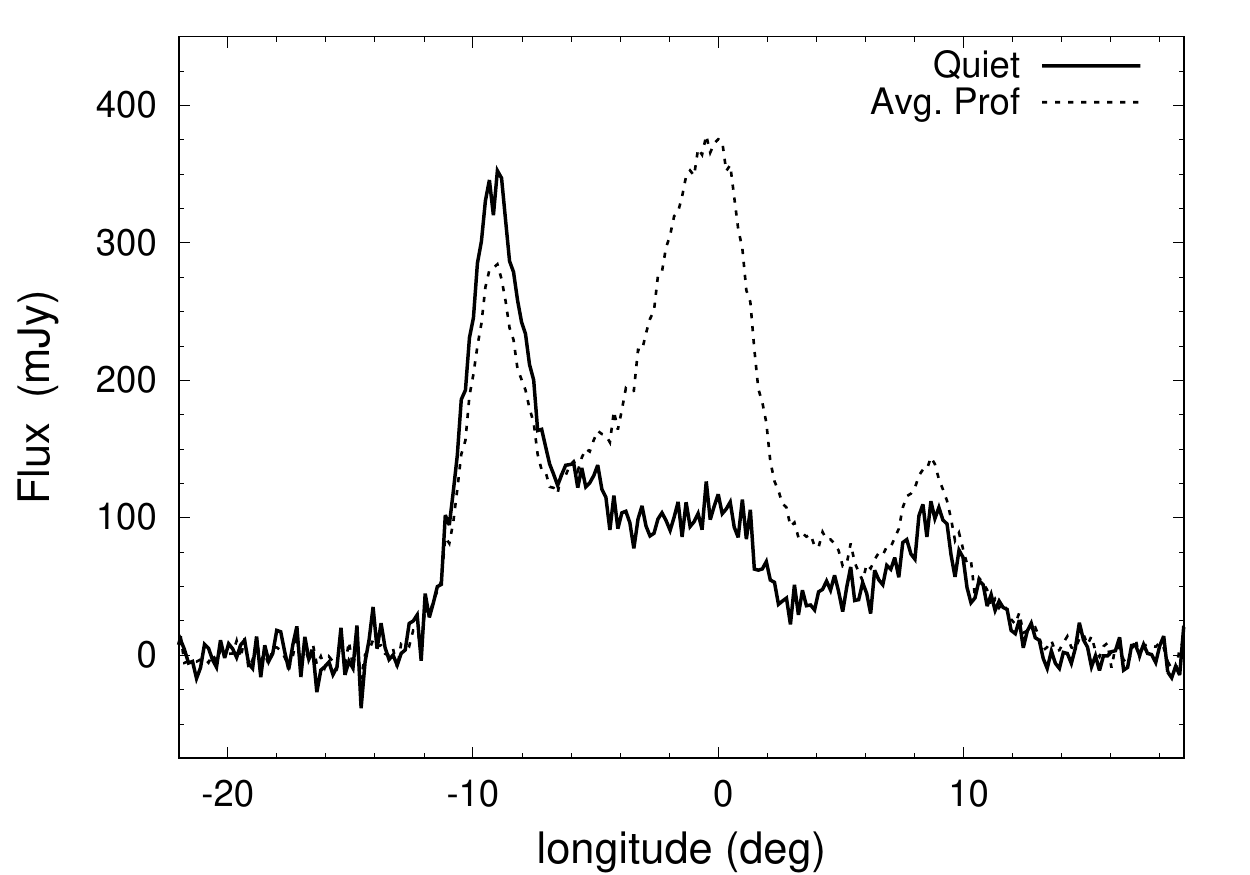}{0.3\textwidth}{(b)}
         }
\gridline{\fig{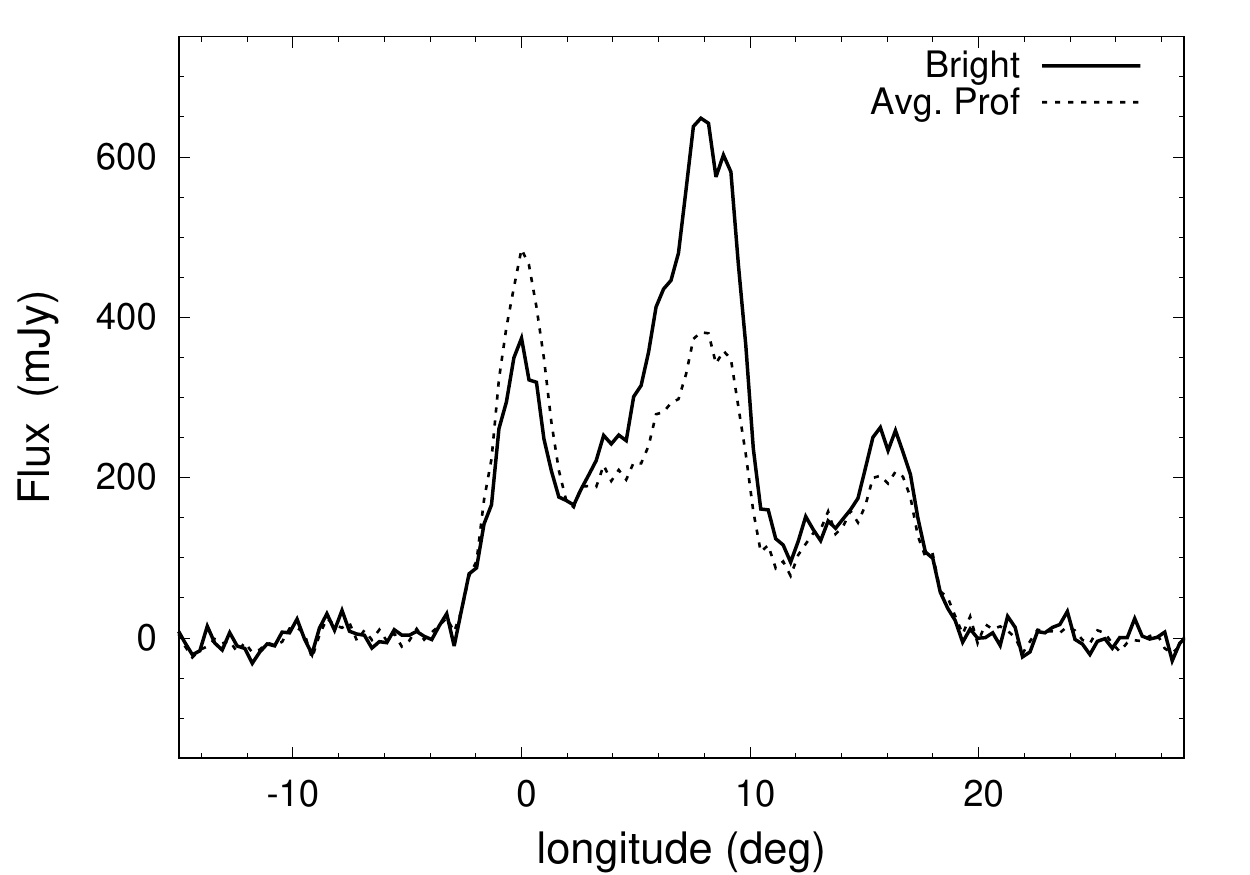}{0.3\textwidth}{(c)}
          \fig{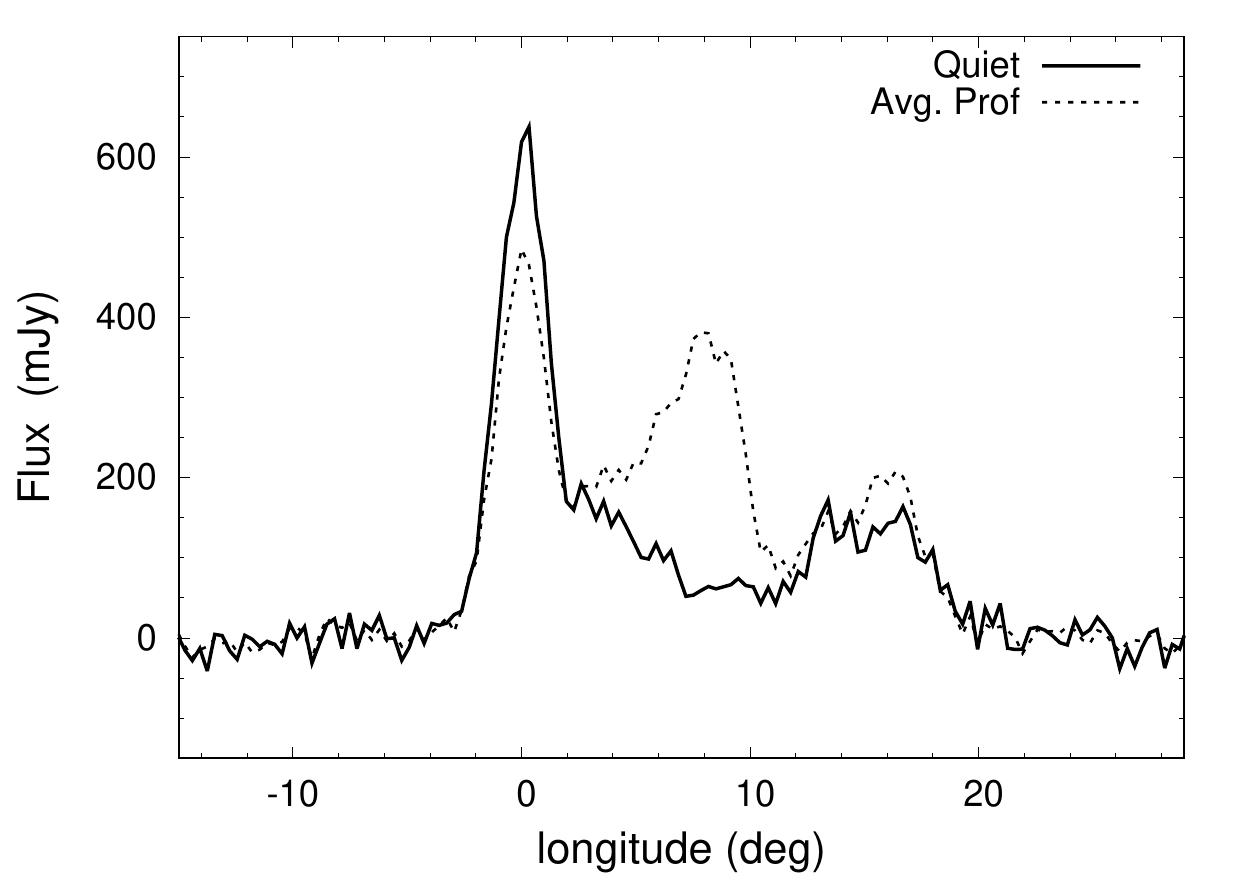}{0.3\textwidth}{(d)}
         }
\caption{The average profiles in PSR B1758--29, (a) Bright mode at 325 MHz,
(b) Quiet mode at 325 MHz, (c) Bright mode at 610 MHz and (d) Quiet mode at 610
MHz.
\label{fig:appB1758}}
\end{figure}

\begin{figure}
\gridline{\fig{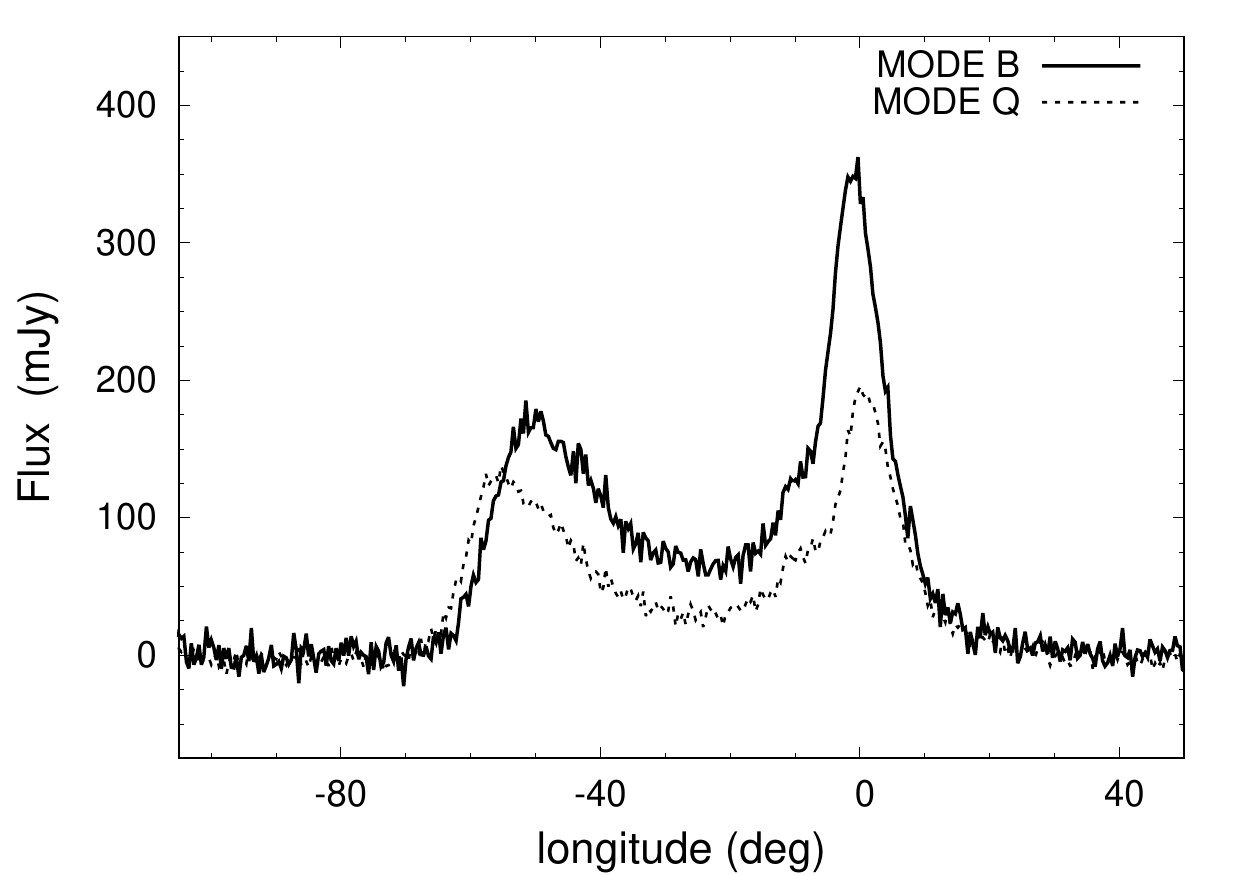}{0.3\textwidth}{}
         }
\caption{The average profiles of PSR B1839--04 at 610 MHz during the Bright and
Quiet modes.
\label{fig:appB1839}}
\end{figure}

\begin{figure}
\gridline{\fig{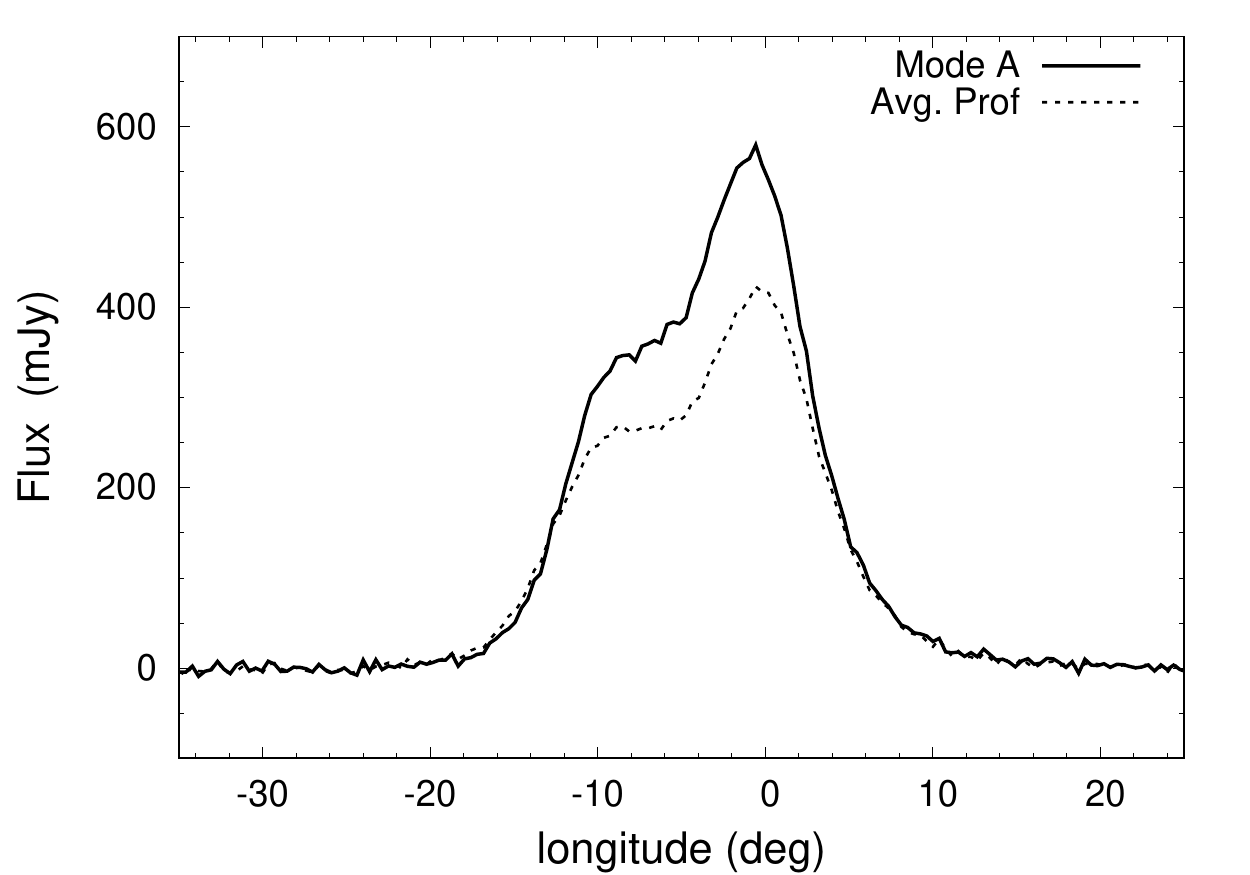}{0.3\textwidth}{(a)}
          \fig{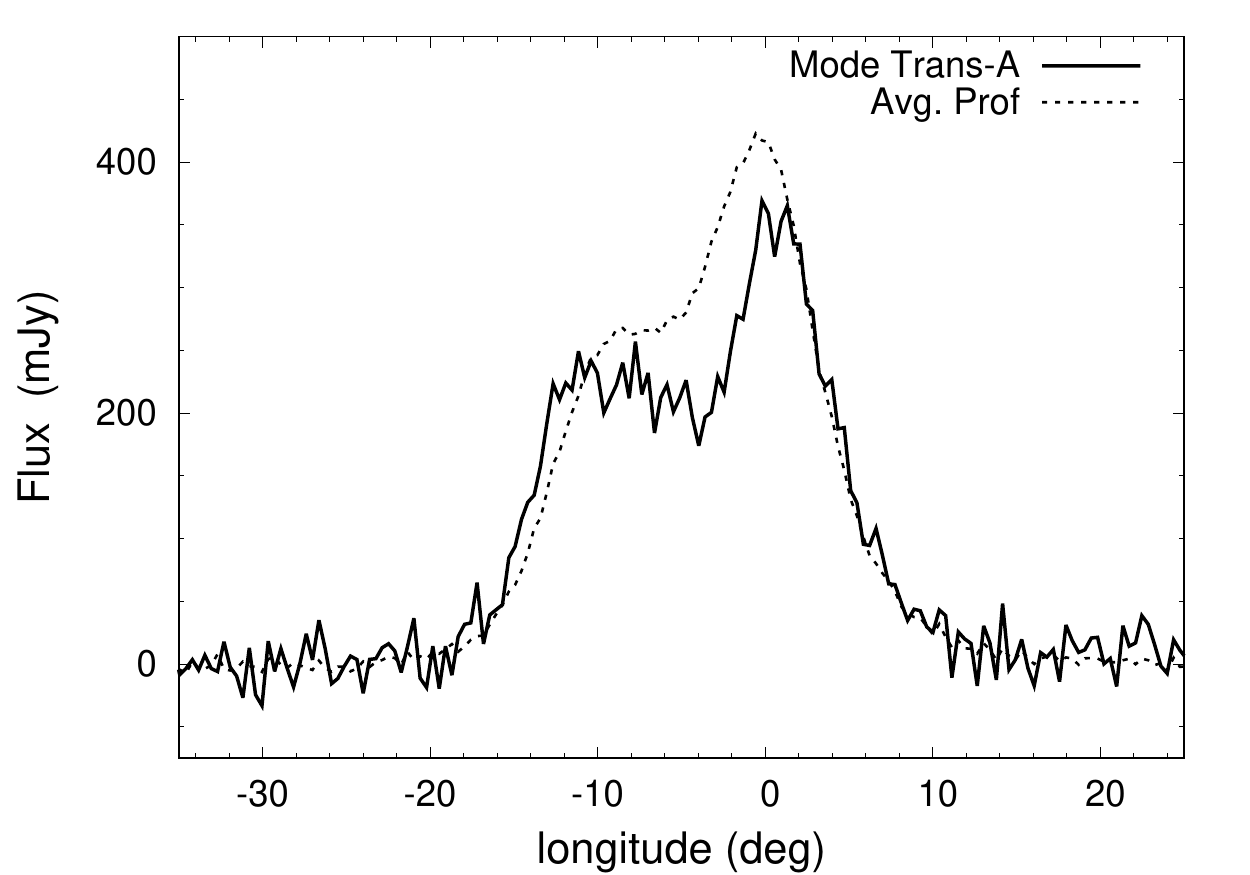}{0.3\textwidth}{(b)}
          \fig{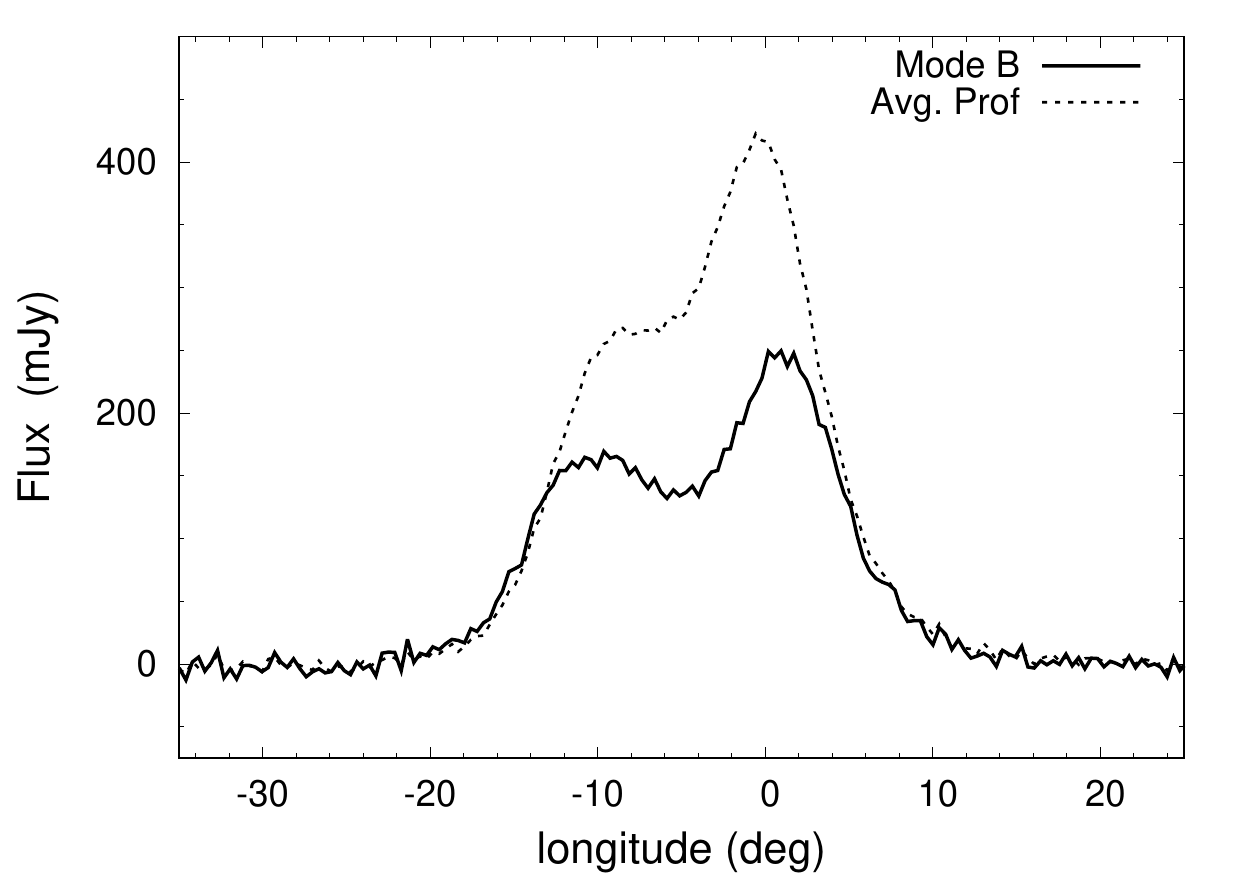}{0.3\textwidth}{(c)}
         }
\gridline{\fig{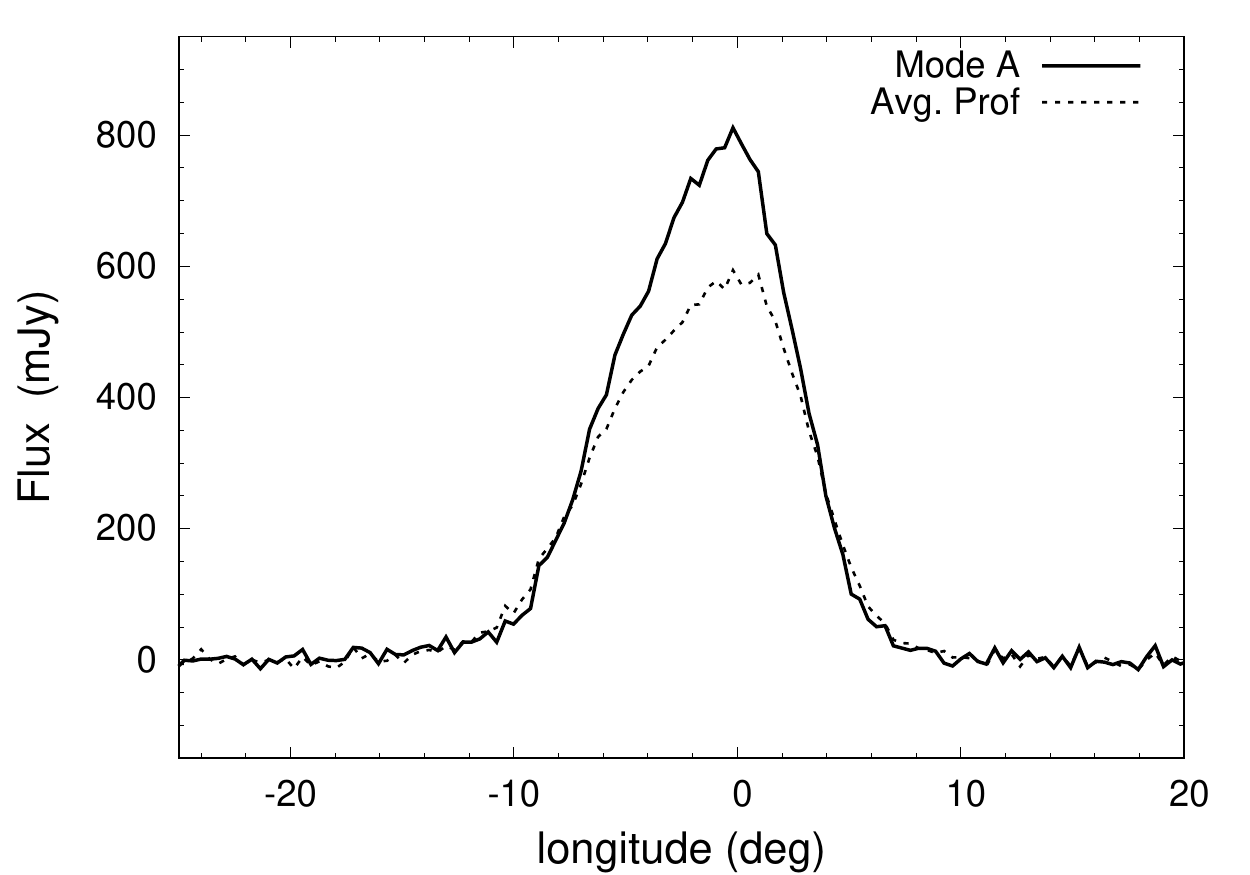}{0.3\textwidth}{(d)}
          \fig{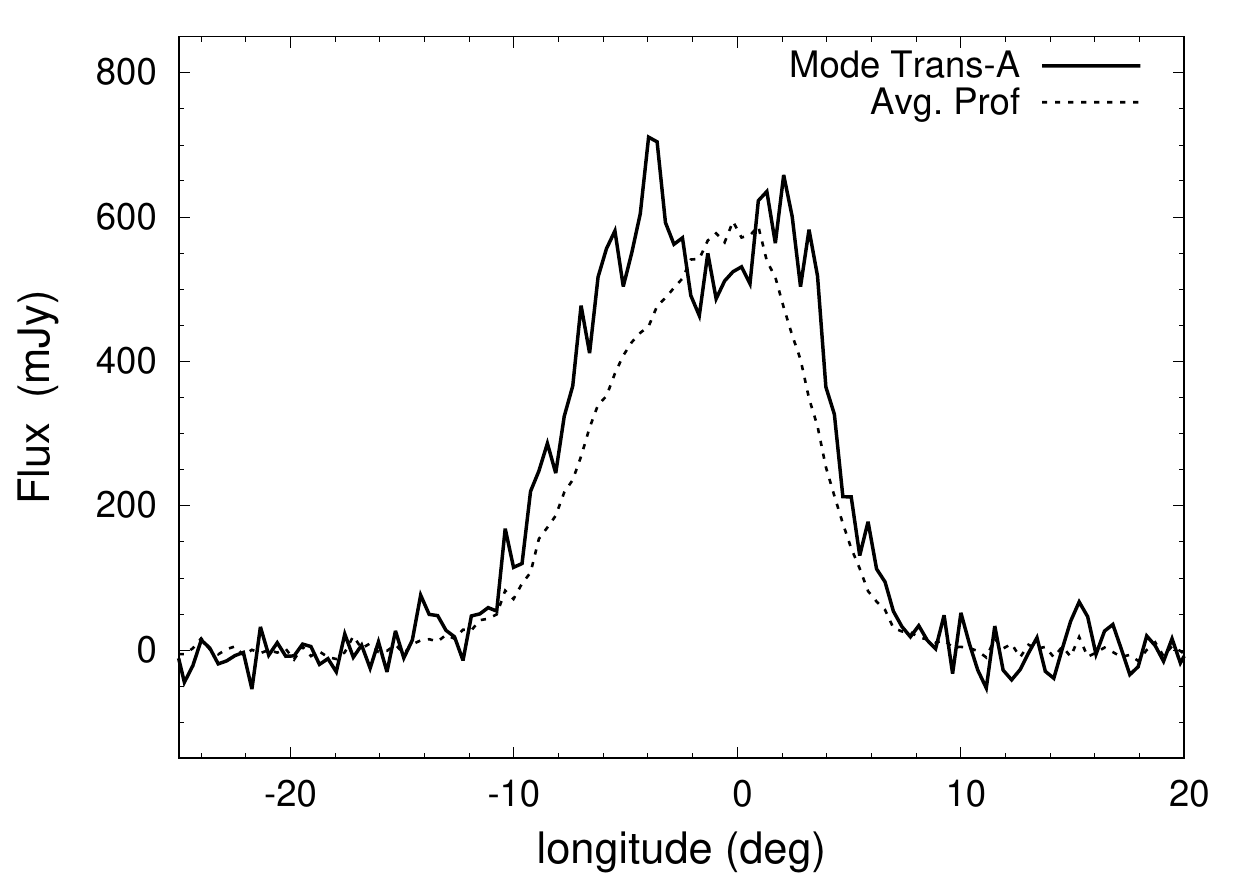}{0.3\textwidth}{(e)}
          \fig{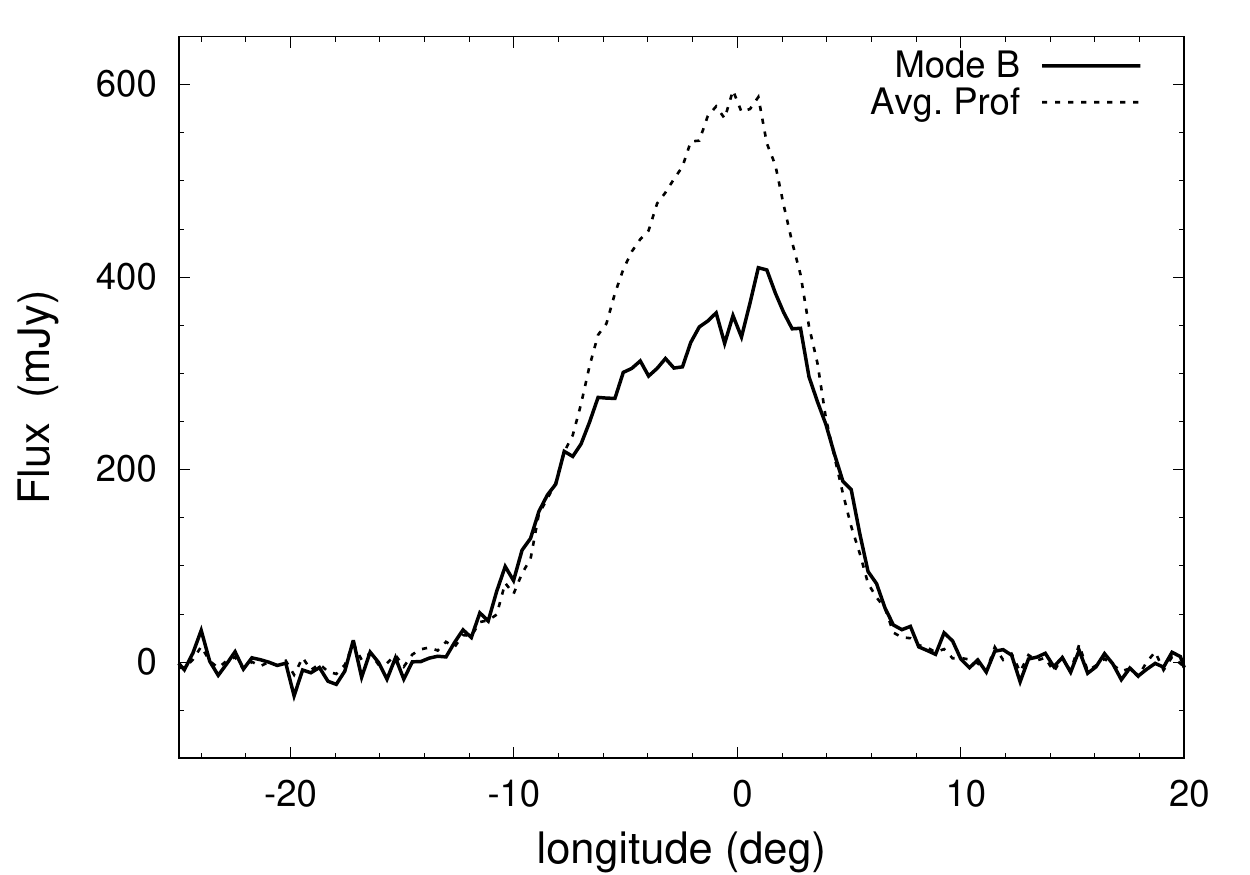}{0.3\textwidth}{(f)}
         }
\caption{The average profiles in PSR B1819--22, (a) Mode A at 325 MHz, (b) Mode
A-Trans at 325 MHz, (c) Mode B at 325 MHz, (d) Mode A at 610 MHz, (e) Mode 
A-Trans at 610 MHz, and (f) Mode B at 610 MHz.
\label{fig:appB1819}}
\end{figure}

\begin{figure}
\gridline{\fig{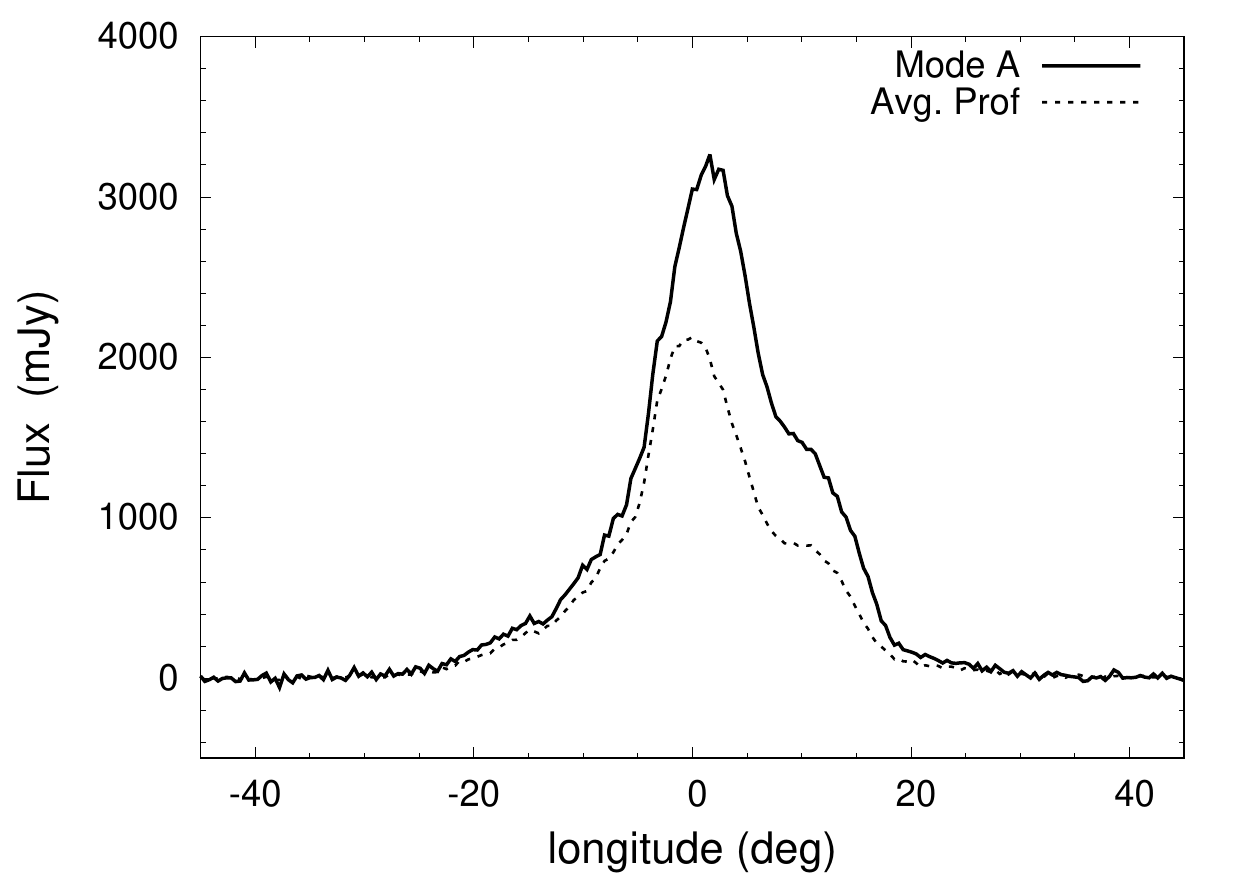}{0.3\textwidth}{(a)}
          \fig{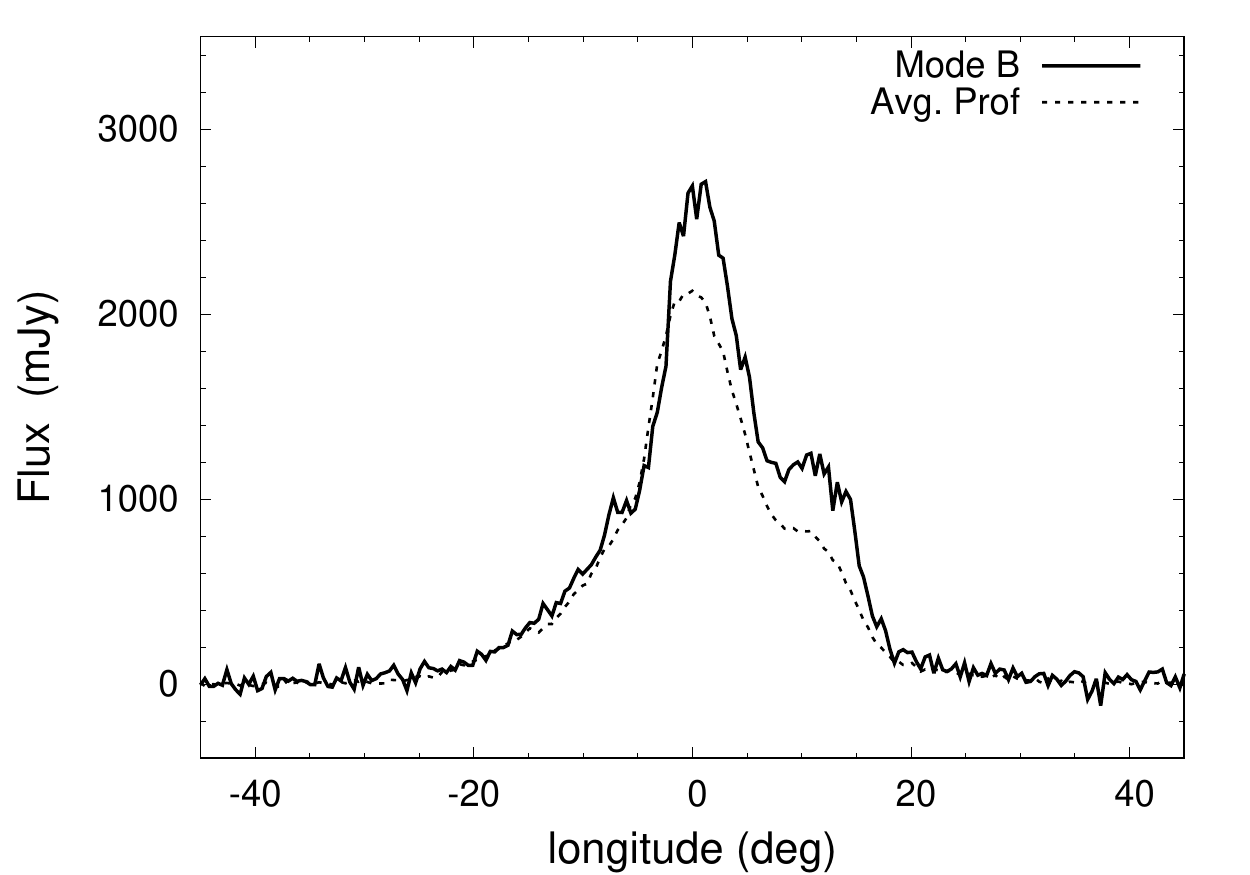}{0.3\textwidth}{(b)}
          \fig{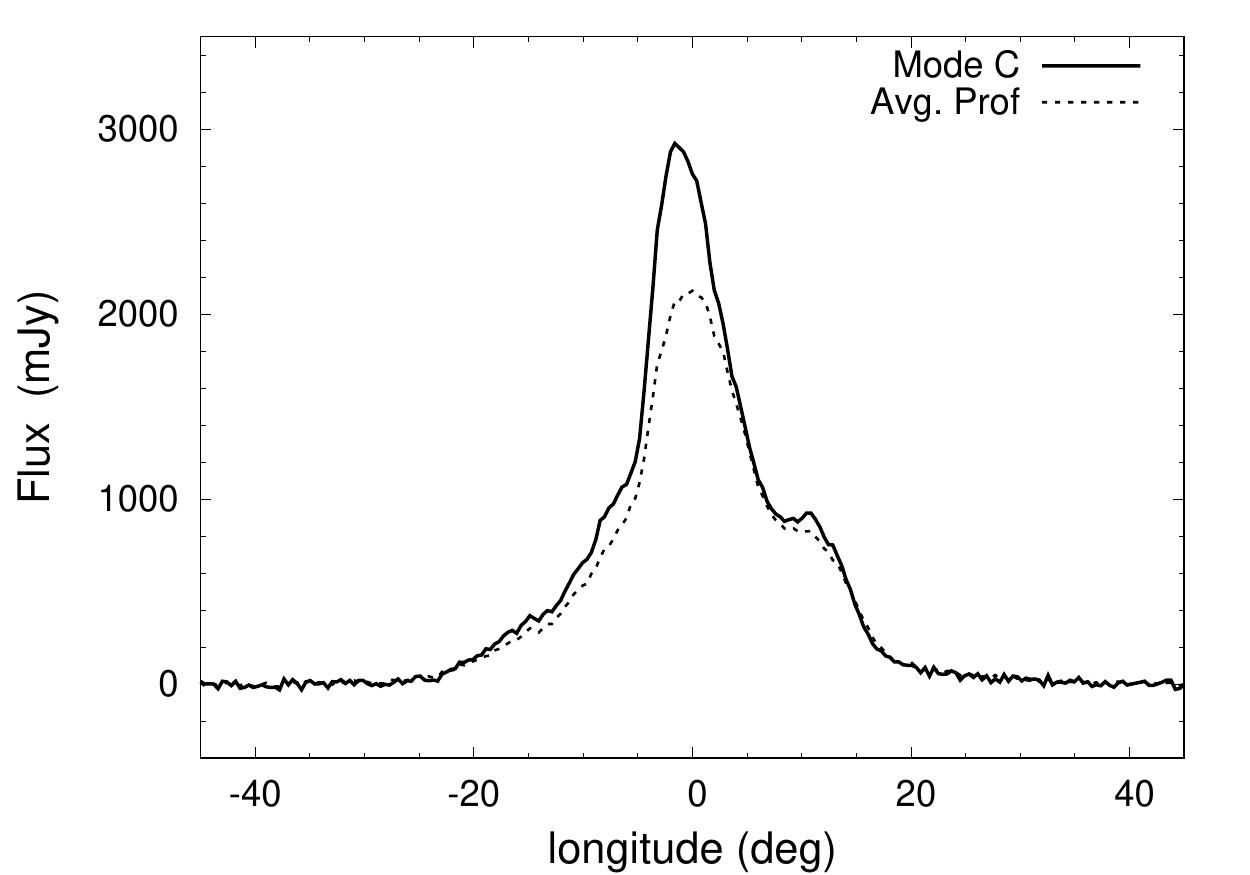}{0.3\textwidth}{(c)}
         }
\gridline{\fig{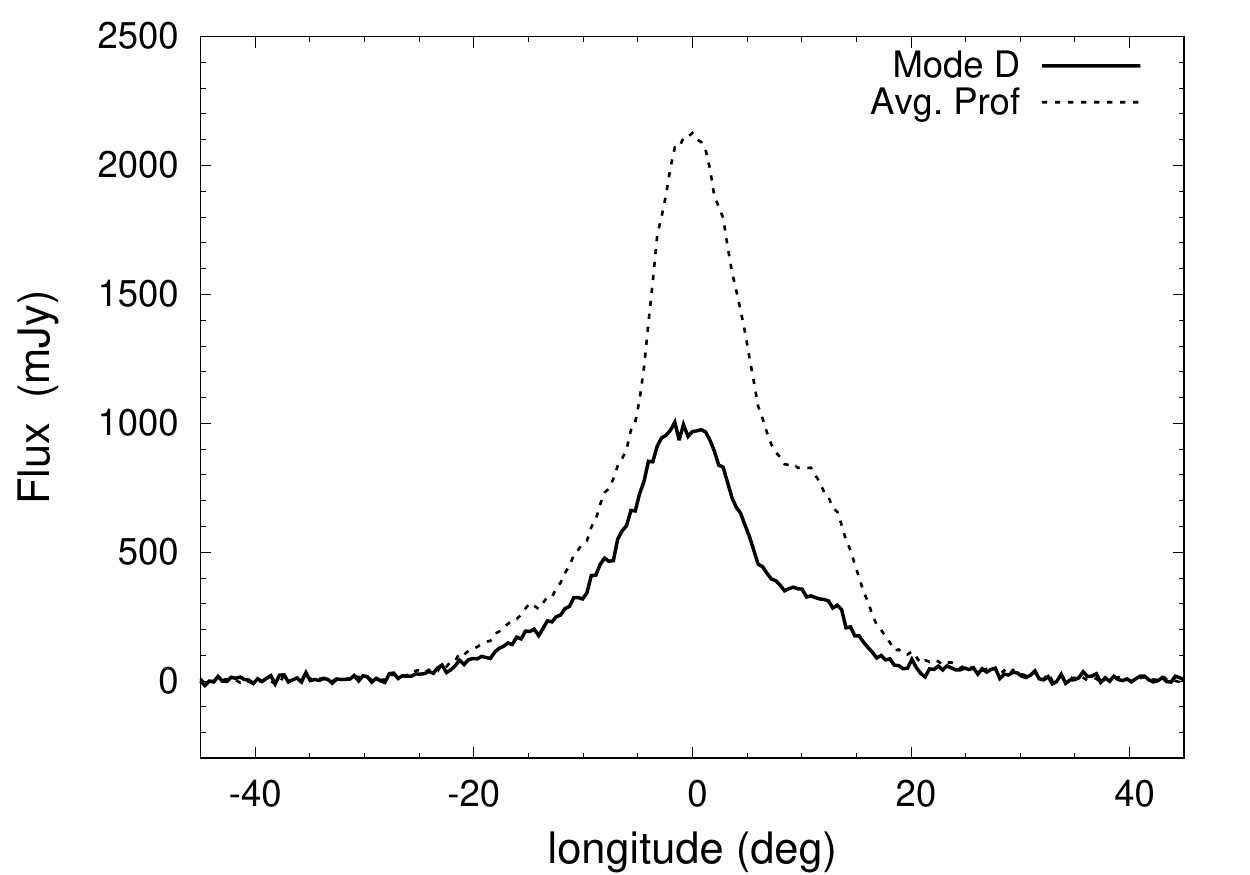}{0.3\textwidth}{(d)}
          \fig{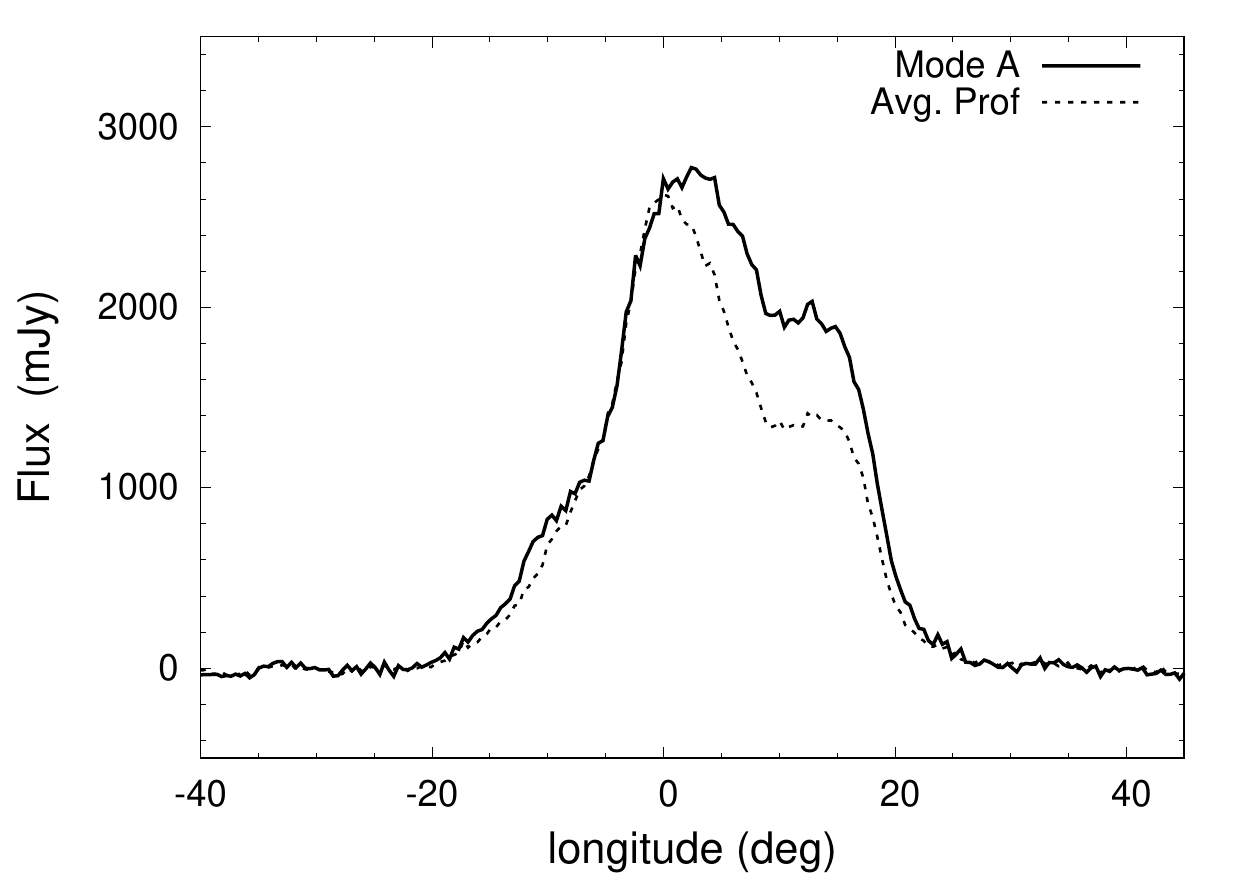}{0.3\textwidth}{(e)}
          \fig{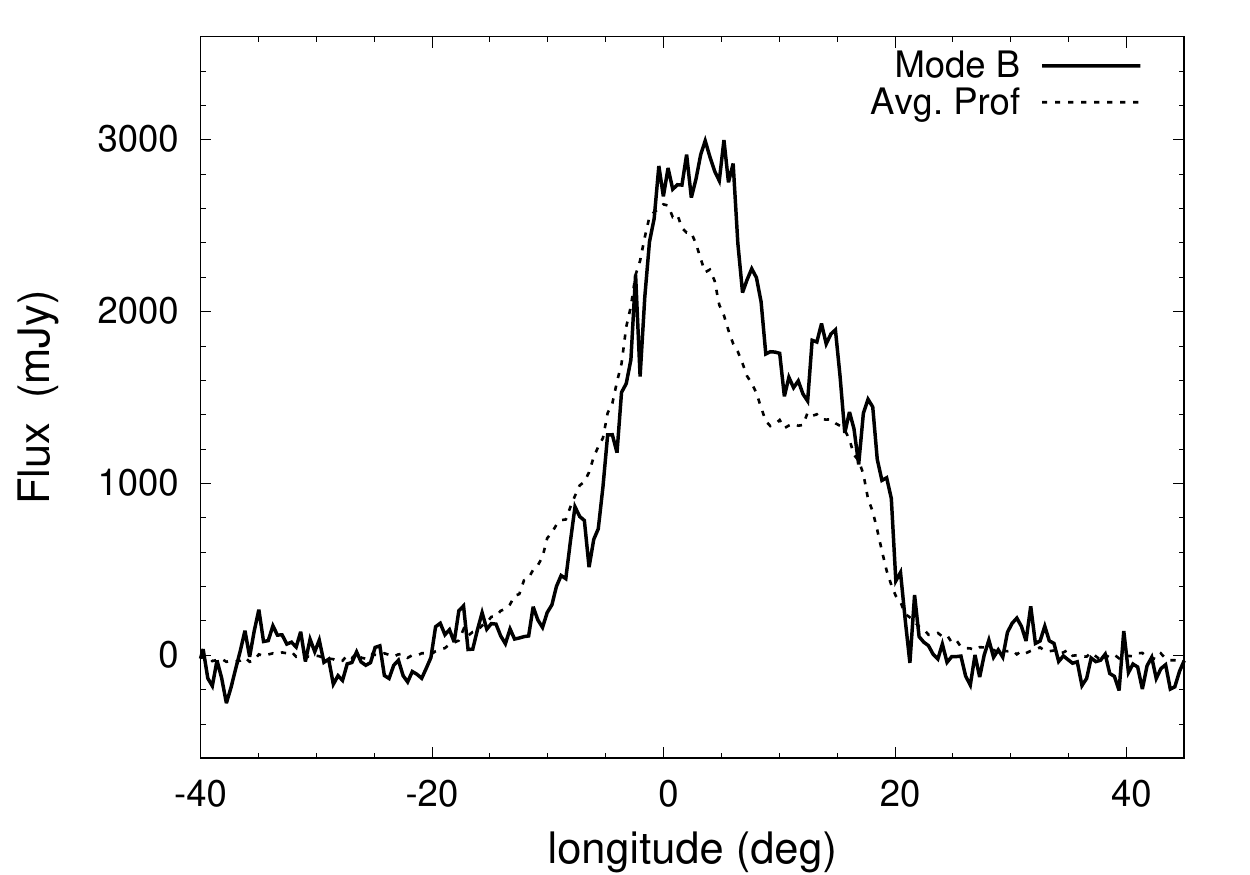}{0.3\textwidth}{(f)}
         }
\gridline{\fig{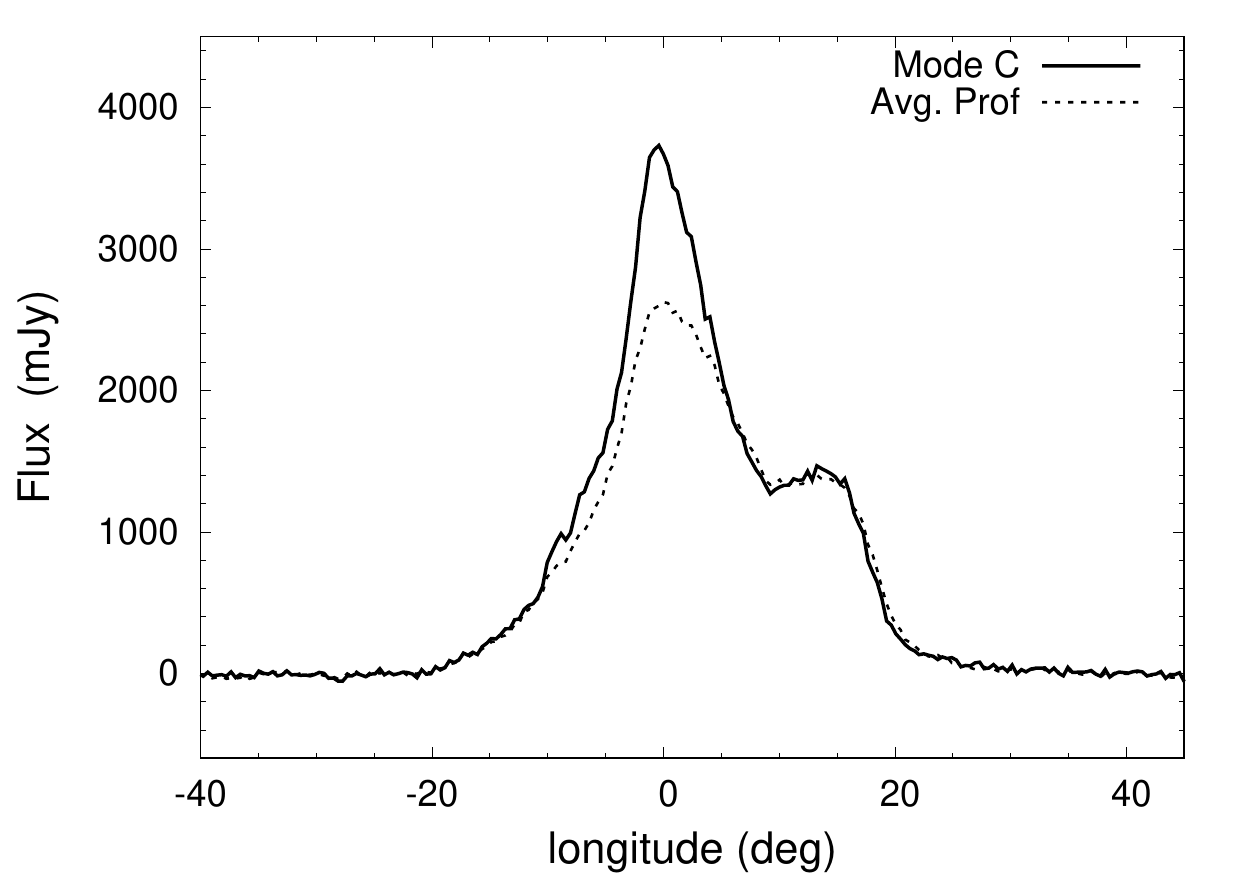}{0.3\textwidth}{(g)}
          \fig{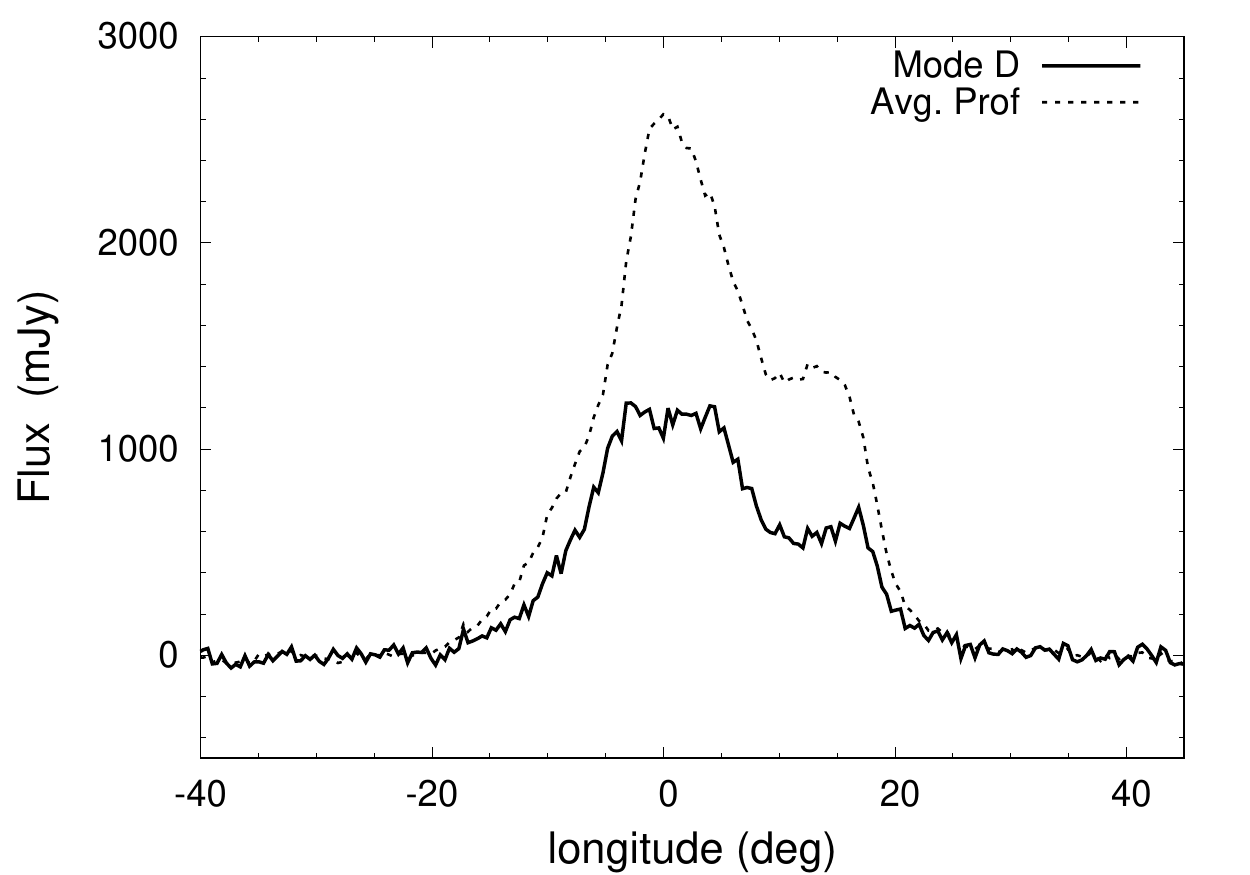}{0.3\textwidth}{(h)}
         }

\caption{The average profiles in PSR B1944+17, (a) Mode A at 325 MHz, (b) Mode 
B at 325 MHz, (c) Mode C at 325 MHz, (d) Mode D at 325 MHz, (e) Mode A at 610 
MHz, (f) Mode B at 610 MHz, (g) Mode C at 610 MHz, and (h) Mode D at 610 MHz.
\label{fig:appB1944}}
\end{figure}

\begin{figure}
\gridline{\fig{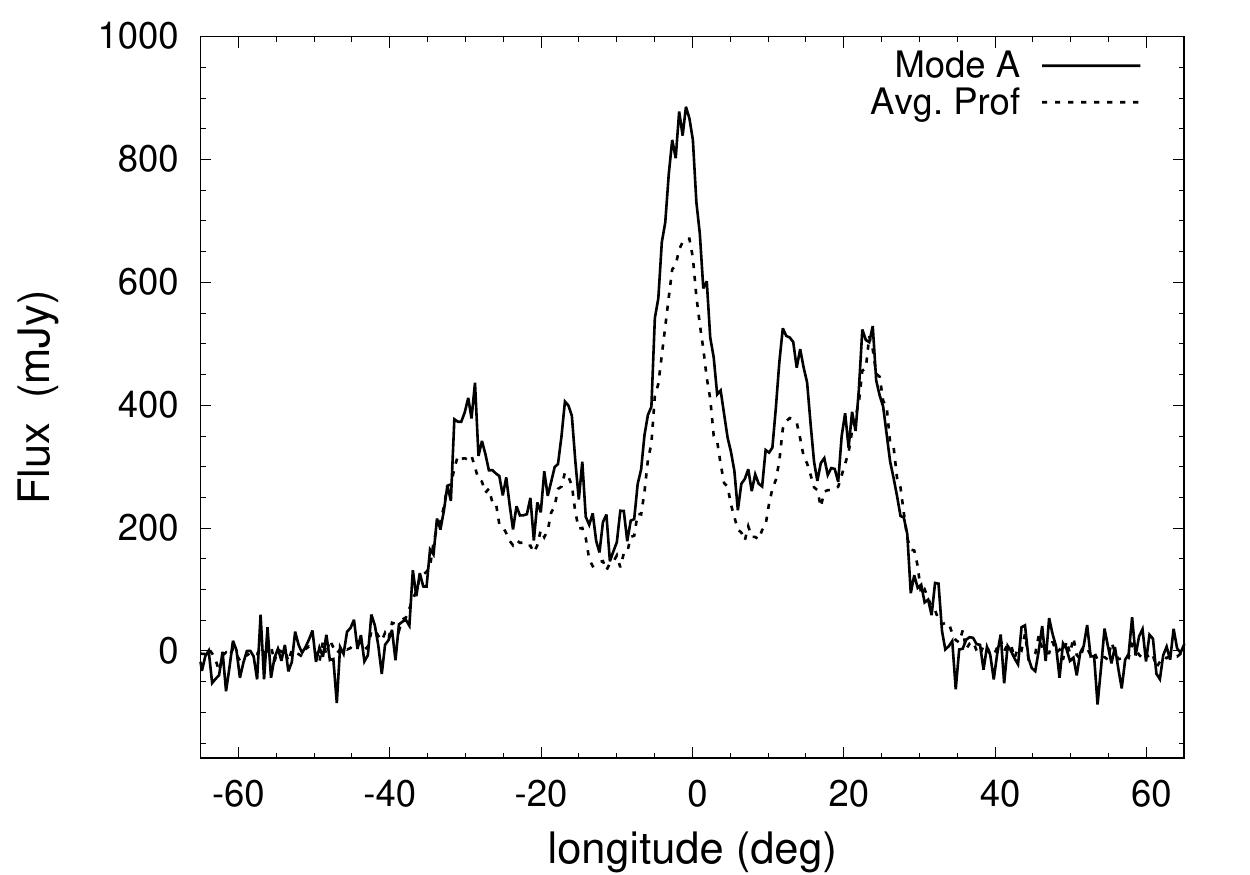}{0.3\textwidth}{(a)}
          \fig{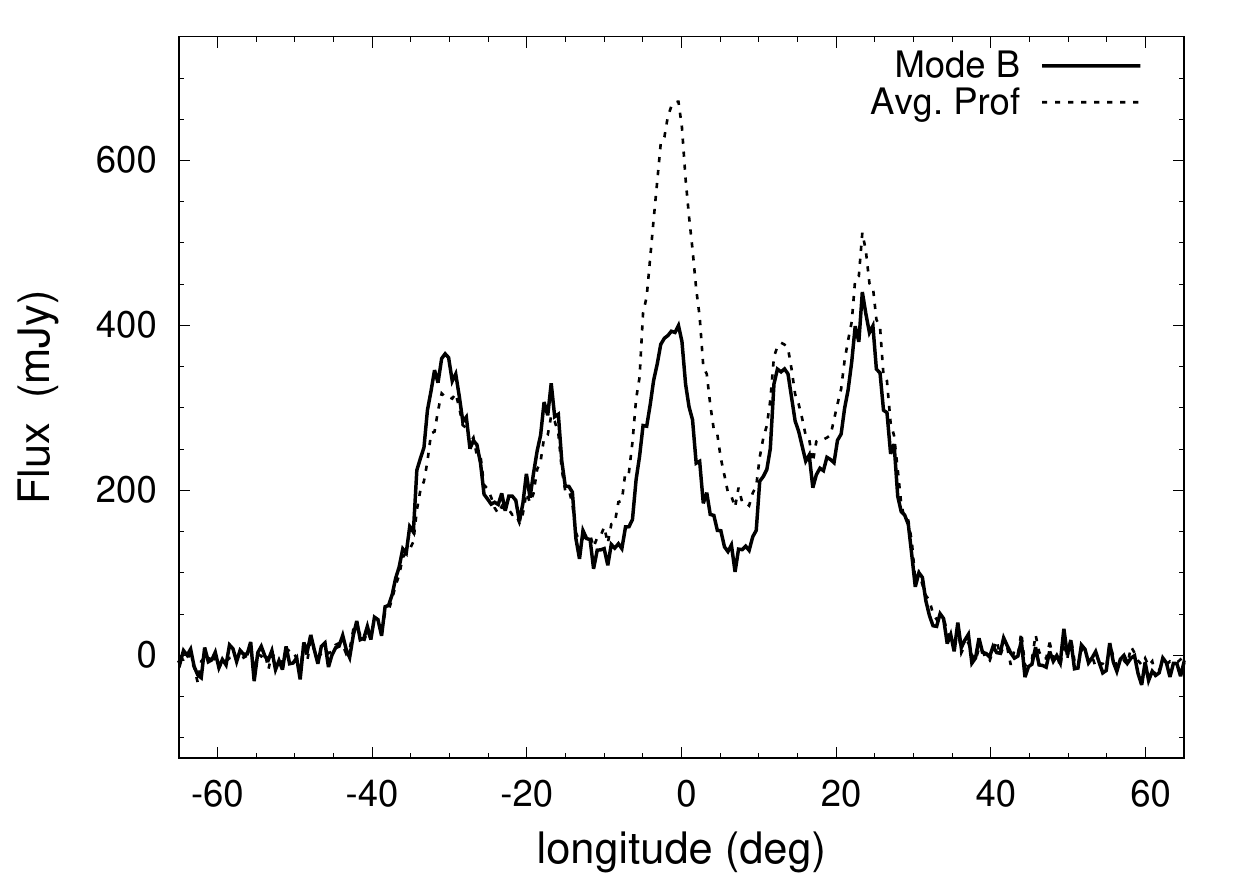}{0.3\textwidth}{(b)}
          \fig{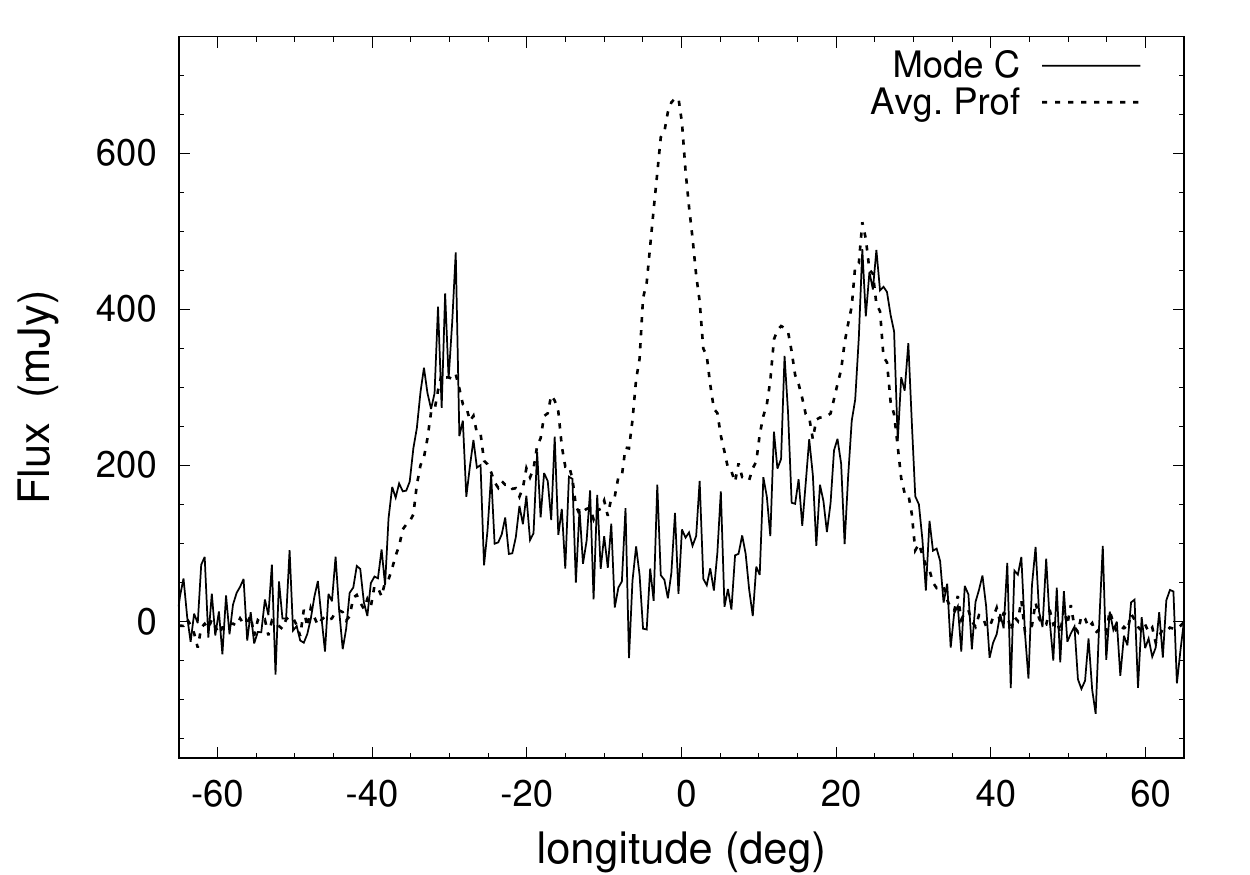}{0.3\textwidth}{(c)}
         }
\gridline{\fig{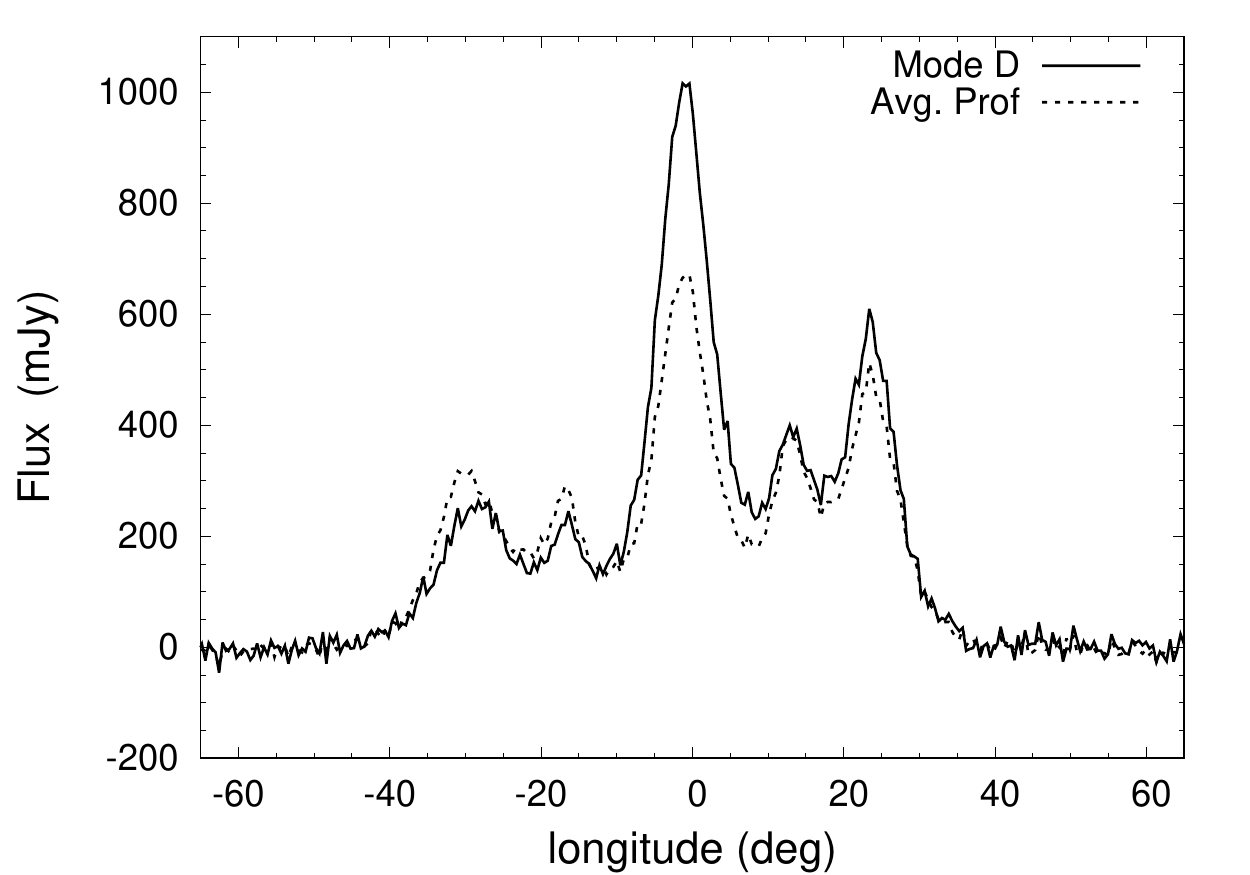}{0.3\textwidth}{(d)}
          \fig{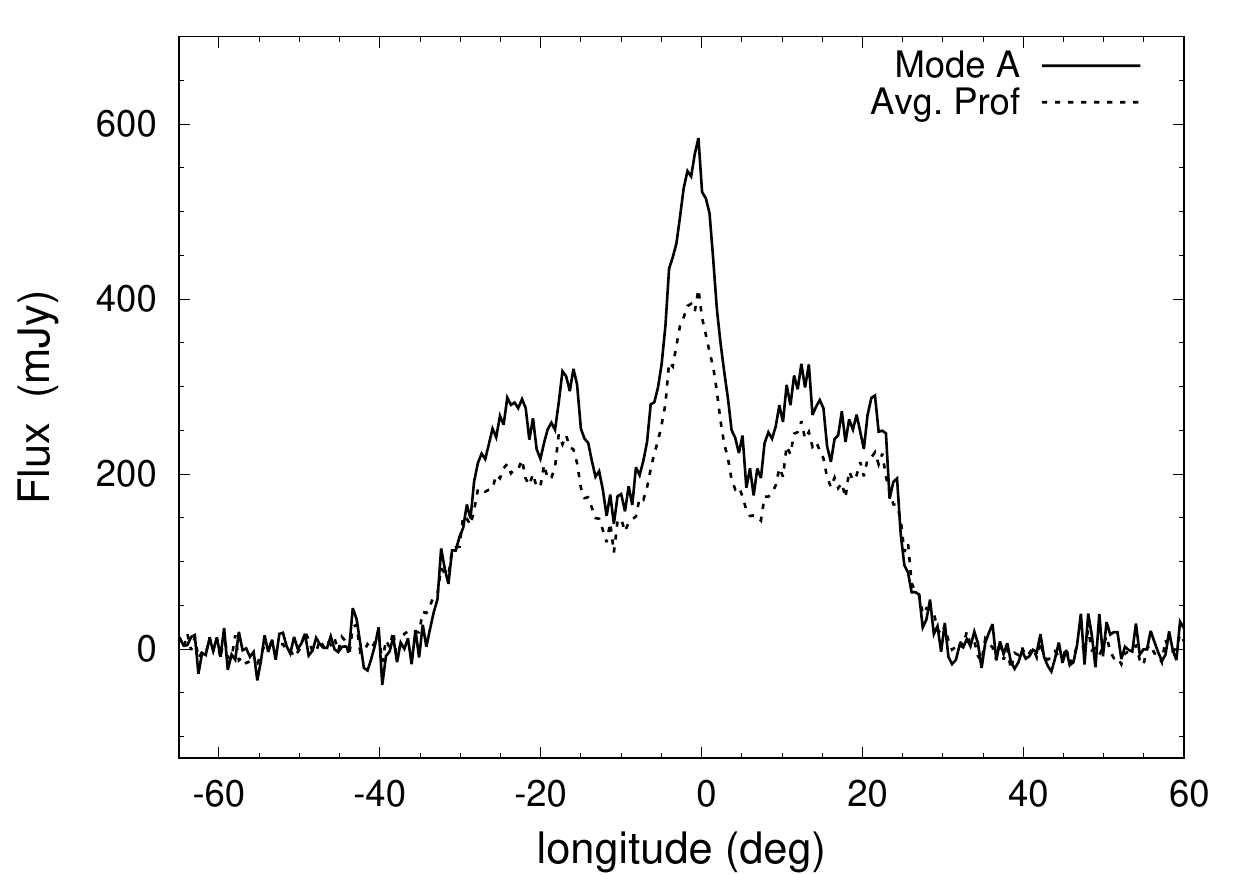}{0.3\textwidth}{(e)}
          \fig{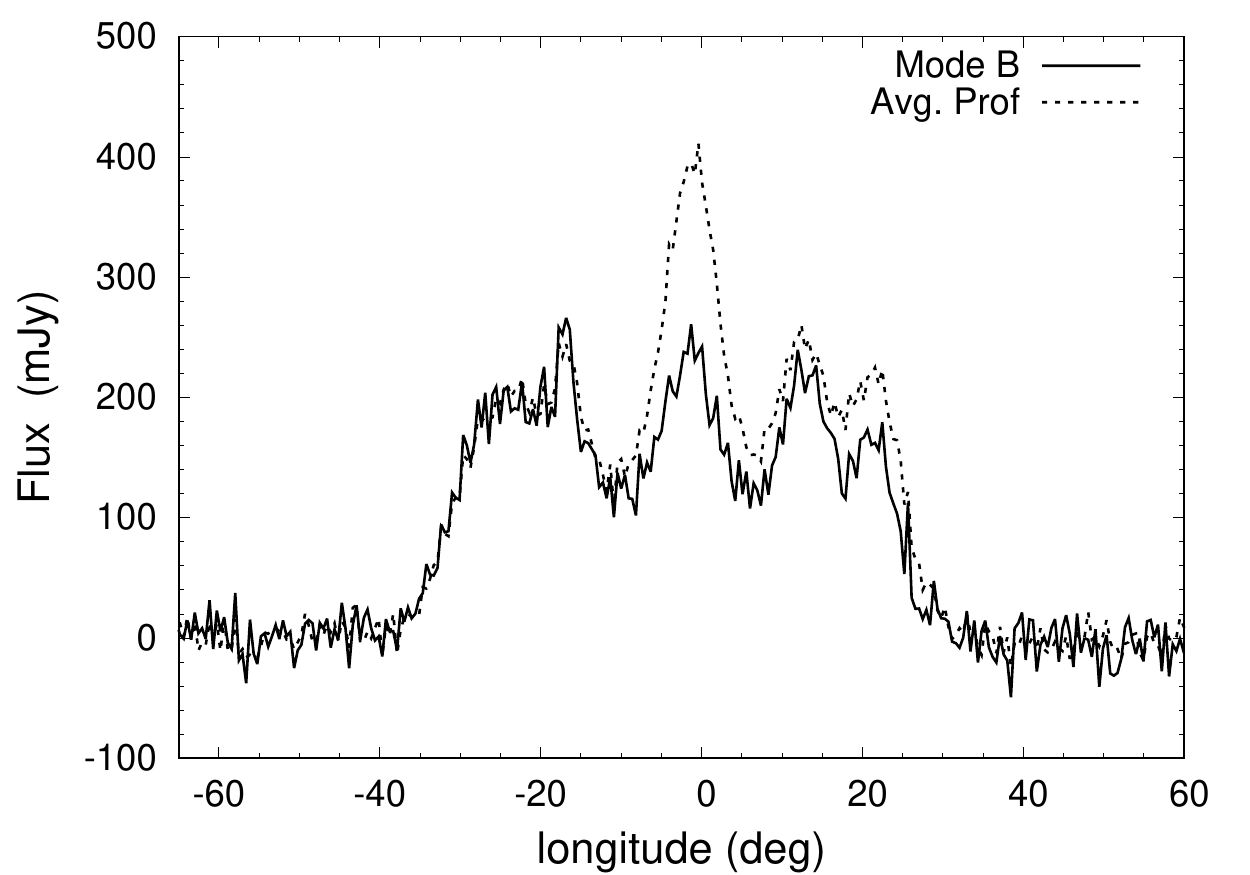}{0.3\textwidth}{(f)}
         }
\gridline{\fig{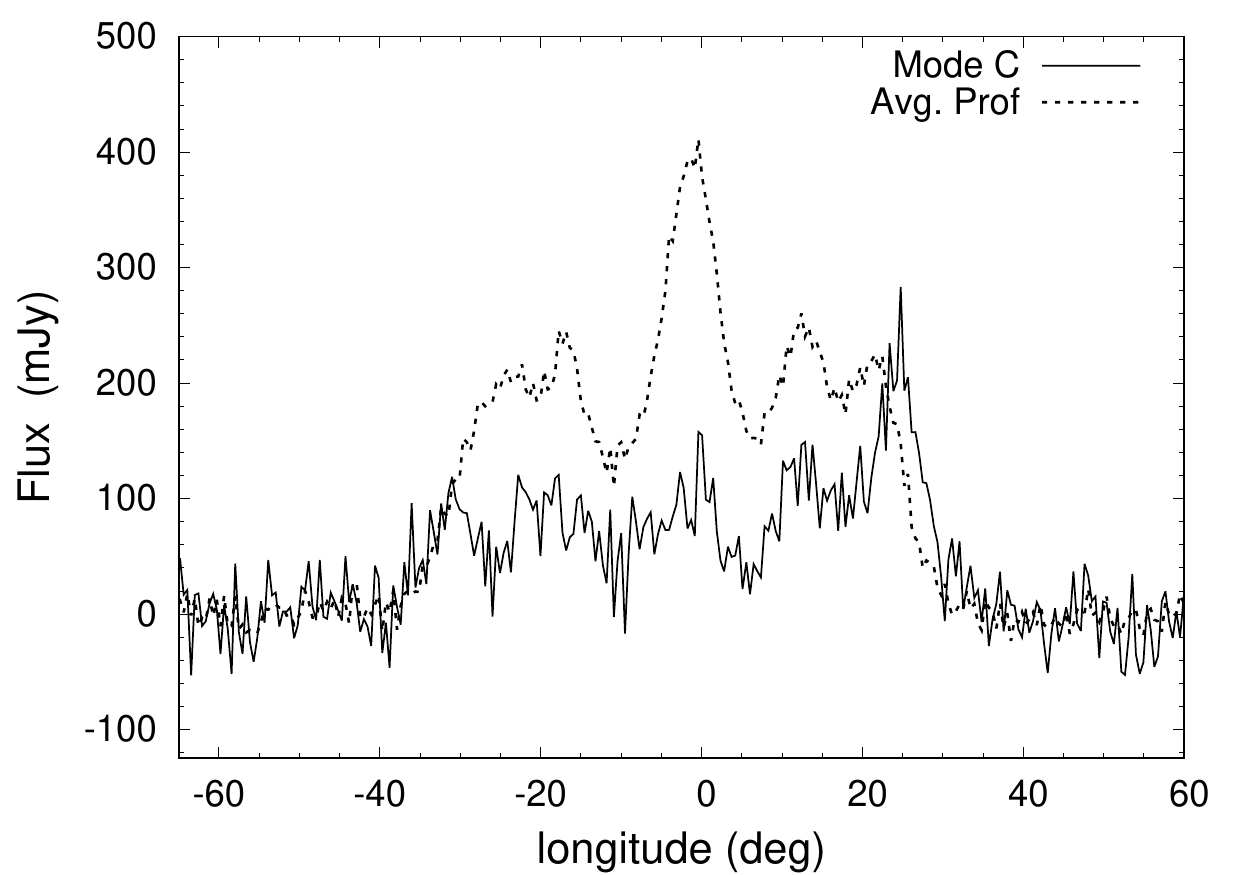}{0.3\textwidth}{(g)}
          \fig{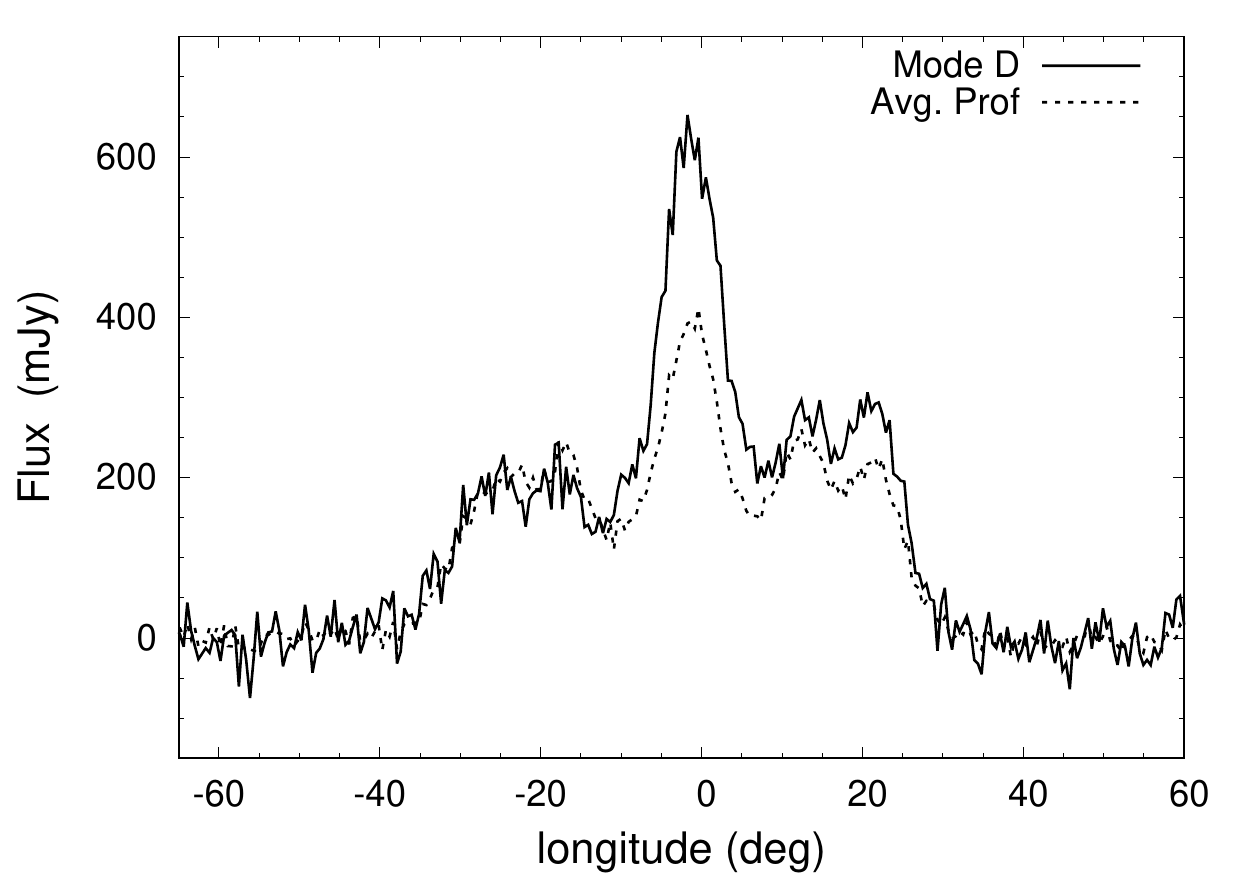}{0.3\textwidth}{(h)}
         }

\caption{The average profiles in PSR B2003--08, (a) Mode A at 325 MHz, (b) Mode
B at 325 MHz, (c) Mode C at 325 MHz, (d) Mode D at 325 MHz, (e) Mode A at 610
MHz, (f) Mode B at 610 MHz, (g) Mode C at 610 MHz, and (h) Mode D at 610 MHz.
\label{fig:appB2003}}
\end{figure}

\begin{figure}
\gridline{\fig{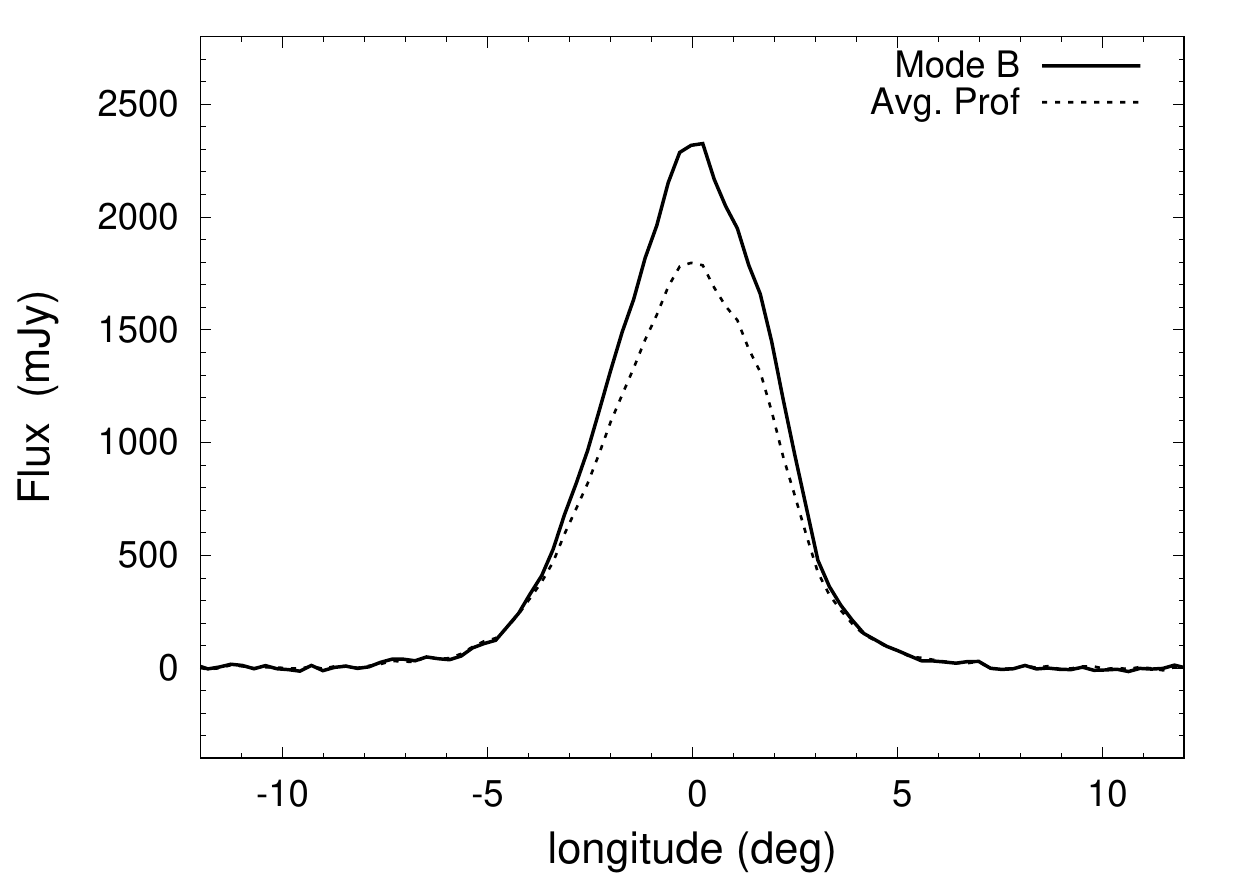}{0.3\textwidth}{(a)}
          \fig{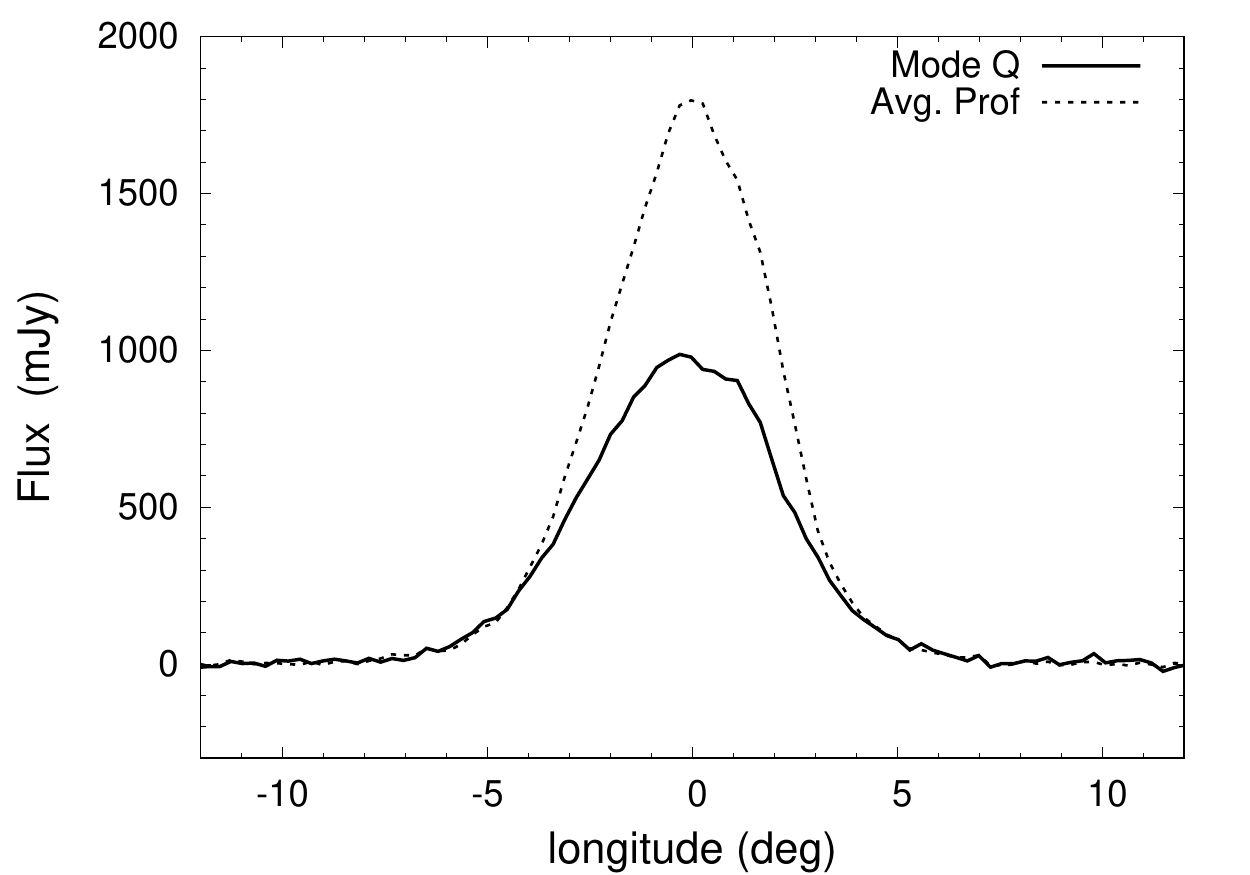}{0.3\textwidth}{(b)}
         }
\caption{The average profiles of PSR B2303+30 at 325 MHz during the (a) Bright 
and (b) Quiet modes.
\label{fig:appB2303}}
\end{figure}

\end{document}